\documentclass[preprint,12pt]{elsarticle}

\usepackage{graphicx}
\usepackage{amsmath}
\usepackage{amssymb}
\usepackage{amsthm}

\usepackage{caption}
\usepackage{subcaption}
\usepackage{xcolor}

\usepackage{multicol}
\usepackage{multirow}
\usepackage{adjustbox}

\begin{document}

\begin{frontmatter}



\title{Data-Driven Closure of Projection-Based Reduced Order Models for Unsteady Compressible Flows}


\author[inst1]{Victor Zucatti}
\author[inst1]{William Wolf}

\affiliation[inst1]{organization={University of Campinas},
            city={Campinas},
            postcode={13086-860},
            country={Brazil}}

\begin{abstract}
A data-driven closure modeling based on proper orthogonal decomposition (POD) temporal modes is used to obtain stable and accurate reduced order models (ROMs) of unsteady compressible flows. Model reduction is obtained via Galerkin and Petrov-Galerkin projection of the non-conservative compressible Navier-Stokes equations. The latter approach is implemented using the least-squares Petrov-Galerkin (LSPG) technique and the present methodology allows pre-computation of both Galerkin and LSPG coefficients. Closure is performed by adding linear and non-linear coefficients to the original ROMs and minimizing the error with respect to the POD temporal modes. In order to further reduce the computational cost of the ROMs, an accelerated greedy missing point estimation (MPE) hyper-reduction method is employed. A canonical compressible cylinder flow is first analyzed and serves as a benchmark. The second problem studied consists of the turbulent flow over a plunging airfoil undergoing deep dynamic stall. For the first case, linear and non-linear closure coefficients are both low in intrusiveness, capable of providing results in excellent agreement with the full order model. Regularization of calibrated models is also straightforward for this case. On the other hand, the dynamic stall flow is significantly more challenging, specially when only linear coefficients are used. Results show that non-linear calibration coefficients outperform their linear counterparts when a POD basis with fewer modes is used in the reconstruction. However, determining a correct level of regularization is more complicated with non-linear coefficients. Hyper-reduced models show good results when combined with non-linear calibration and an appropriate sized POD basis. 
\end{abstract}




\end{frontmatter}


\section{Introduction}

Complex engineering flows can be accurately solved through numerical simulations in current supercomputers. However, such flows can easily require hundreds of millions of degrees of freedom to produce meaningful results \cite{ricciardiJSV2021} and, consequently, the improvement in computational performance of the past decades is still far from sufficient to produce real-time solutions to several problems of interest. On the other hand, the large amount of data currently being generated by high-fidelity simulations and experiments makes surrogate modeling a viable alternative for the investigation of complex flows. These methods are called reduced order models (ROMs) when data mining techniques such as proper orthogonal decomposition (POD) \cite{Sirovich_snap1, Bergmann_02} are employed to reduce the problem dimensionality. Recently, non-linear order reduction techniques such as deep convolutional autoencoders \cite{kevin_04_autoencoder} and kernel-POD \cite{rebeca_01_kPOD} have become a very active research area. Despite the advances in reduced order modeling, the construction of accurate and long-term stable ROMs for complex non-linear problems, such as those found in transport phenomena, still remains a challenging task.

The implementation of techniques for order reduction is often combined with physics-based or data-driven models. The first approach is usually performed by projection of a reduced basis in the governing equations being solved. Galerkin projection \citep{Rowley2004} is by far the most common physics-based technique, but Petrov-Galerkin techniques such as the least-squares Petrov-Galerkin (LSPG) \cite{Kevin01,grimberg2020stability} have also been successfully employed. In the last few years, data-driven ROMs based on regression \cite{lui_wolf_2019,Brunton3932} have been also proposed and they usually have no physical constrains from the governing equations. A comparison of projection-based and data-driven ROMs for unsteady flows involving natural convection can be found in \cite{zucatti_01}.

The need for closure modeling is inherent to the simulation of unsteady turbulent flows that are usually solved by formulations such as detached eddy simulation (DES) and large eddy simulation (LES), which require modeling of subgrid scales. Reduced order models suffer from similar problems; for example, numerical errors arise from high-frequency filtering due to POD-basis truncation. Moreover, numerical integration and derivation errors may arise from the wavenumber interaction of the non-linear operators appearing in the Navier-Stokes equations. These sources of errors lead to unstable and inaccurate ROMs \cite{Rowley2004,Cazemier1998,Noack_01_pressure}. Closure modeling of ROMs is a very active research field and, consequently, a plethora of methods have been proposed in the last twenty years. However, these methods lack application to more realistic problems involving a broad range of temporal and spatial scales. In most applications, ROMs with closure methods are tested in canonical problems, but some exceptions can be seen in \cite{grimberg2020stability,grimberg_novo,zucatti_jcp,Kevin01,lui_wolf_2019}. 

Closure modeling can be divided in physics-based and data-driven. The first category relies primarily on physical intuition such as the concept of an energy cascade fundamental to turbulence. Artificial viscosity closure models \cite{bergmann_enablers,iliescu_01_burgers,iliescu_02_turbulent} fall into this category and the main idea of these methods is to add the lacking dissipation effects of the truncated POD modes and subgrid scales effects. Several approaches are proposed in literature and they can be as simple as changing the viscosity coefficient \cite{bergmann_enablers,iliescu_01_burgers} or more sophisticated such as Smagorinsky type models \cite{iliescu_02_turbulent,iliescu_01_burgers} which vary in space and time. Another physics-based closure model is based on the underlying streamline-upwind Petrov–Galerkin (SUPG) applied to ROMs \cite{iliescu_05_supg}. In this case, stabilization parameters based on the underlying finite element discretization and the POD basis truncation are derived with the appropriate theoretical support given by finite element error analysis. Unfortunately, the performance of this type of closure has been shown to be generally unsatisfactory when modeling complex flows requiring several POD modes \cite{iliescu_02_turbulent} and would probably be ineffective if applied to chaotic turbulent flows.

Data-driven closure models, also known as calibration methods, try to model all ROM inconsistencies as a whole (top-down approach) using the full order model (FOM) data (e.g., simulation snapshots or POD temporal modes) as opposed to attempting to model error sources individually (bottom-up approach). For this category, the underlying principle is that a ROM should be capable of recovering most of the information of the FOM from which it originated. Therefore, it is only natural to exploit the data to its fullest extent before attempting to extrapolate outside the training window. For example, this can be achieved by adding corrective (calibration) coefficients to the original ROM,  capable of improving the model by taking the information provided by POD temporal modes into consideration \cite{galletti_01,couplet2005,Bourguet_02,Bourguet_01,zucatti_jcp,Balajewicz2016}. Alternative methods correct the ROM by directly using the snapshots \cite{grimberg2020stability,grimberg_scitech,mou_compMethods}. Hybrid closure models relying on both physical insight (e.g., energy conservation) and data are also possible \cite{iliescu_03_calibration,iliescu_06_data_physics}. Furthermore, it is relevant to point out that minimization problems, commonly appearing when a data-driven approach is employed, can be ill-conditioned and require some form of regularization. In \cite{elmajd_01,elmajd_02}, some regularization strategies for data-driven calibration of projection-based ROMs are discussed.

In the present work, we use calibration based on POD temporal modes to improve the accuracy and stability of Galerkin and Petrov-Galerkin ROMs. In fact, a more accurate model can be obtained by adding corrective terms to the original model and minimizing the error between the POD and ROM temporal modes. This translates to solving  a non-linear optimization problem. In \cite{couplet2005}, an approximation transforming the original non-linear problem into a less expensive linear optimization is proposed and it is also adopted here. A comparison of the linear and non-linear optimization calibration methods is provided in \cite{favier_01} for a low Reynolds cylinder wake flow and a separated flow around an ONERA-D airfoil with snapshots obtained from PIV. Briefly, the linear optimization approach was shown to be much faster while being almost as accurate as its non-linear counterpart. Additionally, this calibration methodology was used in reduced order modeling of unsteady flows around airfoils at low Reynolds numbers \cite{Bourguet_01,Bourguet_02} and more recently at moderate Reynolds numbers \cite{zucatti_02_scitech,zucatti_jcp}. 

Data-driven closure models have been mostly used with linear calibration coefficients. However, more general (non-linear) coefficients can be considered and have been tested to some extent \cite{mou_compMethods,iliescu_03_calibration,mou_fluids,masterthesisMou,iliescu_04_redeNeural}. In \cite{masterthesisMou}, this calibration methodology not only is used to obtain linear calibration terms but also quadratic and cubic operators. Methodology details are given and results are presented for the one-dimensional Burgers equation. 

In this work, calibration of projection-based ROMs is assessed using linear and non-linear coefficients. For the latter approach, calibration is employed for third and fourth-order tensors justified by the Galerkin and LSPG projections, respectively. The model coefficients are pre-computed as result of adopting the non-conservative form of the compressible Navier-Stokes equations. A greedy hyper-reduction method is applied in the pre-computational stage to obtain completely grid-independent ROMs. Methods are tested on two compressible flow problems being a canonical compressible flow past a cylinder and the turbulent flow over an airfoil undergoing deep dynamical stall. In both cases, regularization is required and performed using a Tikhonov methodology combined with an L-curve approach for appropriate parameter choice \cite{elmajd_01,elmajd_02}. To the best of the authors' knowledge, a pre-computed formulation of the compressible Navier-Stokes equations has never been applied with the LSPG method. Also, we believe that calibration of the fully-discrete ROMs based on error minimization of the temporal modes has never been performed with non-linear coefficients.

\section{Reduced order modeling}

\subsection{Proper orthogonal decomposition}
\label{subsection:POD}

Proper orthogonal decomposition (POD) is a fundamental step in the calculation of projection-based ROMs. Here, POD is applied to compute a low-dimensional subspace of the primitive flow variables $\zeta$, $u$, $v$ and $p$, which represent specific volume, velocity components and pressure, respectively. These variables are chosen since the flows of interest in this work are solved using a non-conservative compressible formulation as detailed in Section \ref{NS:equations}. In the present notation, the unsteady flows can be decomposed as 
\begin{equation} 
	\mathbf{q} (\mathbf{x},t) = \mathbf{\bar{q} (x)} + \mathbf{q'}(\mathbf{x},t)= \mathbf{\bar{q} (x)}  + \sum_{i = 1}^{M} \mathbf{\Phi_i(x)}a_i(t)
	\label{velocity} \mbox{ ,}
\end{equation}

\noindent
where $\mathbf{q}=\left\{\zeta,u,v,p\right\}^{\top} \in \mathbb{R}^N \times [0, T]$, $\mathbf{\bar{q}(x)} \in \mathbb{R}^N$ is the mean flow, $\mathbf{\Phi} \in \mathbb{R}^{N \, \times \, M}$ with $\mathbf{\Phi_i} \in \mathbb{R}^N$ being the orthonormal spatial eigenfunctions, $\mathbf{a} = [a_1, \ldots, a_M]^{\top} \in \mathbb{R}^M \times [0,T]$ represents the temporal modes and $\left\{\cdot\right\}^{\top}$ is the transpose of $\left\{\cdot\right\}$. The parameter $M$ represents the number of data sets extracted from the numerical simulation, $N$ is the number of grid points multiplied by the number of flow variables, $T$ represents the period of snapshot extraction and $i$ represents the mode index.

The POD consists of looking for the deterministic functions $\{ \mathbf{\Phi_i} \}_{i=1}^{M}$ that are most similar in an averaged sense to the realizations $\mathbf{q} (\mathbf{x},t)$. Here, we employ the technique introduced by \cite{Sirovich_snap1} called ``snapshot POD'' where the following Fredholm integral eigenvalue problem is solved as
\begin{equation}
    \int_{0}^{T} C_{ij} \, a_i (t') \, dt' = \lambda_i \, a_i (t) \mbox{ ,}
    \label{eq:fredholm}
\end{equation}
\noindent
and where the temporal covariance matrix $\mathbf{C} \in \mathbb{R}^{M \, \times \, M}$ is defined by
\begin{equation}
    C_{ij} = \frac{1}{T} \int_{\Omega} \mathbf{q} (\mathbf{x},t_i) \, \mathbf{q} (\mathbf{x},t_j) \, d \mathbf{x} 
    \approx \frac{1}{T} \langle \mathbf{q} (\mathbf{x},t_i) , \mathbf{q} (\mathbf{x},t_j) \rangle_{\mathbf{\Pi}}
    \mbox{ .}
\end{equation}

\noindent
In the previous equation, $\mathbf{\Pi} \in \mathbb{R}^{N \, \times \, N}$ is a symmetric positive definite matrix defining the inner-product $\langle \mathbf{q} (\mathbf{x},t_i) , \mathbf{q} (\mathbf{x},t_j) \rangle_{\mathbf{\Pi}}=\mathbf{\langle q_i , q_j \rangle_{\Pi} = q_i^T \Pi q_j}$. In the present application, the matrix $\mathbf{\Pi}$ is diagonal with non-zero elements defined as $\mathbf{\Pi_{ii} = A_i}$, where $\mathbf{A_i}$ is the area associated to the $i$-th vector element.  
The covariance matrix $\mathbf{C}$ is symmetric positive semidefinite and, therefore, allows the use of singular value decomposition to compute the singular values and singular vectors which are, in turn, related to the eigenvalues $\lambda_i$ and eigenvectors (modes) of the POD reconstruction. Such modes are calculated so that the reconstruction is optimal in the sense of truncated mean quadratic error. The idea of writing a temporal covariance matrix (snapshot POD) comes from the fact that the solution cost grows rapidly for large computational grids. This is an issue especially in multidimensional problems.

In Eq. \ref{eq:fredholm}, $a_i$ represent the $i-th$ time-dependent orthogonal POD eigenfunctions, also called temporal modes. 
The reader is referred to \cite{Bergmann_01, Rowley2004, Sirovich_snap1, Bergmann_02, Noack_01_pressure} for more details on the calculation of these modes. The spatial basis functions $\mathbf{\Phi_i}$ can be calculated from the realizations $\mathbf{q}$ and the temporal modes $\mathbf{a}$ with
\begin{equation}
    \mathbf{\Phi_i} (\mathbf{x}) = \frac{1}{T \lambda_i} \int_{0}^{T} q (\mathbf{x},t) \, a_i (t) \, dt \mbox{ .}
\end{equation}
In summary, the spatial modes are used in the Galerkin and LSPG projections to reconstruct a system of ordinary differential equations that, in turn, will determine the evolution of the temporal modes. With the calculation of both terms, it is possible to reconstruct the fluctuation fields appearing in Eq. \ref{velocity} as $\mathbf{q} \approx \mathbf{\hat{q}} =  \mathbf{\bar{q}}  + \sum_{i = 1}^{m} \mathbf{\Phi_i} a_i$. In practical ROM applications, instead of employing the full set $M$ of POD modes for the fluctuation field reconstruction, one seeks $m \ll M$. Therefore, only the most energetic POD modes are used in the ROM.

\subsection{Projection methods}

Let us consider the system of non-linear partial differential equations $\mathbf{F(q)} \in \mathbb{R}^N$ defined in a connected open region $\Omega \subset \mathbb{R}^{N_g}$ with a boundary $\Gamma$
\begin{equation}
	\begin{cases}
		\mathbf{F(q}) = \frac{d \mathbf{q}}{dt} - \mathbf{G(q)} = \mathbf{0} \quad in \quad \Omega \\
		\mathbf{q}(\mathbf{x},0) = \mathbf{q_0} \\
		\mathbf{q = q_b} \quad on \quad \Gamma \mbox{ .}
	\end{cases}
	\label{dynamical_system}
\end{equation}

\noindent
In the system above, $\mathbf{q}$ is a function of space and time, $N_g$ is the number of grid points, and the non-linear operator $\mathbf{G(q)}$ is given by the convective and diffusive operators appearing in the mass, momentum and energy equations, herein referred to as Navier-Stokes equations. Let $\mathbf{\Phi} \in \mathbb{R}^{N \, \times \, m}$ with $\{ \mathbf{\Phi_i} \}_{i=1}^{m}$ defining an orthonormal basis obtained by POD. The state variable $\mathbf{q}$ is then approximated as the linear combination of this basis vector as
\begin{equation}
	\mathbf{q} \approx \mathbf{\hat{q}} = \mathbf{\bar{q}} + \sum_{i = 1}^{m} \mathbf{\Phi_i} \, a_i 
	\mbox{ ,}
	\label{eq:q_decomposed}
\end{equation}
where the explicit dependencies on space and time are omitted for simplicity.

After approximation, the set of equations is written as $ \mathbf{F} (\mathbf{q}) \approx \mathbf{R} (\mathbf{\hat{q}}) \neq \mathbf{0}$ for the physical problem being solved. Here, $\mathbf{R(\hat{q}) = R} (\dot{\mathbf{a}},\mathbf{a})$ is the residual after order reduction and spatial discretization. A solution is sought by enforcing the residual $\mathbf{R(\hat{q})}$ orthogonality as
\begin{equation}
	\mathbf{\langle \Psi_i , R} (\mathbf{\bar{q}} + \sum_{j=1}^{m} \mathbf{\Phi_j} a_j ) \rangle_{\mathbf{\Pi}} = \mathbf{0}
	\mbox{ ,} \quad \quad \mbox{i} = 1, \ldots, m \mbox{ ,}
\end{equation}
\noindent
where $\{\mathbf{\Psi_i} \}_{i=1}^{m}$ is the test basis. A projection method is generally called Galerkin (Petrov-Galerkin) when the test and solution bases are equal (different),  i.e., $\mathbf{\Psi = \Phi}$ ($\mathbf{\Psi \neq \Phi}$). The POD solution basis should satisfy the boundary conditions of Eq. \ref{dynamical_system} and, in Ref. \cite{Bergmann_01}, the ability to obtain homogeneous Dirichlet or Neumann boundary conditions from the snapshots is discussed.

\subsubsection{Galerkin projection}

Galerkin projection is the most popular alternative for reduced order modeling of time dependent problems. This can be attributed to its implementation simplicity and solid mathematical foundation. Applying the Galerkin projection method ($\mathbf{\Psi = \Phi}$) to Eq. \ref{dynamical_system} we obtain

\begin{equation}
    \dot{a}_i = \langle \mathbf{\Phi_i}, \mathbf{\hat{G}} (\mathbf{\bar{q}} + \sum_{j=1}^{m} \mathbf{\Phi_j} a_j )\rangle_{\mathbf{\Pi}}
    \mbox{ ,} \quad \quad \mbox{i} = 1, \ldots, m \mbox{ ,}
	\label{galerkin_ode}
\end{equation}

\noindent
where $\mathbf{\hat{G} (\hat{q})}$ is the approximation of operator $\mathbf{G (q)}$ after order reduction and spatial discretization. Moreover, the initial conditions are determined by selection and projection of a single snapshot $\mathbf{q_0}$ in the vector basis
\begin{equation}
	a_i (0) = \langle \mathbf{\Phi_i (x)}, \mathbf{q_0} \rangle_{\mathbf{\Pi}}
	\mbox{ ,} \quad \quad \mbox{i} = 1, \ldots, m \mbox{ .}
	\label{galerkin_ode_a0} 
\end{equation}

The previous system of ordinary differential equations represents the ROM associated to the FOM and can be solved using a time-marching method. The right-hand side of Eq. \ref{galerkin_ode} should not scale with the full-order model so as to achieve reduced computational cost. Following the POD-Galerkin approach, the ROM obtained for the non-conservative form of the compressible Navier-Stokes equations (see details in Section \ref{NS:equations}) can be written as
\begin{equation} 
	\dot{a}_i 
	+ e_i 
	+ A_{ij} a_j  
	+ N_{ijk} a_j a_k = 0_i
	\mbox{ ,} \quad \quad \mbox{i, j, k} = 1, \ldots, m \mbox{ ,}
	\label{galerkin_ode_cav}
\end{equation}
where the time-independent ODE coefficients $\mathbf{e} \in \mathbb{R}^m$, $\mathbf{A} \in \mathbb{R}^{m \, \times \, m}$ and $\mathbf{N} \in \mathbb{R}^{m \, \times \, m \, \times \, m}$ can be found in \cite{zucatti_jcp}. A two-step (offline/online) computational approach is possible when the governing equations are linear or have polynomial non-linearities. In the offline stage, a reduced-order system of equations is obtained after projecting the reduced-order basis in the governing equations. This stage is performed only once but can be particularly costly depending on the number of modes $m$ and the maximum tensor order of the model used in the reconstruction given that the inner product $\mathbf{\Pi}$ is grid-size dependent. The online step corresponds to solving the resulting reduced-order system of equations such as the system of ODEs given by Eq. \ref{galerkin_ode_cav}. This step no longer scales with the FOM and, thus, is cheap in general.

In this work, for reasons that will become clearer later,  it is convenient to promptly discretize $\dot{a}_i$ in Eq. \ref{galerkin_ode_cav} by introducing a time-marching method. As a result, the implicit Euler time-discretization of the previous equation yields
\begin{equation}
    f_i^G (a_i^n) = 
    \frac{a_i^n - a_i^{n-1}}{\Delta t} 
    + e_i
    + A_{ij} a_j^n
    + N_{ijk} a_j^n a_k^n = 0_i
    \mbox{ ,} \quad \quad \mbox{i, j, k} = 1, \ldots, m \mbox{ ,}
    \label{eq:g_fatorado}
\end{equation}

\noindent
where $\mathbf{f^G (a^n)}$ is the non-linear algebraic system of equations of the Galerkin ROM.

\subsubsection{Least-squares Petrov-Galerkin}

A Petrov-Galerkin approach uses different test and trial bases ($\mathbf{\Psi \neq \Phi}$). Particularly, the least-squares Petrov-Galerkin (LSPG) is an alternative to the Galerkin method showing encouraging results \cite{Kevin01,grimberg2020stability,grimberg_novo} when used to generate ROMs for non-linear time-dependent problems such as those usually found in fluid flows.
A thorough theoretical comparison of both LSPG and Galerkin can be found in \cite{Kevin02_comp}. The LSPG method \cite{Kevin01,Quarteroni2016} seeks to minimize the fully discrete residual $\mathbf{\hat{R}(a^n)} \in \mathbb{R}^N$ (i.e., the residual after temporal and spatial discretization) at each $n$-th time-step as
\begin{equation}
	\operatorname*{\textrm{minimize}}_{\mathbf{a^n}} \mathcal{F}  \mathbf{(a^n)}
	\mbox{ .}
	\label{eq:regression_obj}
\end{equation}

\noindent
The objective function $\mathcal{F} \mathbf{(a^n)}$ is defined in the following special form
\begin{equation}
    \mathcal{F}  \mathbf{(a^n)} =
    \frac{1}{2} \mathbf{ \| \hat{R} (a^n) \|_{\Pi}^{2}} =
    \frac{1}{2} \mathbf{\langle \hat{R} (a^n), \hat{R} (a^n) \rangle_{\Pi}}
    \mbox{ ,}
\end{equation}

\noindent
or equivalently
\begin{equation}
     \mathbf{a^{n}} = 
     \operatorname*{\textrm{arg min}}_{\mathbf{\alpha}^n \, \in \, \mathbb{R}^m}
     \| \mathbf{\hat{R}} (\mathbf{\alpha^n)} \|_{\mathbf{\Pi}}^{2}
     \mbox{ .}
     \label{eq:regression_arg_min}
\end{equation}

In this work, the fully discrete residual $\mathbf{\hat{R} (a^n)}$ represents the $2D$ non-conservative compressible Navier-Stokes equations (Eqs. \ref{eq:nonconsNS}) after the state variable $\mathbf{q}$ is approximated by the POD basis vector, finite-difference spatial discretization and implicit time discretization. As a result, the residual associated with its time discretization by the implicit Euler scheme is
\begin{equation}
    \mathbf{\hat{R} (a^n)} = 
    \mathbf{\Phi} \frac{\mathbf{a^n - a^{n-1}}}{\Delta t} + \mathbf{\hat{G} (a^n)}
    \mbox{ ,}
\end{equation}

\noindent
where $\Delta t$ is the time step, $\mathbf{\hat{G}}$ is the same operator as in Eq. \ref{galerkin_ode} and initial conditions are also given by Eq. \ref{galerkin_ode_a0}.

Optimality conditions are derived from Taylor's theorem and can be determined by examining the gradient $\nabla \mathcal{F} \mathbf{(a^n})$ and Hessian $\Delta \mathcal{F} \mathbf{(a^n})$ matrices \cite{book_nonlinearOpt}. The derivatives of $\mathcal{F}  \mathbf{(a^n})$ can be expressed in terms of the Jacobian $\mathbf{J(a^n) = \frac{\partial \hat{R} (a^n)}{\partial a^n}}$. Applying the first-order necessary condition $\mathbf{\nabla}  \mathcal{F}(\mathbf{a^n}) = 0$ yields
\begin{equation}
    \nabla \mathcal{F} (\mathbf{a^n}) = 
    \mathbf{\langle J(a^n) , \hat{R}(a^n) \rangle_\Pi} =
    \mathbf{\Bigg \langle \frac{\partial \hat{R}(a^n)}{\partial a^n} , \hat{R}(a^n) \Bigg \rangle_\Pi = 0}
    \mbox{ ,}
    \label{eq:lspg_ṕrojection}
\end{equation}
which is equivalent to Eq. \ref{eq:regression_obj}, provided that convergence to the same local minimum occurs \cite{grimberg2020stability}. Therefore, in the LSPG method, the discrete test basis $\mathbf{\Psi (a^n)} \in \mathbb{R}^{N \, \times \, m}$ is given by
\begin{equation}
	\mathbf{\Psi (a^n)} = 
	\mathbf{J (a^n)} = 
	\mathbf{\frac{\partial \mathbf{\hat{R}(a^n})}{\partial \mathbf{a^n}}}
	\mbox{ .}
\end{equation}

The offline/online approach discussed previously is also possible when using the LSPG method \cite{phdthesisGrimberg,phdthesisZahr}.
However, pre-calculating the LSPG coefficients can quickly become a cumbersome task, specially when compared to the Galerkin method, given the complexity introduced by the test basis $\mathbf{\Psi (a^n)}$.
In fact, the LSPG ROM not only has additional terms but also comes with the downside of higher-orders tensors that can lead to  prohibitive computational costs when several POD modes are required in the model construction. This cost is associated with the multiple spatial terms arising from the inner product from Eq. \ref{eq:lspg_ṕrojection} that contains the Jacobian of the residual, which is extensive for the discrete compressible Navier-Stokes equations.
Similarly, decomposing the state variable $\mathbf{q}$ following Eq. \ref{eq:q_decomposed} contributes to further intensify this issue.
For the implicit Euler time marching scheme, the LSPG ROM of the non-conservative compressible Navier-Stokes equations can be written as
\begin{equation}
    \begin{split}
    f_i^{PG} (a_i^n) =
    \frac{a_i^{n} - a_i^{n-1}}{\Delta t} 
    + (e_i^{\prime} 
    + A_{ij}^{\prime} a_j^{n} 
    + B_{ij}^{\prime} a_j^{n-1}
    + N_{ijk}^{\prime} a_j^{n} a_k^{n}
    + L_{ijk}^{\prime} a_j^{n} a_k^{n-1}) \, + &\\
    \Delta t
    (e_i^{\prime \prime} 
    + A_{ij}^{\prime \prime} a_j^{n}
    + N_{ijk}^{\prime \prime} a_j^{n} a_k^{n}
    + Q_{ijkl}^{\prime \prime} a_j^{n} a_k^{n} a_l^{n}) 
    = 0 \mbox{ ,} \quad \quad \mbox{i, j, k, l} = 1, \ldots, m \mbox{ ,} &
    \end{split}
    \label{eq:lspg_fatorado}
\end{equation}

\noindent
where $\mathbf{f^{PG}}$ describes the non-linear algebraic system of equations of the LSPG ROM and $\mathbf{e}^{\prime} \in \mathbb{R}^{m}$, $\mathbf{A}^{\prime} \in \mathbb{R}^{m \, \times \, m}$, $\mathbf{B}^{\prime} \in \mathbb{R}^{m \, \times \, m}$, $\mathbf{N}^{\prime} \in \mathbb{R}^{m \, \times \, m \, \times \, m}$, $\mathbf{L}^{\prime} \in \mathbb{R}^{m \, \times \, m \, \times \, m}$, $\mathbf{e}^{\prime \prime} \in \mathbb{R}^{m}$, $\mathbf{A}^{\prime \prime} \in \mathbb{R}^{m \, \times \, m}$, $\mathbf{N}^{\prime \prime} \in \mathbb{R}^{m \, \times \, m \, \times \, m}$ and $\mathbf{Q}^{\prime \prime} \in \mathbb{R}^{m \, \times \, m \, \times \, m \, \times \, m}$ are the pre-computable coefficients defining the algebraic system of equations.
In particular, the non-conservative form of the compressible Navier-Stokes equations has quadratic non-linearities that lead to the emergence of the fourth-order tensor $Q_{ijkl}^{\prime \prime}$ in Eq. \ref{eq:lspg_fatorado}. Consequently, not only is the LSPG ROM more computationally expensive during the offline stage but also it has a significant overhead after pre-computation (online stage) compared to the Galerkin ROM. While the  governing equations produce third-order tensors in the Galerkin model (Eq. \ref{galerkin_ode_cav}),  they lead to fourth-order tensors in the LSPG.

\subsection{Hyper-reduction}

Although an order reduction method is applied, models generated using projection-based techniques can fail to produce relevant computational time gains. Pre-computation of Galerkin/Petrov-Galerkin coefficients (offline/online decoupling procedure) is impossible for problems containing strong non-linearities (i.e. non-polynomial) or non-affine parameter dependence. On the other hand, even when pre-computation is possible, the offline stage can be very costly depending on the number of modes used in the reconstruction and complexity of the coefficients. In particular, Petrov-Galerkin methods can be very computationally demanding during the offline stage. Non-linear methods \cite{kevin_04_autoencoder,rebeca_01_kPOD} recently being applied in a reduced order modeling context also require further approximation and techniques are being developed to specifically address this issue. In all cases, ROMs could benefit from an additional layer of approximation provided by hyper-reduction methods. This is typically performed by sampling the spatial modes using a greedy method before projection.

Many studies \cite{Willcox_02_gappy, Kevin01,zucatti_jcp} adopt a gappy POD  framework \cite{Sirovich_gappy} when using hyper-reduction. In this approach, given a subset of indices $J = \{ j_1, \dots , j_s \} \subset \{ 1, \dots , N \}$, a solution vector $\mathbf{\hat{q}}$ is approximated by sampling the spatial modes using a mask projection matrix $\mathbf{P} = [\mathbf{e_{j_1}, \dots , e_{j_{s}}}]^T \in \mathbb{R}^{N \times s}$ to construct the estimated solution $\Tilde{\mathbf{q}} \approx \mathbf{P^T \Phi a}$. Here, $s$ is the number of indices retained from the original vector of size $N$ and $\mathbf{e_{j_k}}$ denotes the vector with a $1$ in the $j_k$-th coordinate and $0$'s elsewhere. Ideally, the number of sampled entries should be kept to a minimum but produce an optimal approximation of the temporal modes of the modified mask problem. In other words, the error $\epsilon$ between the gapless solution $\mathbf{\hat{q}}$ and $\Tilde{\mathbf{q}}$, defined as 
\begin{equation}
    \epsilon = \| \mathbf{P^T \hat{q}} - \mathbf{P^T \Phi a} \|_{\mathbf{\Pi}}^2 \,\,\mbox{ ,}
\end{equation}
\noindent
should be minimal for a given mask matrix $\mathbf{P}$ sampling $s$ vector entries. This is equivalent to the minimization problem
\begin{equation}
    \operatorname*{\textrm{minimize}}_{\mathbf{P}} 
    \| \mathbf{I} - \langle \mathbf{\tilde{\Phi}, \tilde{\Phi}} \rangle_{\mathbf{\Pi}} \|_2^2 \,\,\mbox{ ,}
    \label{eq:gappy_min}
\end{equation}

\noindent
where $\mathbf{\tilde{\Phi}} =  \mathbf{P^T \Phi}$. In this work, hyper-reduction is performed using the POD basis $\mathbf{\Phi}$ for  the reconstruction of the solution vector $\mathbf{\hat{q}}$. However, a basis built from the FOM residual or fully discrete residual vector $\mathbf{\bar{R}}$ could also be adopted.

The optimal solution of Eq. \ref{eq:gappy_min} for a set of given size can be an intractable combinatorial optimization problem even for relatively small problems.
Consequently, this problem is usually solved for a sub-optimal set of indices using a greedy algorithm such as the missing point estimation (MPE) method \cite{Willcox_03_MPE}. In an iteration of this method, the condition number 
\begin{equation}
    c\left(\mathbf{\langle \Tilde{\Phi}, \Tilde{\Phi} \rangle_{\mathbf{\Pi}} }\right) \equiv 
    \frac{\lambda_{max} \left(\mathbf{\langle \Tilde{\Phi}, \Tilde{\Phi} \rangle_{\mathbf{\Pi}}}\right)}{\lambda_{min} \left(\mathbf{\langle \Tilde{\Phi}, \Tilde{\Phi} \rangle_{\mathbf{\Pi}}}\right)}
\end{equation}
of the approximated identity matrix $\langle \mathbf{\Tilde{\Phi} , \Tilde{\Phi}} \rangle_{\mathbf{\Pi}} \approx \mathbf{I}$ (i.e., inner product of the mask POD basis) is evaluated for all the remaining candidate points $J_c = \{1, \ldots, N \} - J_e$ and eliminates the index with the highest condition number. Here, $J_e$ corresponds to the set of eliminated indices and is initially an empty set. This procedure is repeated until a user specified near-optimal set of points is left. Computational cost of the MPE method may quickly become prohibitive, although the method is cheaper than the optimal solution. Furthermore, the condition number can serve as an a priori non-sharp error indicator of the hyper-reduced problem.

In the present work, we adopt the accelerated greedy MPE technique for hyper-reduction. This procedure \cite{Willcox01} allows for substantial time gains compared to the naive MPE while still using the same underlying principles. Similarly to the condition number problem previously discussed, the greedy point selection is equivalent to picking the spatial mode entry responsible for the largest growth in the condition number of the modified eigenvalue problem. In the accelerated version of the MPE, the costly modified symmetric eigenvalue problem is not actually solved, but the index selection occurs by analyzing properties of the candidate indices.

\subsection{Non-conservative compressible Navier-Stokes equations}
\label{NS:equations}

Consider the $2D$ non-conservative compressible Navier-Stokes equations presented by \cite{iollo_01} as
\begin{subequations}
    \begin{align}
    \begin{split}
        \zeta_t = & \zeta (u_x + v_y) - u \zeta_x - v \zeta_y   \mbox{ ,}
    \end{split}\\
    \begin{split}
        u_t = & - u u_x - v u_y - \zeta p_x + \frac{Ma}{Re} \zeta \bigg[ \bigg( \frac{4}{3} u_x - \frac{2}{3} v_y \bigg)_x + (v_x + u_y)_y  \bigg]   \mbox{ ,}
    \end{split}\\
    \begin{split}
        v_t = & - u v_x - v v_y - \zeta p_y + \frac{Ma}{Re} \zeta \bigg[ \bigg( \frac{4}{3} v_y - \frac{2}{3} u_x \bigg)_y + (v_x + u_y)_x \bigg]   \mbox{ , and}
    \end{split}\\
    \begin{split}
        p_t = & - u p_x - v p_y - \gamma p (u_x + v_y) + \frac{\gamma Ma}{Re Pr} [(p \zeta)_{xx} + (p \zeta)_{yy}] \\
        & + \frac{(\gamma - 1) Ma}{Re} \bigg[ u_x \bigg( \frac{4}{3} u_x - \frac{2}{3} v_y \bigg) + v_y \bigg( \frac{4}{3} v_y - \frac{2}{3} u_x \bigg) + (v_x + u_y)^2 \bigg]  \mbox{ ,}
    \end{split}
    \end{align}
    \label{eq:nonconsNS}
\end{subequations}
\noindent
where $\zeta = 1 / \rho$ is the specific volume, $u$ and $v$ are the $x$ and $y$ velocity components, respectively, and $p$ is the pressure. In the set of Eqs. \ref{eq:nonconsNS}, $Pr$, $Re$ and $Ma$ denote the reference Prandtl, Reynolds and Mach numbers, respectively. Subscripts denote partial derivatives and $\gamma$ is the specific heat ratio. 

The quadratic non-linearities of the non-conservative compressible Navier-Stokes equations allow the pre-computation of Galerkin and LSPG coefficients. This set of equations was previously used for construction of ROMs in Refs. \cite{iollo_01,Balajewicz2016,zucatti_jcp,zucatti_02_scitech}. However, to the authors knowledge, this work is the first to perform pre-computation of LSPG coefficients for Eq. \ref{eq:nonconsNS}. In CFD applications, the compressible Navier-Stokes equations are usually expressed in conservative form but this would add tensors of even higher orders than those computed in this work. 
This issue has been typically remedied by using an additional approximation layer such as hyper-reduction \cite{Saifon01,Willcox_04_DEIM}. Further approximation can be detrimental to the models and it has motivated corrective techniques such as the empirical quadrature procedure (EQP) \cite{yano2019discontinuous}, stabilization by entropy preservation \cite{chan2019entropy} and the energy-conserving sampling and weighting (ECSW) methods \cite{ecws,grimberg2020stability}.

\subsection{Linear least-squares calibration of projection-based ROMs}
\label{calibration}

In general, calibration methods attempt to improve stability and accuracy of time-dependent reduced order models by adding corrective terms to the non-linear system of algebraic (or differential) equations obtained by projection-based or data-driven models. Considering the tensors appearing in the LSPG formulation, the corrective terms would be given by
\begin{equation}
    f_i^n
    + e_i^c 
    + A_{ij}^c a_j^{n}
    + N_{ijk}^c a_j^{n} a_k^{n}
    + Q_{ijkl}^c a_j^{n} a_k^{n} a_l^{n} = 0
    \mbox{ ,} \quad \quad \mbox{i, j, k, l} = 1, \ldots, m \mbox{ ,}
\end{equation}

\noindent
where $\mathbf{f}$ could be either the Galerkin ROM $\mathbf{f^{G}}$ or the LSPG ROM $\mathbf{f^{PG}}$. The terms $\mathbf{e^c} \in \mathbb{R}^m$, $\mathbf{A^c} \in \mathbb{R}^{m \, \times \, m}$, $\mathbf{N^c} \in \mathbb{R}^{m \, \times \, m \, \times \, m}$ and $\mathbf{Q^c} \in \mathbb{R}^{m \, \times \, m \, \times \, m \, \times \, m}$ are the calibration operators added to the initial ROM. In this section, calibration is presented for a specific set of calibration terms (first to fourth-order tensors), but can be applied without loss of generality for a different set of calibration tensors. For example, the same reasoning is valid when only linear terms (i.e., $\mathbf{e^c}$ and $\mathbf{A^c}$) are considered. In fact, the choice for linear calibration terms is prevalent in the literature. However, calibration by non-linear operators have been tested to a smaller degree with higher-order tensors \cite{iliescu_03_calibration,mou_compMethods,mou_fluids,masterthesisMou} and deep residual neural networks \cite{iliescu_04_redeNeural}.

In a linear least-squares calibration, the goal may be to minimize the error $E_1^{ROM}$ between temporal modes obtained by solving the system of non-linear equations of the ROM $\mathbf{a^{ROM}} (t)$ and the original POD temporal modes $\mathbf{a} (t)$ in a user specified training window $0\leq t \leq T$ as
\begin{equation}
    E_1^{ROM} = \sum_{i=1}^{m} \int_0^T
    \left(a_i (t) - a_i^{ROM} (t) \right)^2 dt
    \mbox{ .}
    \label{eq:opt_nonlinear}
\end{equation}
In the above equation, $\mathbf{a} (t)$ refers to the temporal modes obtained directly by the POD snapshot method in Eq. \ref{eq:fredholm}, while the term $\mathbf{a^{ROM}} (t)$ represents the temporal modes obtained by the solutions of Eqs. \ref{eq:g_fatorado} and \ref{eq:lspg_fatorado} for the Galerkin and LSPG models, respectively.
A second error norm $E_2^{ROM}$ can be also defined by the ODE system and is frequently preferred when performing calibration \cite{zucatti_jcp,iliescu_03_calibration,favier_01,mou_fluids,mou_compFluidDyn,mou_compMethods}
\begin{equation}
    E_2^{ROM} = \sum_{i=1}^{m} \int_0^T 
    \left(\dot{a}_i (t) - \dot{a}^{ROM}_i (t)\right)^2 dt
    \mbox{ .}
    \label{eq:error_ode}
\end{equation}

However, given that the LSPG model is naturally an algebraic system of equations, error norm $E_1^{ROM}$ is used in this work and it was also applied in \cite{Bourguet_01,Bourguet_02} in the context of Galerkin-ROM calibration. The non-linear optimization problem that arises when minimizing the error norm $E_1^{ROM}$ from Eq. \ref{eq:opt_nonlinear} can be transformed to a linear one by the approximation $\mathbf{a^{ROM}}(t) \approx \tilde{\mathbf{a}}(t)$, as discussed in \cite{couplet2005}. The new error norm $E_1$ is expressed as
\begin{equation}
    E_1 = \sum_{i=1}^{m} \int_0^T 
    \left(a_i (t) - \tilde{a}_i (t)\right)^2 dt
    \mbox{ ,}
    \label{eq:opt_linear}
\end{equation}

\noindent
where $\tilde{a}_i (t_n)$ is given by 
\begin{subequations}
\begin{equation}
    \frac{\tilde{a}_i^n - a_i^{n-1}}{\Delta t_s} - (e_i + A_{ij} \tilde{a}_j^n + N_{ijk} \tilde{a}_j^n \tilde{a}_k^n) = 0
    \mbox{ ,} \quad \quad \mbox{i, j, k} = 1, \ldots, m \mbox{, \,\,\,\, or} 
    \label{eq:galerkin_cal}
\end{equation}
\begin{equation}
    \begin{split}
    &\frac{\tilde{a}_i^{n} - a_i^{n-1}}{\Delta t_s} 
    + (e_i^{\prime} 
    + A_{ij}^{\prime} \tilde{a}_j^{n} 
    + B_{ij}^{\prime} a_j^{n-1}
    + N_{ijk}^{\prime} \tilde{a}_j^{n} \tilde{a}_k^{n}
    + L_{ijk}^{\prime} \tilde{a}_j^{n} a_k^{n-1}) \, + \\
    & \Delta t_s
    (e_i^{\prime \prime} 
    + A_{ij}^{\prime \prime} \tilde{a}_j^{n}
    + N_{ijk}^{\prime \prime} \tilde{a}_j^{n} \tilde{a}_k^{n}
    + Q_{ijkl}^{\prime \prime} \tilde{a}_j^{n} \tilde{a}_k^{n} \tilde{a}_l^{n})
    = 0
    \mbox{ ,} \quad \quad \mbox{i, j, k, l} = 1, \ldots, m \mbox{ .}
    \end{split}
    \label{eq:lspg_fatorado_cal}
\end{equation}
\label{eq:lspg_galerkin_cal}
\end{subequations}

\noindent
In the equations above, $\Delta t_s$ is the time step between snapshots and $\mathbf{a^{n-1}}$ is the POD temporal mode at $t_{n-1}$. A pre-computable approximation of the ROM temporal modes can be obtained for all time instants using Eq. \ref{eq:galerkin_cal} (Galerkin) or \ref{eq:lspg_fatorado_cal} (LSPG). Also, in both cases, time discretization is performed using the implicit Euler method and for the first time step $\tilde{a}_i^1 = a_i^1$. Suppression of the non-linear constraint to obtain an affine operator may seem extreme at first, but this can be understood as a measure of how the non-calibrated ROM alters the solution in each time step. If successful, calibration should systematically compensate for any major deviation from the POD temporal modes.

Here, $E_1$ will be used to indicate the non-calibrated affine approximation error while $E_1^c$ will indicate the same error with calibration terms $\mathbf{e^c} \in \mathbb{R}^m$, $\mathbf{A^c} \in \mathbb{R}^{m \, \times \, m}$, $\mathbf{N^c} \in \mathbb{R}^{m \, \times \, m \, \times \, m}$ and $\mathbf{Q^c} \in \mathbb{R}^{m \, \times \, m \, \times \, m \, \times \, m}$ taken into consideration
\begin{equation}
\begin{split}
    & E_1^c (\mathbf{e^c,A^c,N^c,Q^c})
    = \sum_{i=1}^{m} \int_0^T \Big[
    a_i (t) - \tilde{a}_i (t) \\
    & - \int_{t}^{t+\Delta t_s} \Big(e_i^c + A_{ij}^c a_j (t^{\prime}) + N_{ijk}^c a_j (t^{\prime}) a_k (t^{\prime}) + Q_{ijkl}^c a_j (t^{\prime}) a_k (t^{\prime}) a_l (t^{\prime}) \Big) dt^{\prime}
    \Big]^2 dt
    \mbox{ .}
\end{split}
    \label{eq:error_E2c}
\end{equation}

Adding a regularization term when minimizing error $E_1^c$ can prevent overfitting of the solution which in turn could lead to a better model. Taking this into account, a functional $\mathcal{L}(\mathbf{e^c, A^c, N^c, Q^c})$ is defined as
\begin{equation}
    \mathcal{L} (\mathbf{e^c,A^c,N^c,Q^c}) = 
    E_1^c(\mathbf{e^c, A^c, N^c, Q^c}) + \tilde{\theta} (\mathbf{\|e^c\|_F^2 + \|A^c\|_F^2 + \|N^c\|_F^2 + \|Q^c\|_F^2}) \mbox{ ,}
    \label{eq:calibration}
\end{equation}
\noindent
with
\begin{equation}
    \tilde{\theta} = \frac{1 - \theta}{\theta} \frac{E_1}{\mathbf{\|e\|_F^2 + \|A\|_F^2 + \|N\|_F^2 + \|Q\|_F^2}} \mbox{ ,}
    \label{eq:func_L}
\end{equation}
where the second term on the right hand side of Eq. \ref{eq:calibration} corresponds to the regularization term and $\| \cdot \|_\mathbf{F}$ represents the Frobenius norm. The parameter $\theta \in (0 , 1]$, and values of $\theta$ close to $0$ add more importance to the original ROM while $\theta$ close to $1$ adds a higher weight to the prediction quality of the calibrated model along the training window. It is also relevant to point out that the calibration parameter $\theta$ can be regarded as a regularization parameter in the Tikhonov regularization methodology \cite{Bourguet_01}.

Minimizing the functional $\mathcal{L}$ gives rise to the linear system
\begin{equation}
    \mathcal{L} (\mathbf{K^c}) = 
    \sum_{i=1}^{m} \int_{0}^{T} \left(
    a_i(t) - \tilde{a}_i(t) - 
    \int_{t}^{t+\Delta t_s} \sum_{j=1}^{m^{*}} K_{ij}^{c} a_j^{*} (t^{\prime})  dt^{\prime}
    \right)^2 dt + \tilde{\theta} \mathbf{\|K^c\|_F^2} \mbox{ ,}
    \label{eq:functional_L}
\end{equation}

\noindent
where $\mathbf{a^{*}} = [1 \;\;  \mathbf{a} \;\; \mathbf{a} \hat{\otimes} \mathbf{a} \;\; \mathbf{a} \hat{\otimes} \mathbf{a} \hat{\otimes} \mathbf{a}] \in \mathbb{R}^{m^{*}} \times [0,T]$ is the enriched temporal mode vector. Here, $\otimes$ symbolize the column-wise Kronecker product, which for a column vector $\mathbf{a} = [a_1, \ldots , a_m]^\top \in \mathbb{R}^m$ is given by 

\begin{equation}
    \mathbf{a} \otimes \mathbf{a} = [a_1^2 \;\; a_1 a_2 \; \ldots \; a_1 a_m \;\; a_2 a_1 \; \ldots \; a_2^2 \; \ldots \; a_2 a_m \; \ldots \; a_m^2] \in \mathbb{R}^{m^2}
    \mbox{ ,}
\end{equation}

\noindent
and $\hat{\otimes}$ corresponds to the (modified) Kronecker product without duplicate terms (for instance, $a_1 a_2 = a_2 a_1$ and $a_1 a_1 a_2 = a_1 a_2 a_1 = a_2 a_1 a_1$). Similarly, the different calibration coefficients are grouped such that $\mathbf{K^c = [e^c, \; A^c, \; \hat{N}^c, \; \hat{Q}^c]} \in \mathbb{R}^{m \, \times \, m^{*}}$. Moreover, $\hat{\mathbf{N}}^c \in \mathbb{R}^{m \, \times \, m(m+1)/2}$ and $\hat{\mathbf{Q}}^c \in \mathbb{R}^{m \, \times \, m(m+1)(m+2)/6}$ are, respectively, the matricized equivalents of tensors $\mathbf{N}^c$ and $\mathbf{Q}^c$ without redundant coefficients. Failure to eliminate duplicate terms leads to a dependent (ill-posed) linear system of equations. For the calibration terms being considered, the dimension $m^{*}$ is given by 
\begin{equation}
    m^{*} = 1 + m + \frac{m (m + 1)}{2} + \frac{m (m+1) (m+2)}{6} \mbox{ .}
\end{equation}

For a given parameter $\theta$, the optimality condition $\partial \mathcal{L} / \partial K_{ij}^c = 0$ leads to solving $m$ linear systems of size $m^{*}$ or, for $i = 1, \dots , m$
\begin{equation}
    \mathbf{D^T  K_i^c = b^i} \mbox{ ,}
\end{equation}
\noindent
where $\mathbf{K_i^c}$ is the $i^{th}$ row of $\mathbf{K^c}$,
\begin{equation}
    D_{ij} = \int_{0}^{T} 
    \left( \int_{t}^{t+\Delta t_s} a_i^{*} (t^{\prime}) dt^{\prime} \right)
    \left( \int_{t}^{t+\Delta t_s} a_j^{*} (t^{\prime}) dt^{\prime} \right)
    dt + \tilde{\theta} \delta_{ij}
\end{equation}

\noindent
and
\begin{equation}
    b_i^j = \int_{0}^{T} 
    \bigg(a_j (t) - \tilde{a}_j (t) \bigg) 
    \bigg( \int_{t}^{t+\Delta t_s} a_i^{*} (t^{\prime}) dt^{\prime} \bigg)
    dt \mbox{ .}
\end{equation}
Matrix $\mathbf{D} \in \mathbb{R}^{m^{*} \, \times \, m^{*}}$ is calculated only once while vector $\mathbf{b^i}$ must be evaluated for each mode during the calibration phase. Further details and a thorough discussion can be found in \cite{bourguet_these}.

An appropriate value for the calibration (regularization) parameter $\theta$ needs to be chosen. This can be done systematically with the L-curve method and is discussed by \cite{elmajd_01,elmajd_02} in a reduced order modeling context. This method is based on the examination of the curve representing the norm of the regularized solution $\mathbf{\|K^c\|_F}$ versus a corresponding residual norm ($E_1^c$ in this paper). Typically, this curve exhibits an ``L" shape such as the blue curves in Fig. \ref{fig:cylinder_lcurves} if the problem is ill-conditioned or contains noise. The corner of the L-curve usually represents a fair trade-off between the  solution norm $\mathbf{\|K^c\|_F}$ (ordinate) and minimization of the error norm $E_1^c$ (abscissa). For a typical well-behaved L-curve, the identification of the corner can be performed by curvature maximization and  corresponds to the maximum curvature point \cite{elmajd_02}.

Usually the L-curve is obtained after minimizing the functional $\mathcal{L(\mathbf{K^c})}$ in Eq. \ref{eq:functional_L} for different values of $\theta \in (0,1]$ \cite{zucatti_jcp,zucatti_02_scitech,elmajd_01}. In this work, an iterative heuristic approach is preferred. To begin with, a first set of calibration terms is determined for a suitably chosen calibration parameter $\theta_1$ and the non-calibrated ROM. Afterwards, these calibration coefficients are added to the original ROM coefficients. Next, a second set of calibration terms is obtained using the modified reduced order coefficients as a replacement for the non-calibrated coefficients combined with a second regularization parameter $\theta_2$. At each iteration, the error norm $E_1^c$ and the norm of the calibration coefficients (recovered after subtracting the original terms) are computed. The iterations only come to a halt when an appropriate L-curve is obtained. In practice, the regularization parameters are chosen heuristically. If too big, the calibration coefficients could already be to the left side of the L-curve corner. On the other hand, too small regularization parameters may lead to a substantial number of iterations. Moreover, this approach allows to systematically take into consideration the errors associated with the time integration of the calibration coefficients. As a consequence, calibration coefficients obtained by this procedure could perform better compared to their counterparts obtained by solving a single optimization problem.

\section{Results and Discussion}

This section presents results of Galerkin and LSPG ROMs for two test cases: a compressible cylinder flow at low Reynolds number, and a turbulent flow over a plunging airfoil under deep dynamic stall. An assessment of linear and non-linear calibration is presented for the ROMs computed with and without hyper-reduction.

\subsection{Cylinder flow}

A compressible cylinder flow is studied to perform an initial assessment of the Galerkin and Petrov-Galerkin ROMs. The effects of linear and non-linear calibration are analyzed for both models. The flow has a freestream Mach number $M_{\infty}=0.4$ and the Reynolds number based on the cylinder diameter is set as $Re_D=150$. Results obtained from the FOM are sampled for $1{,}120$ snapshots with constant non-dimensional time step $\Delta t_{s} = 0{,}125$. The first $560$ snapshots, which represent $70$ non-dimensional time units based on freestream velocity, are used to construct a reduced order basis by the snapshot POD method introduced in section \ref{subsection:POD}. The computational O-grid employed in the simulation has $420 \times 599$ points in the streamwise and wall-normal directions, respectively.

The optimality property of the POD method is expected to produce a basis where only a small number of modes is necessary to reconstruct the input data and, thus, the benefit of adding further modes is expected to rapidly decay. Usually, the number of POD modes used in the ROM is chosen according to the relative information content $\mbox{RIC} (m) = \sum_{i = 1}^{m} \lambda_i / \sum_{i = 1}^{M} \lambda_i$ and should satisfy a predefined threshold. For the problem at hand, only four modes ($98{,}87\%$ of the model RIC) are used in the ROM reconstructions. The implicit Euler method is the chosen time-marching scheme and spatial derivatives of the POD basis are computed using a $10th$-order explicit central finite difference scheme for both Galerkin and LSPG methods. Although Zucatti {\em et al.} \cite{zucatti_01} show that high-resolution methods can be beneficial, calibration should be able to compensate for this source of error. Some tests were also made using a $2nd$-order central finite difference scheme and resulted in ROMs producing the same outcome but requiring slightly more invasive calibration operators (i.e., with higher $\| \cdot \|_{\mathbf{F}}$ norms) compared to the $10th$-order method. Also, higher-order spatial derivatives imply in little pre-computation overhead to the models. Finally, reduced order models obtained either by Galerkin or LSPG schemes are expressed, after spatial and temporal discretization, as a set of coupled non-linear algebraic system of equations which are solved using the Levenberg-Marquardt algorithm.

The performance of both linear and non-linear calibration coefficients is evaluated. Here, linear calibration (using only constant and linear coefficients) translates to minimizing the functional $\mathcal{L}$ of Eq. \ref{eq:functional_L} for $\mathbf{e^c}$ and $\mathbf{A^c}$ ($\mathbf{K^c = [e^c, A^c]}$). Similarly, non-linear calibration (considering linear and non-linear coefficients) alludes to the same minimization problem but with higher order tensors ($\mathbf{N^c}$ and $\mathbf{Q^c}$) also being taken into consideration. Specifically, non-linear calibration of the LSPG ROMs also adds operators $\mathbf{N^c}$ and $\mathbf{Q^c}$ ($\mathbf{K^c = [e^c, A^c, \hat{N}^c, \hat{Q}^c]}$) but non-linear calibration of the Galerkin ROMs only adds $\mathbf{N^c}$ ($\mathbf{K^c = [e^c, A^c, \hat{N}^c]}$). 

\begin{figure}[hbt!]
    \centering
    \begin{subfigure}[H]{0.45\textwidth}
        \centering
        \includegraphics[width=1.\textwidth,trim={0mm 0mm 0mm 0mm},clip]{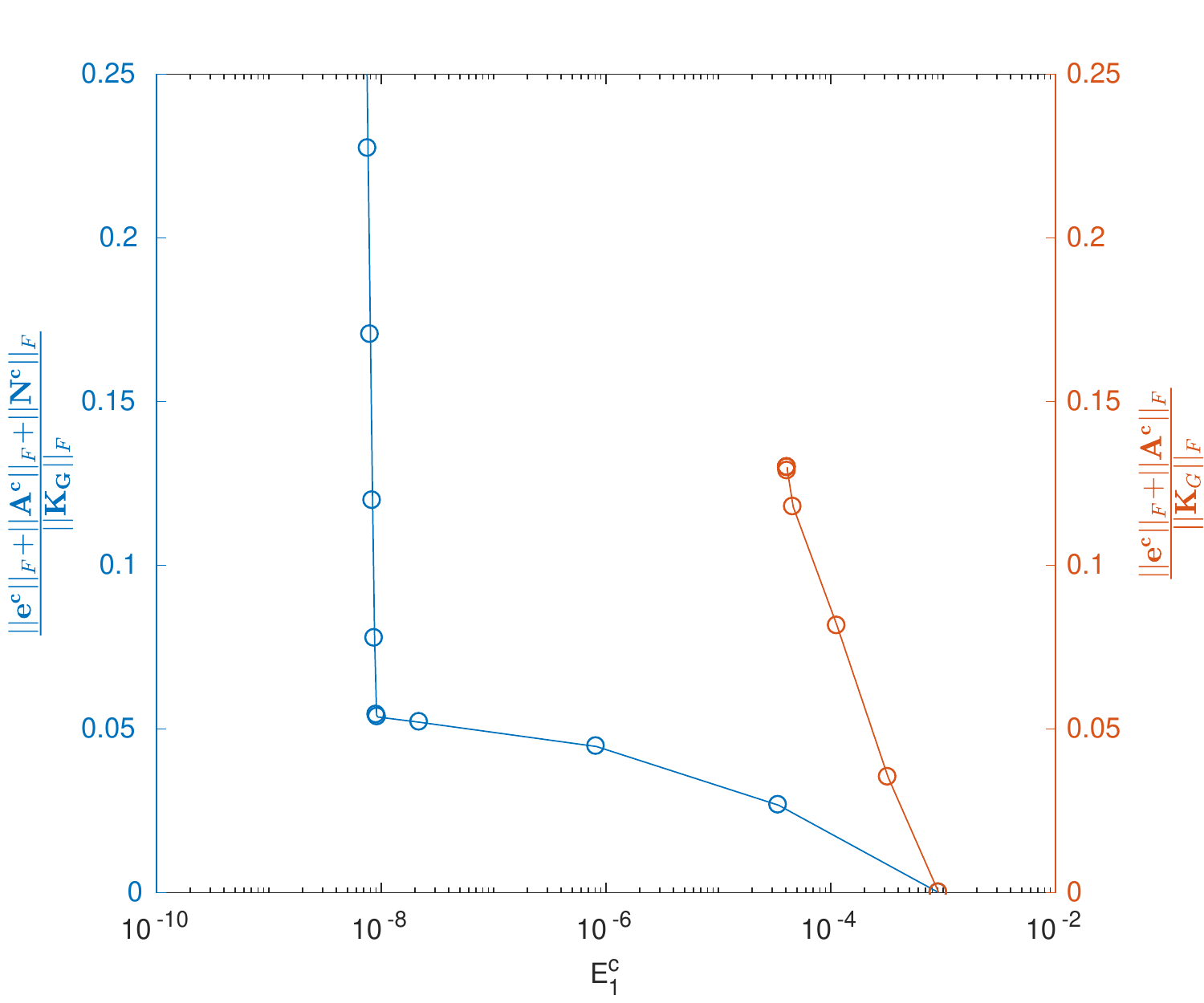}
        \caption{Galerkin L-curves.}
    \end{subfigure}
    ~
    \begin{subfigure}[H]{0.45\textwidth}
        \centering
        \includegraphics[width=1.\textwidth,trim={0mm 0mm 0mm 0mm},clip]{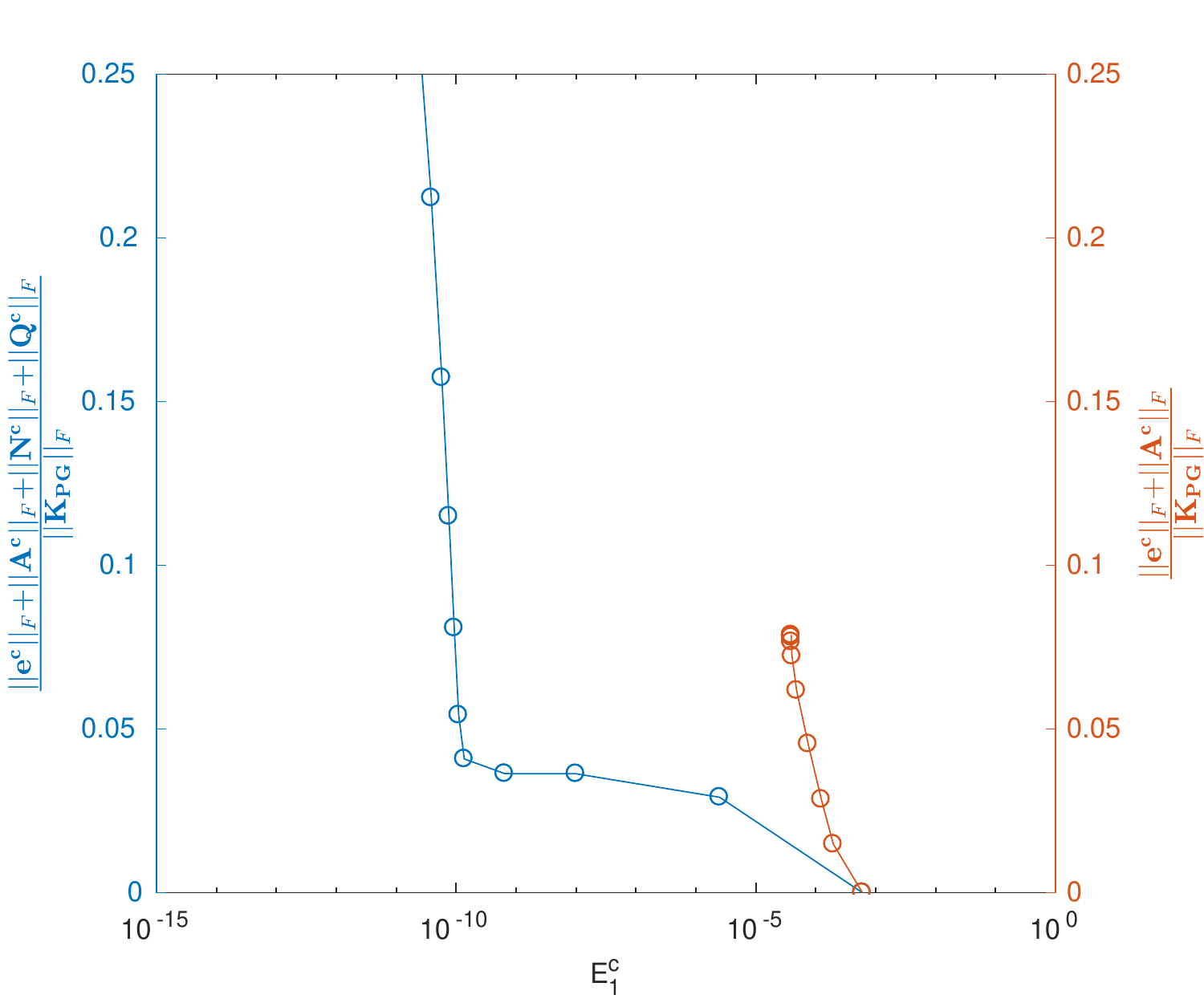}
        \caption{LSPG L-curves.}
    \end{subfigure}
    \caption{Ratio of Frobenius norms computed for calibrated and original ROM coefficients as a function of the approximation error using the iterative procedure described in section \ref{calibration}.}
    \label{fig:cylinder_lcurves}
\end{figure}

As discussed in section \ref{calibration}, calibration may require regularization to avoid adding coefficients resulting from an ill-posed minimization problem. A suitable set of calibration coefficients for each ROM is determined after analyzing the L-curves from Fig. \ref{fig:cylinder_lcurves}. Here, $\mathbf{K_G}$ and $\mathbf{K_{PG}}$ correspond to the grouped calibration terms of the original non-calibrated Galerkin and LSPG ROMs, respectively.
For a similar approximation error $E_1^c$, the sum of the Frobenius norms $\| \cdot \|_F$ of the linear calibration coefficients is larger than that computed for the non-linear calibration. Therefore, the non-linear calibration operators are less intrusive until the L-curve reaches a corner. 
As can be also observed in Fig. \ref{fig:cylinder_lcurves}, the coefficients from calibration are nearly an order of magnitude smaller compared to those from the original ROMs for both the Galerkin and LSPG models. Moreover, the latter method exhibits calibration coefficients which are less intrusive compared to their counterparts from the Galerkin method.
It should be pointed out that $E_1^c$ serves only as an error indicator and a better model performance with respect to other calibrated ROMs is uncertain.

When calibration with non-linear coefficients is performed, a rapid growth of the relative norm is observed without further reduction of $E_1^c$ when a threshold is met. In these cases, the coefficients corresponding to the corner of the L-curve are selected as the suitable calibration operators. On the other hand, linear calibration operators converge in norm for both LSPG and Galerkin methods and coefficients are selected after convergence.

\begin{figure}[hbt!]
    \centering
    \begin{subfigure}[H]{0.45\textwidth}
        \centering
        \includegraphics[width=1.\textwidth,trim={2mm 2mm 10mm 5mm},clip]{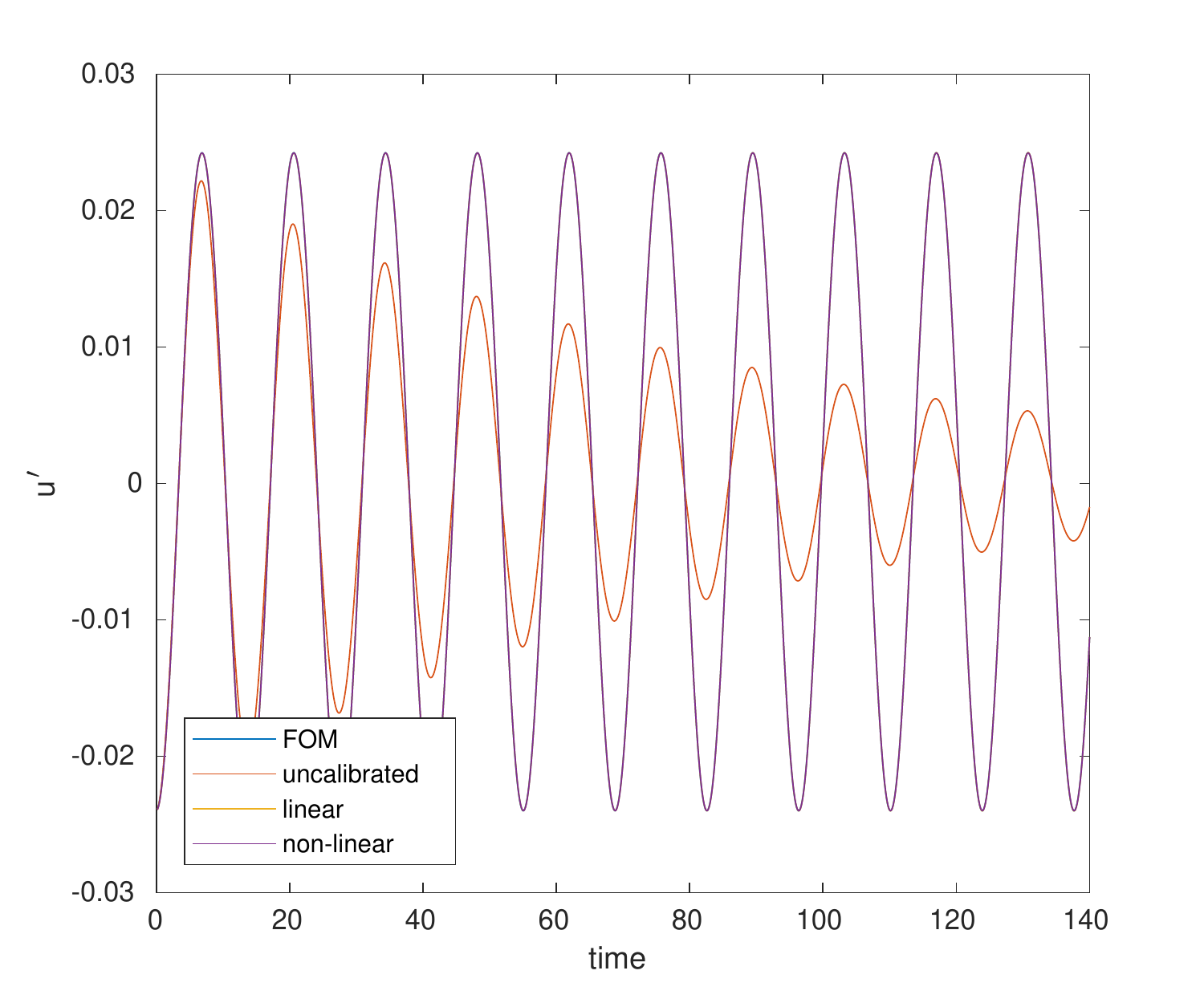}
        \caption{Galerkin models.}
    \end{subfigure}
    ~
    \begin{subfigure}[H]{0.45\textwidth}
        \centering
        \includegraphics[width=1.\textwidth,trim={2mm 2mm 10mm 5mm},clip]{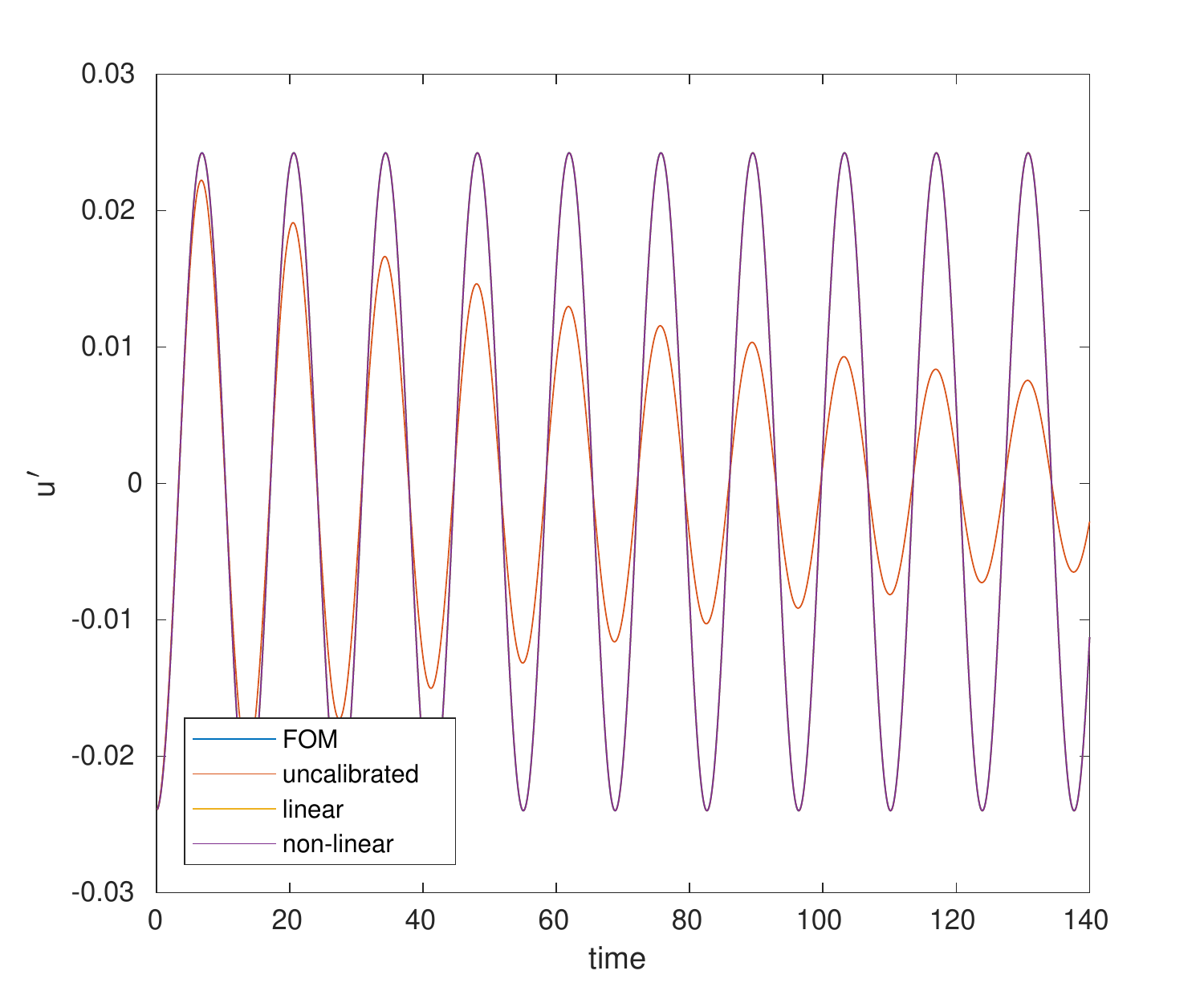}
        \caption{LSPG models.}
    \end{subfigure}
    \caption{Time histories of u-velocity fluctuations computed by the implicit Euler time-marching scheme at a probe located on the cylinder wake.}
    \label{fig:cylinder_probes}
\end{figure}

Figure \ref{fig:cylinder_probes} shows the time histories of u-velocity fluctuations for the FOM, Galerkin and LSPG models with and without calibration. Solutions are presented for the training region ($0 \leq t \leq 70$) and for an extrapolation period for which FOM solutions are available ($70 < t \leq 140$). In this case, calibration of both linear and non-linear terms is evaluated. All calibrated models are visually indistinguishable from the FOM solution in Fig. \ref{fig:cylinder_probes}. In comparison, very dissipative ROM solutions are obtained when the implicit Euler method is employed without calibration. Specifically, the uncalibrated LSPG model is less dissipative than the uncalibrated Galerkin. The present first-order implicit time-marching scheme is generally inappropriate for long-time integration in spite of the good stability properties. 
It is important to remind that each calibrated ROM is computed for a specific time step $\Delta t_s$ and altering this parameter after the model is already generated can be harmful. For example, time-step refinement diminishes the dissipative nature of the implicit Euler method which may lead to unstable solutions. As a positive effect, calibration allows for improved solution accuracy even with simple time-marching schemes.

\begin{figure}[hbt!]
    \centering
    \begin{subfigure}[H]{.48\textwidth}
        \centering
        \includegraphics[width=1.\textwidth,trim={60mm 75mm 181.5mm 75mm},clip]{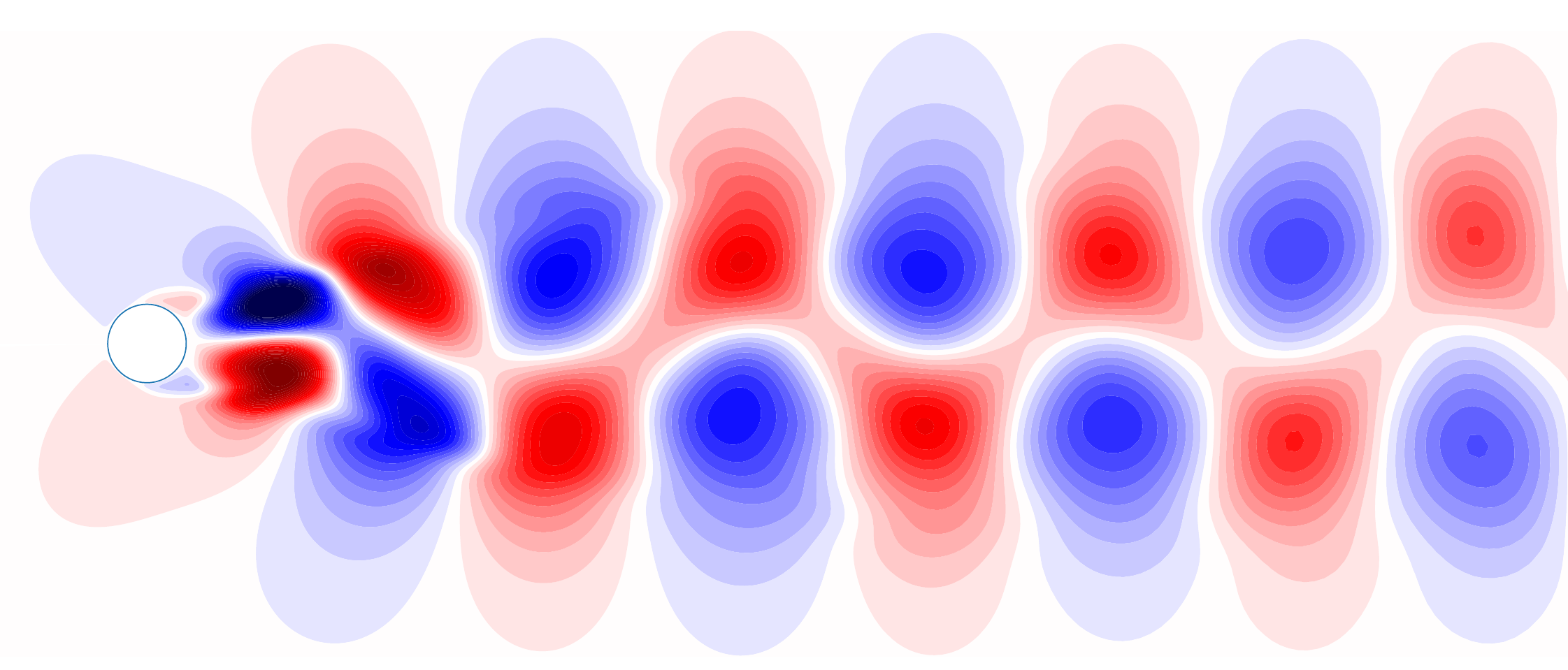}
        \caption{Galerkin ROM with calibrated linear terms.}
    \end{subfigure}
    ~
    \begin{subfigure}[H]{.48\textwidth}
        \centering
        \includegraphics[width=1.\textwidth,trim={60mm 75mm 181.5mm 75mm},clip]{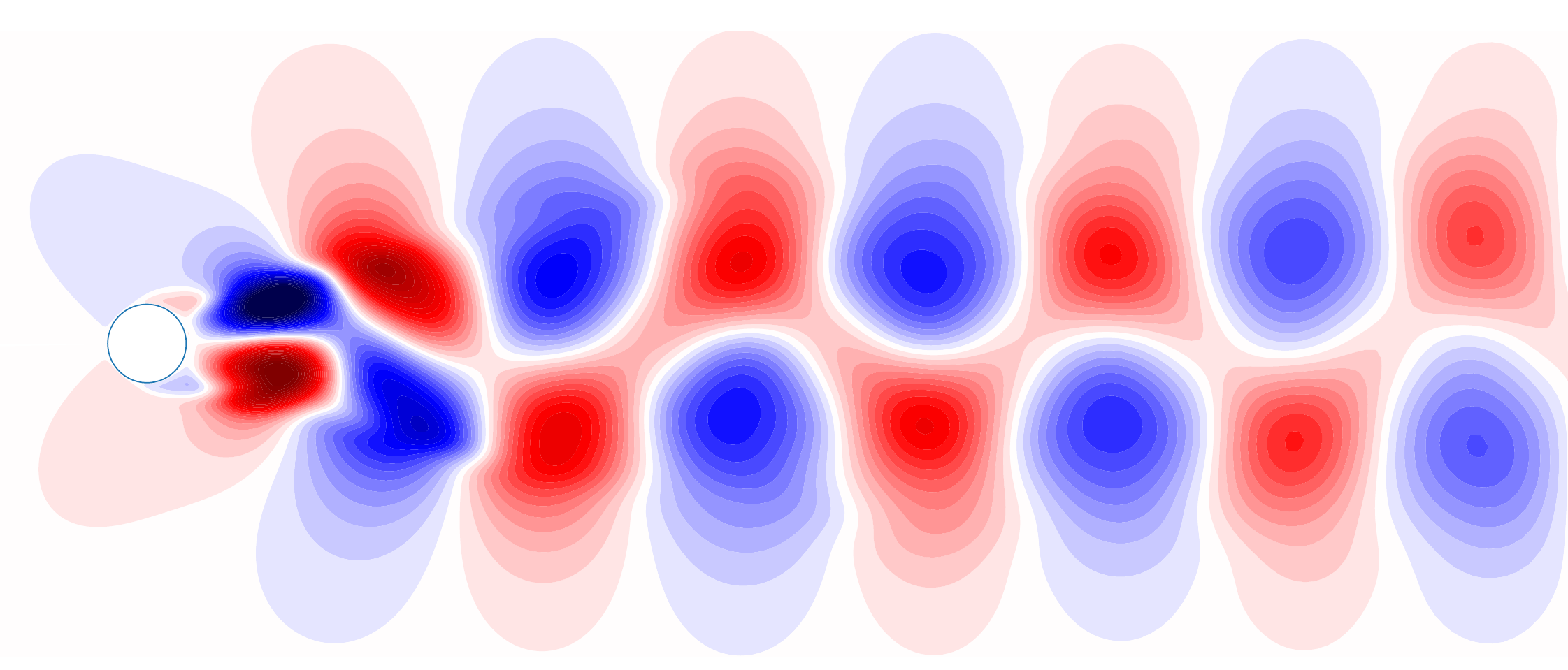}
        \caption{LSPG ROM with calibrated linear terms.}
    \end{subfigure}
    \begin{subfigure}[H]{.48\textwidth}
        \centering
        \includegraphics[width=1.\textwidth,trim={60mm 75mm 181.5mm 75mm},clip]{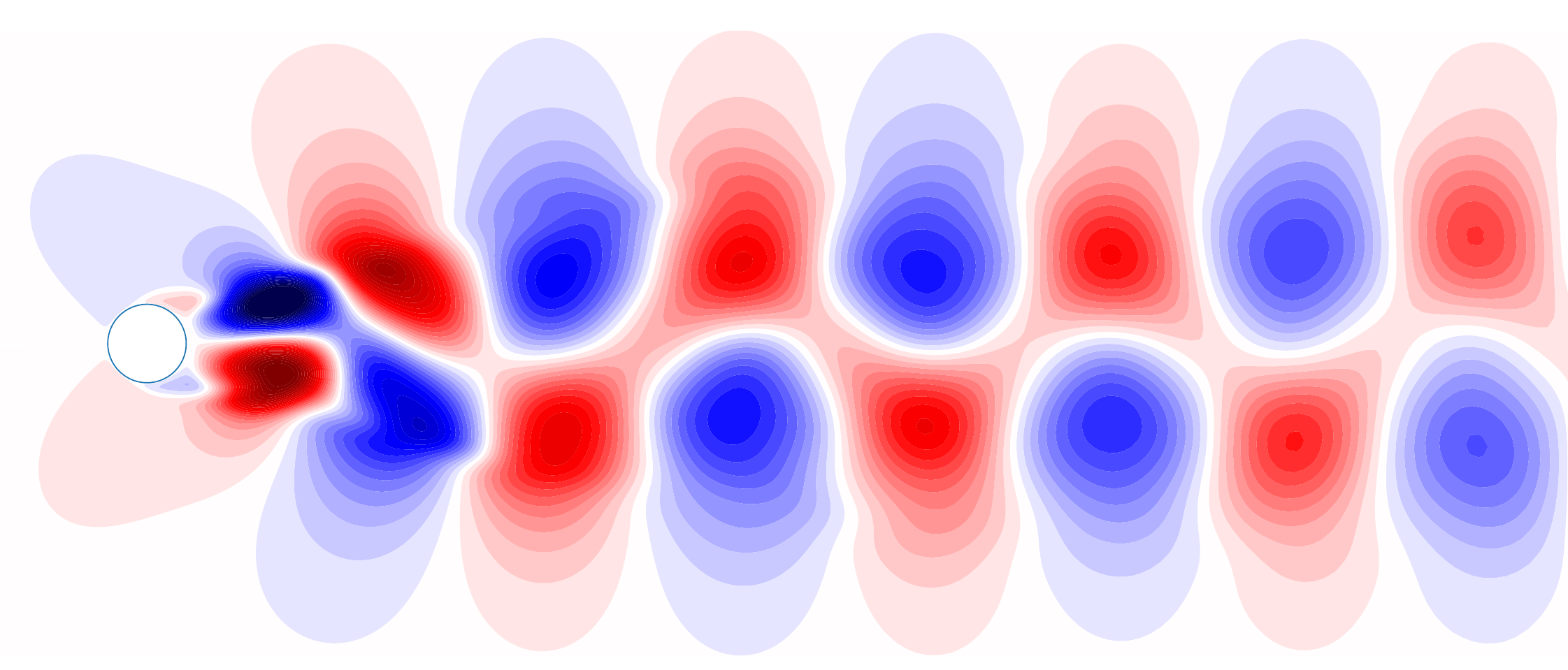}
        \caption{Galerkin ROM with calibrated non-linear terms.}
    \end{subfigure}
    ~
    \begin{subfigure}[H]{.48\textwidth}
        \centering
        \includegraphics[width=1.\textwidth,trim={60mm 75mm 181.5mm 75mm},clip]{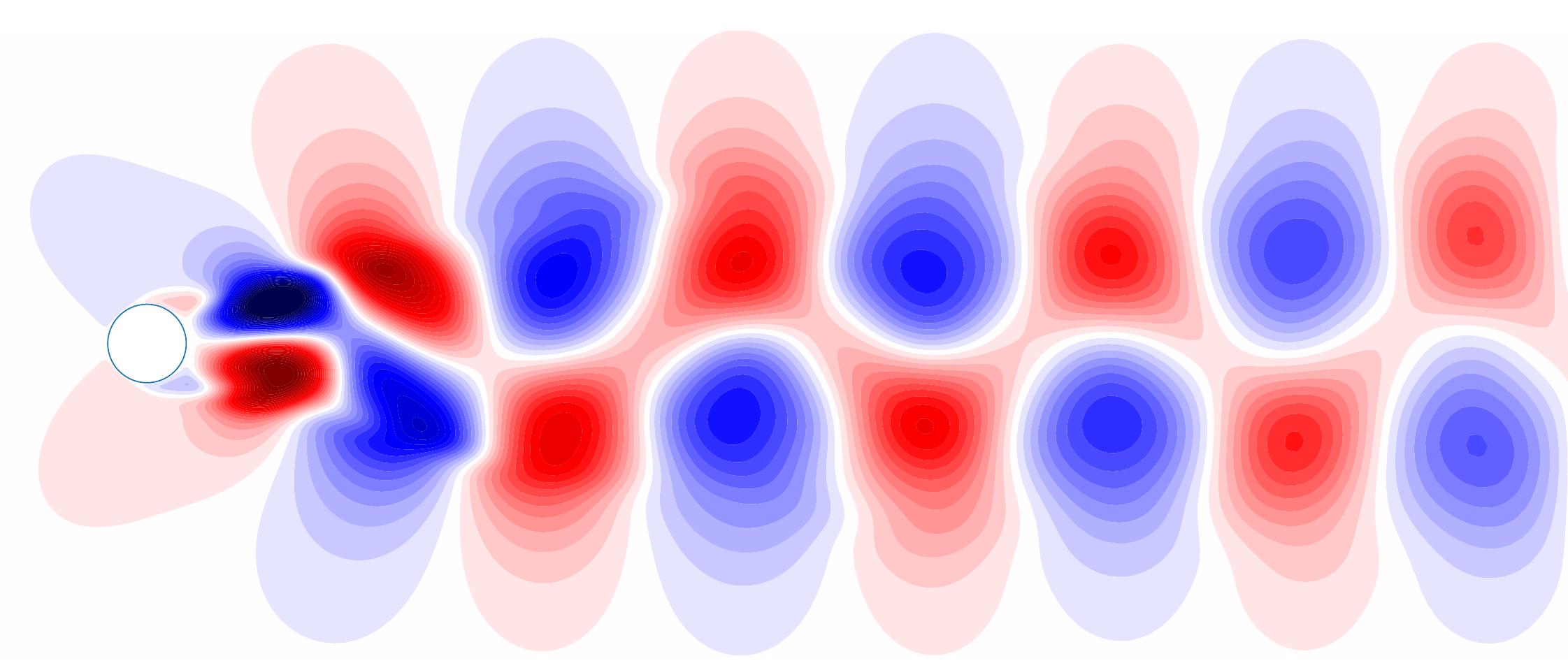}
        \caption{LSPG ROM with calibrated non-linear terms.}
    \end{subfigure}
    \begin{subfigure}[H]{.48\textwidth}
        \centering
        \includegraphics[width=1.\textwidth,trim={60mm 75mm 181.5mm 75mm},clip]{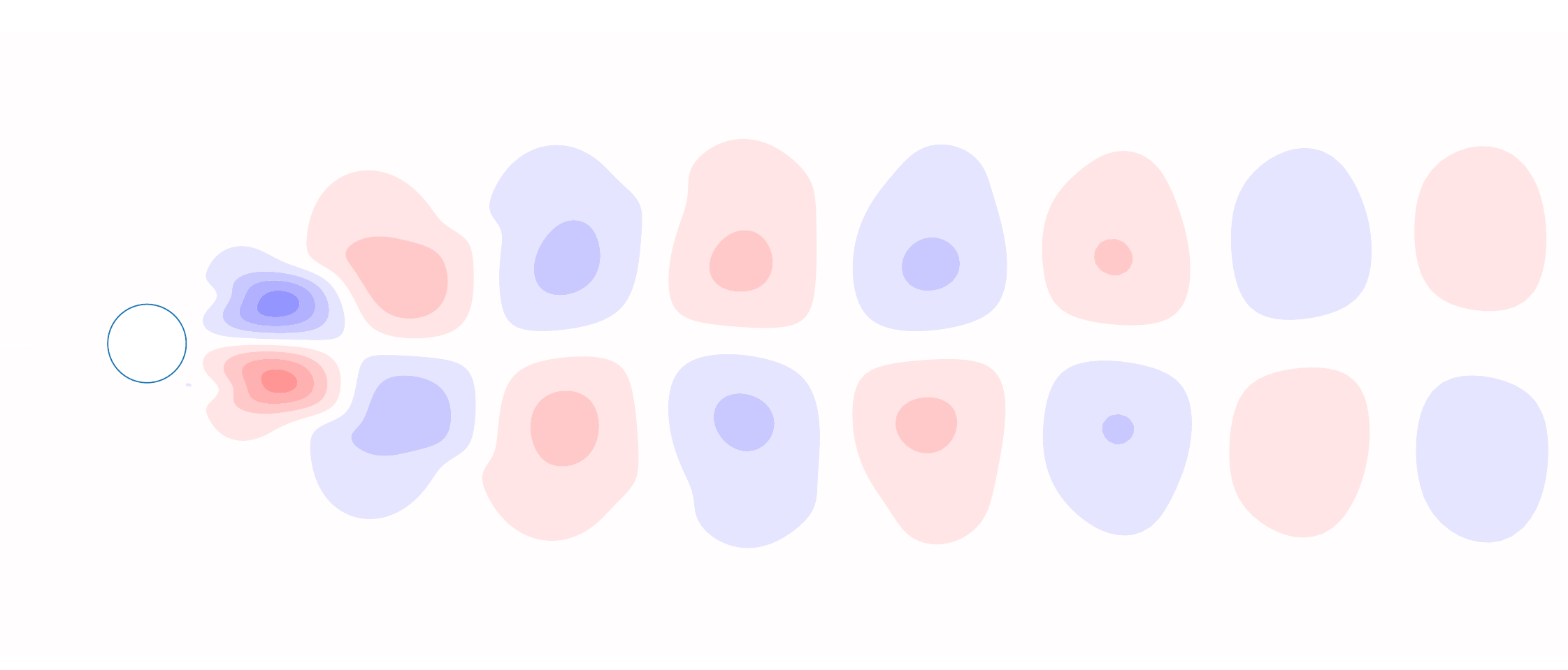}
        \caption{Uncalibrated Galerkin ROM.}
    \end{subfigure}
    ~
    \begin{subfigure}[H]{.48\textwidth}
        \centering
        \includegraphics[width=1.\textwidth,trim={60mm 75mm 181.5mm 75mm},clip]{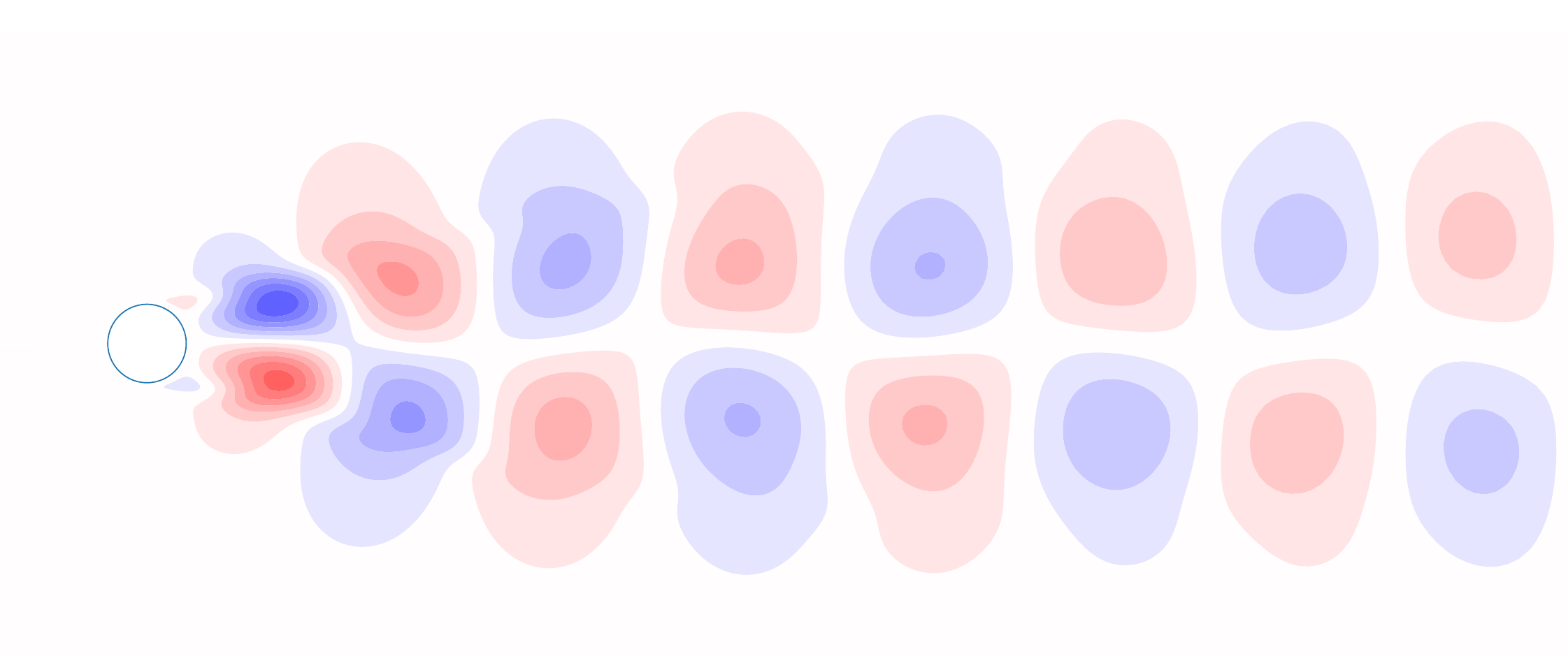}
        \caption{Uncalibrated LSPG ROM.}
    \end{subfigure}
    \begin{subfigure}[H]{.48\textwidth}
        \centering
    \includegraphics[width=1.\textwidth,trim={60mm 75mm 181.5mm 75mm},clip]{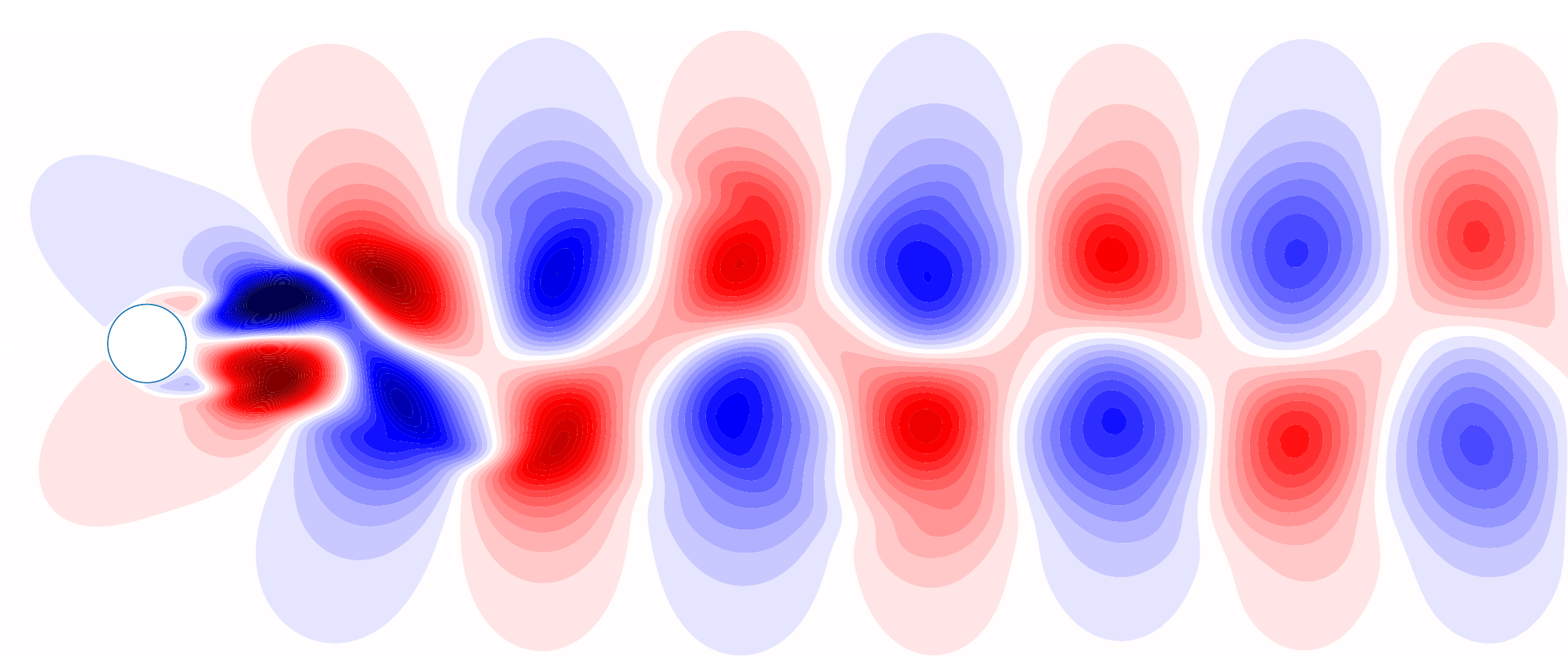}
        \caption{Full-order model.}
    \end{subfigure}
    \caption{Velocity fluctuations $u'$ obtained for different ROMs and the FOM at $t=140$. The color limits are kept the same for all figures $-0.20 < u'< 0.20$.}
    \label{fig:cylinder_snapshots}
\end{figure}

Figure \ref{fig:cylinder_snapshots} shows the u-velocity fluctuation contours for the FOM and ROMs considered. Solutions are computed at $t=140$ and the contour levels are kept equal for all cases to provide a fair comparison. All calibrated ROM solutions present a good comparison to the FOM with only very small discrepancies. On the other hand, the uncalibrated ROMs perform very poorly. Despite the small phase error, one can clearly see the effects of the diminishing magnitudes of the fluctuation field accused by the probes in Fig. \ref{fig:cylinder_probes}. In particular, the magnitudes of the fluctuation field of the uncalibrated Galerkin ROM is generally smaller compared to that of the uncalibrated LSPG ROM.

\subsection{Deep dynamic stall of a plunging airfoil}

This second test case investigates a periodically plunging SD7003 airfoil undergoing deep dynamic stall. The airfoil is subject to a sinusoidal motion with reduced frequency $k = \pi f L / U_{\infty} = 0.5$, where $f$ in the plunging frequency, $L$ is the chord of the airfoil and $U_{\infty}$ in the reference free stream velocity. The motion amplitude is set as $h_0 / L = 0.5$ with a static angle of attack $\alpha_0 = 8^{\circ}$. The chord Reynolds number based on the freestream velocity is $Re = \text{60,000}$ and the freestream Mach number is $M = 0.1$. This flow configuration is interesting for micro air vehicle applications and further details of the flow setup and the LES methodology employed in the present case can be found in \cite{brener_PhysRevFluids}.
\begin{figure}[hbt!]
    \centering
    \begin{subfigure}[H]{0.49\textwidth}
        \centering
        \includegraphics[width=1.\textwidth,trim={20mm 10mm 5mm 90mm},clip]{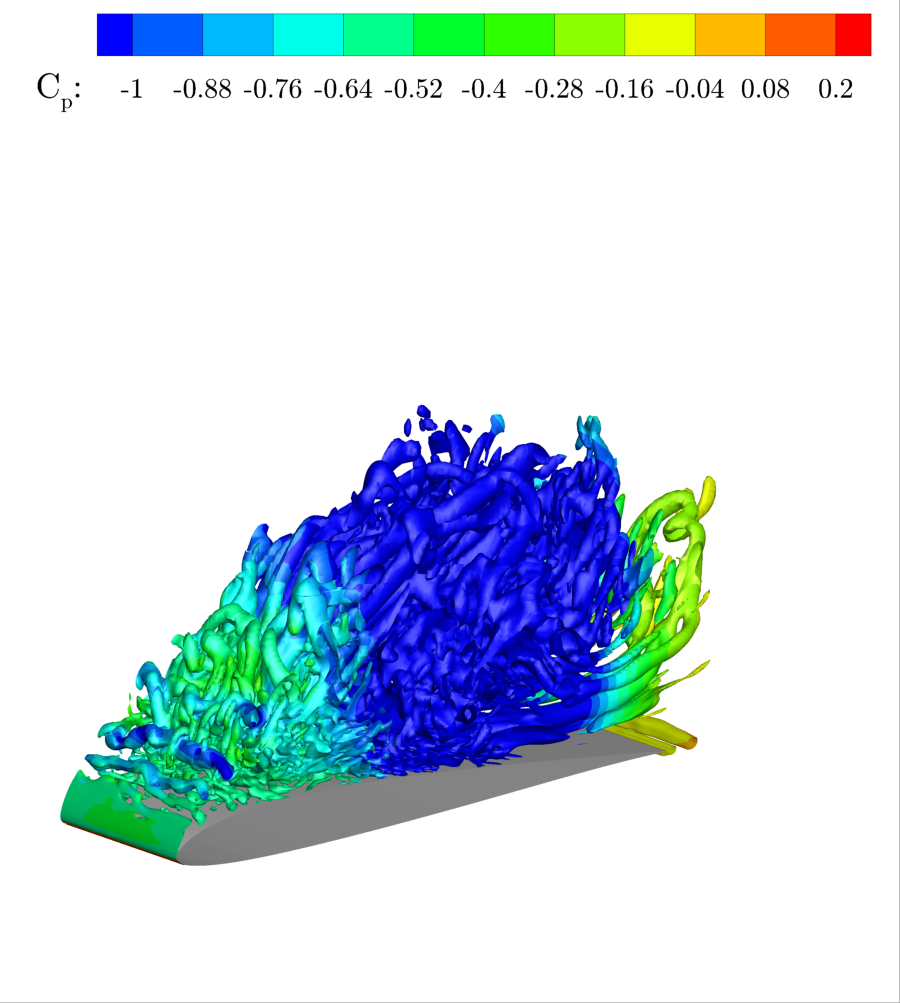}
    \end{subfigure}
    \begin{subfigure}[H]{0.49\textwidth}
        \centering
        \includegraphics[width=1.\textwidth,trim={20mm 10mm 5mm 90mm},clip]{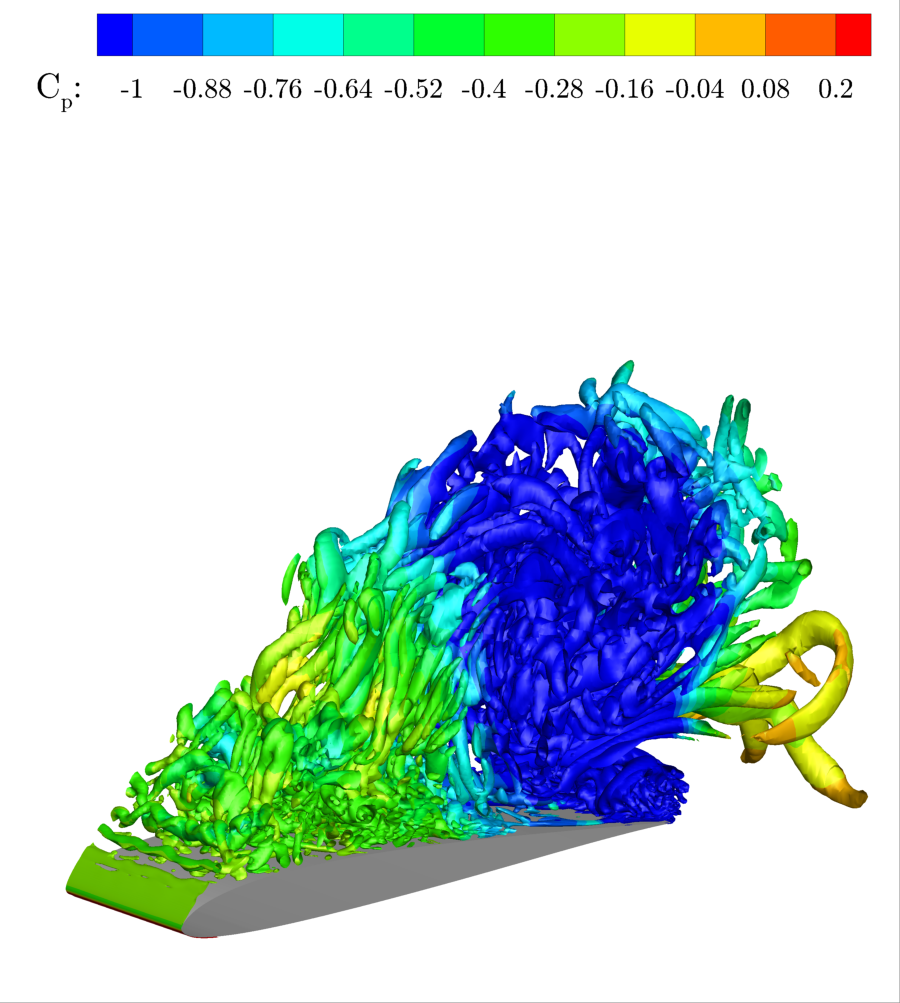}
    \end{subfigure}
    \caption{Iso-surfaces of Q-criterion colored by pressure at two instants of the plunge motion.}
    \label{fig:airfoil_q_crit}
\end{figure}

The mesh configuration is a body-fitted O-grid with $481 \times 351 \times 96$ points in the streamwise, wall-normal and spanwise directions, respectively. The iso-surfaces of Q-criterion colored by pressure are shown in Fig. \ref{fig:airfoil_q_crit}. In the figure, some of the main flow features can be observed, for example the massive leading edge vortex which is transported over the airfoil suction side, and a trailing edge vortex that develops during the dynamic stall process. A detailed analysis of the present flow can be found in \cite{brener_PhysRevFluids}.
\begin{figure}[hbt!]
    \centering
    \begin{subfigure}[H]{0.45\textwidth}
        \centering
        \includegraphics[width=.9\textwidth,trim={0mm 0mm 0mm 0mm},clip]{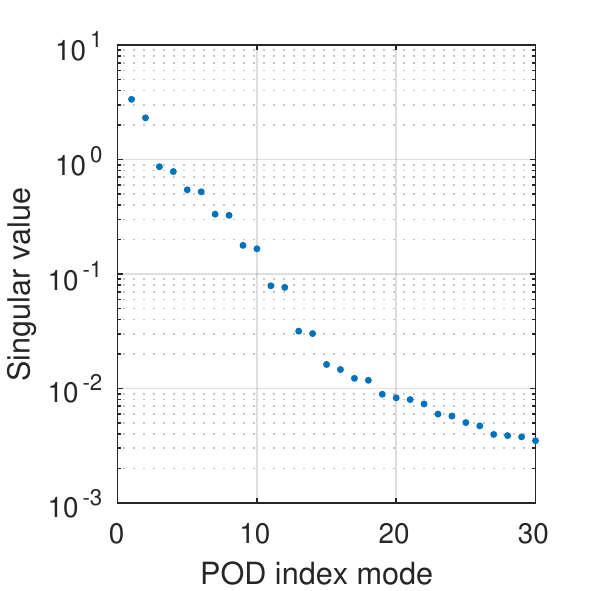}
    \end{subfigure}
    ~
    \begin{subfigure}[H]{0.45\textwidth}
        \centering
        \includegraphics[width=.9\textwidth,trim={0mm 0mm 0mm 0mm},clip]{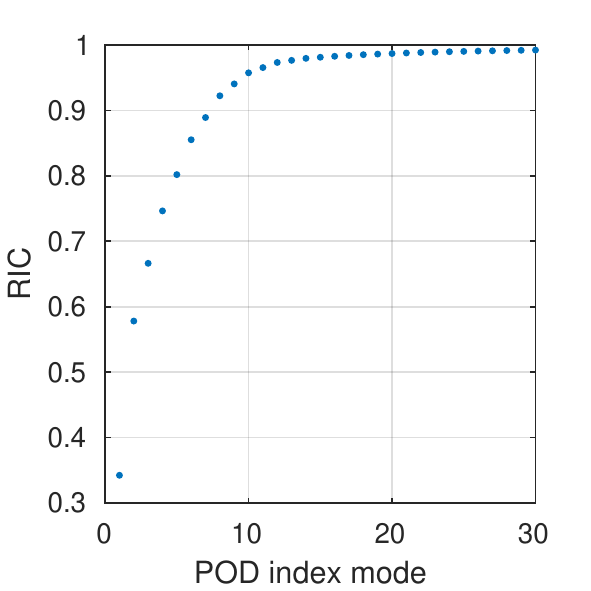}
    \end{subfigure}
    \caption{Spectrum of singular values (left) and relative information content (right).}
    \label{fig:airfoil_sv_ric}
\end{figure}

This problem was studied previously in the context of both data-driven \cite{lui_wolf_2019} and projection-based reduced order modeling \cite{zucatti_02_scitech}. In \cite{lui_wolf_2019}, a ROM is constructed via deep neural networks (DNNs) and a comparison with sparse regression \cite{Brunton3932} is provided. The DNN method was capable of learning and reproducing the transient features of the flow with great accuracy in both training and testing windows. On the other hand, a calibration approach with linear coefficients based in minimizing the error norm given by Eq. \ref{eq:error_ode}, and without pre-computation of the LSPG coefficients, was performed in \cite{zucatti_02_scitech}. Results showed that linear calibration significantly improves stability and accuracy for both Galerkin and LSPG ROMs. The impact of hyper-reduction was also evaluated. In particular, the work developed in \cite{zucatti_02_scitech} was fundamental to the current article and will be further discussed throughout the text when appropriate.

Reduced order modeling is tested using 2,244 spanwise-averaged snapshot solutions (i.e., 2D solutions) of the flow with a constant non-dimensional time step $\Delta t = 0.036$. Half of these snapshots (three plunging cycles) are used to construct a reduced order basis by the snapshot POD method. An accurate reconstruction of the dataset from a small POD basis is expected and, as shown by the present results, the benefits of including additional modes rapidly vanish. The fast magnitude decay of the singular values for the first 30 most energetic modes (out of 1,122) is presented in Fig. \ref{fig:airfoil_sv_ric}. The magnitudes are similar for mode pairs, what indicates that such modes contain similar frequency information, differing almost exclusively with respect to phase. Additionally, the evolution of the RIC for these singular values is also presented in Fig. \ref{fig:airfoil_sv_ric} and correspond to a RIC of $99.23 \%$.

\begin{figure}[hbt!]
    \centering
    \begin{subfigure}[H]{.3\textwidth}
        \centering
        \includegraphics[width=1.\textwidth,trim={100mm 30mm 100mm 30mm},clip]{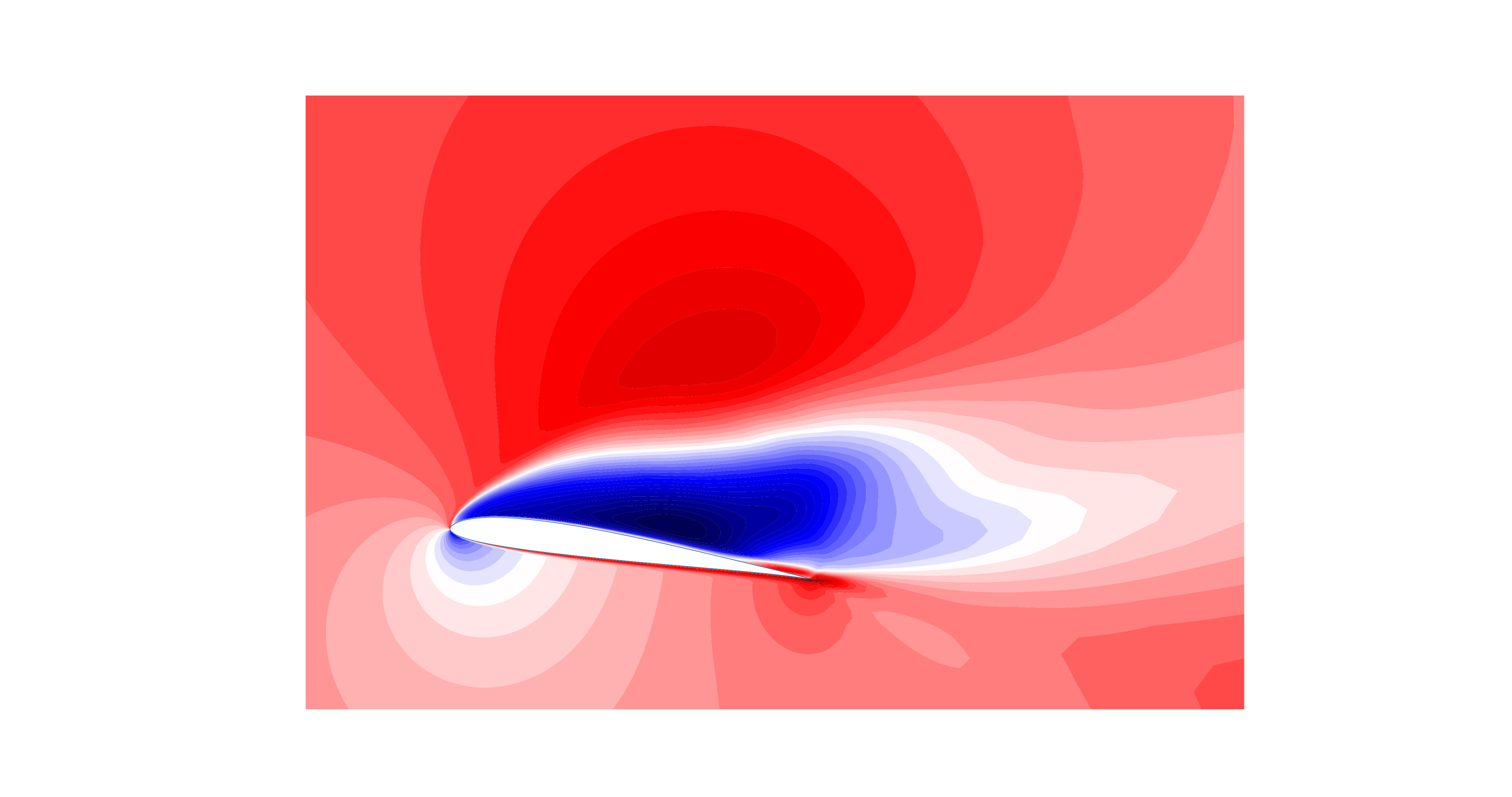}
    \end{subfigure}
    ~
    \begin{subfigure}[H]{.3\textwidth}
        \centering
        \includegraphics[width=1.\textwidth,trim={100mm 30mm 100mm 30mm},clip]{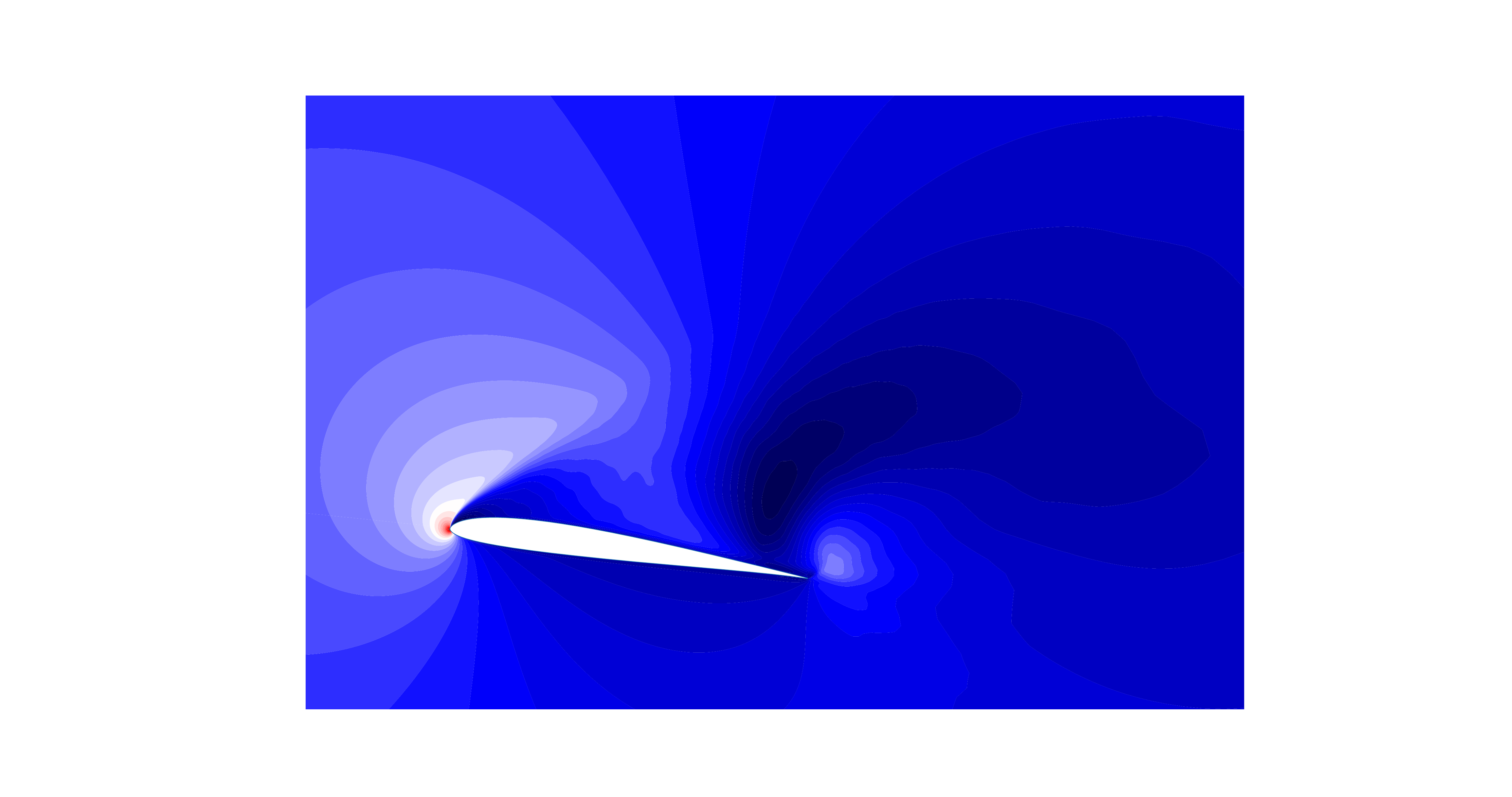}
    \end{subfigure}
    ~
    \begin{subfigure}[H]{.3\textwidth}
        \centering
        \includegraphics[width=1.\textwidth,trim={100mm 30mm 100mm 30mm},clip]{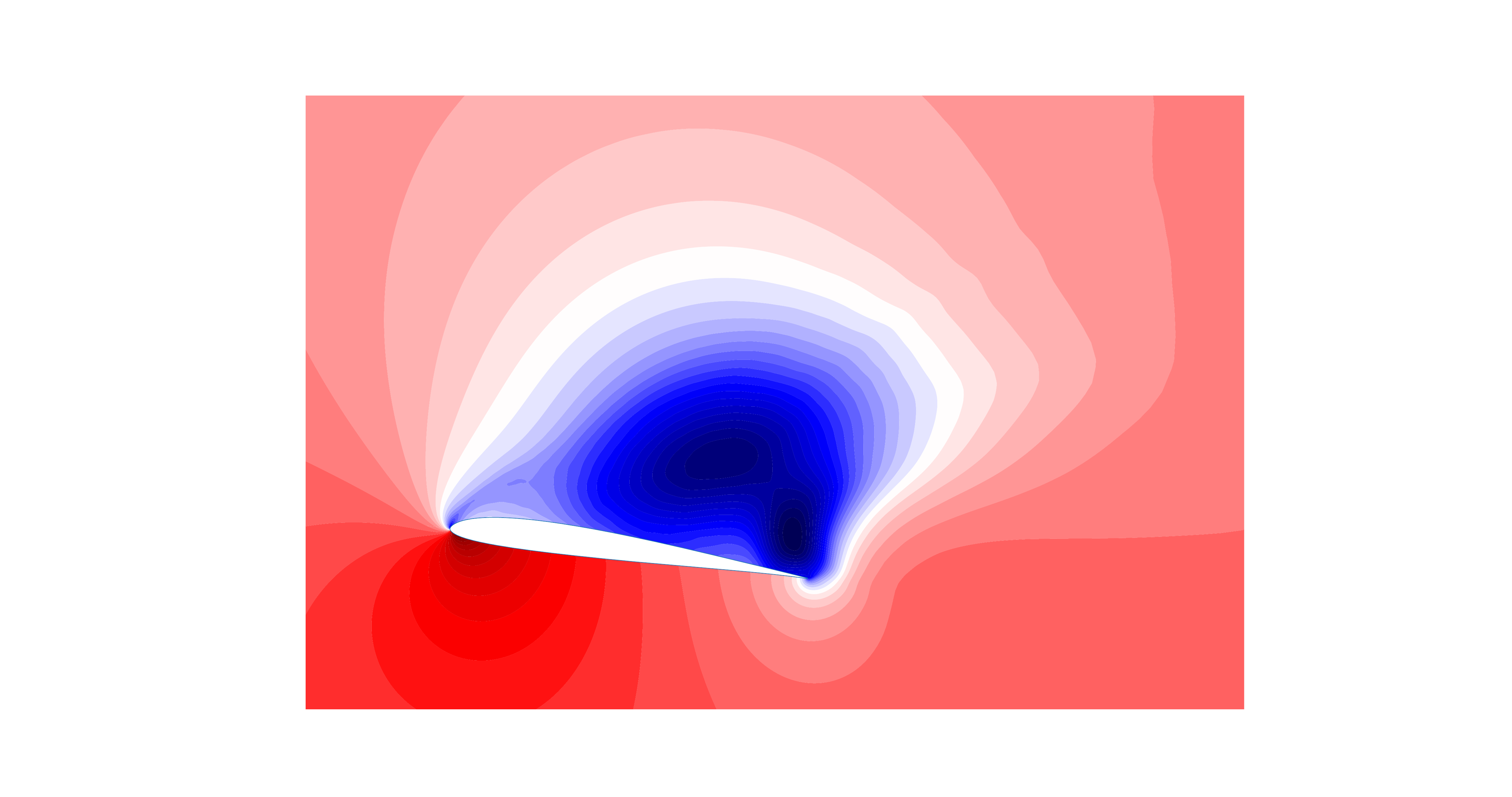}
    \end{subfigure}
    ~
    \begin{subfigure}[H]{.3\textwidth}
        \centering
        \includegraphics[width=1.\textwidth,trim={100mm 30mm 100mm 30mm},clip]{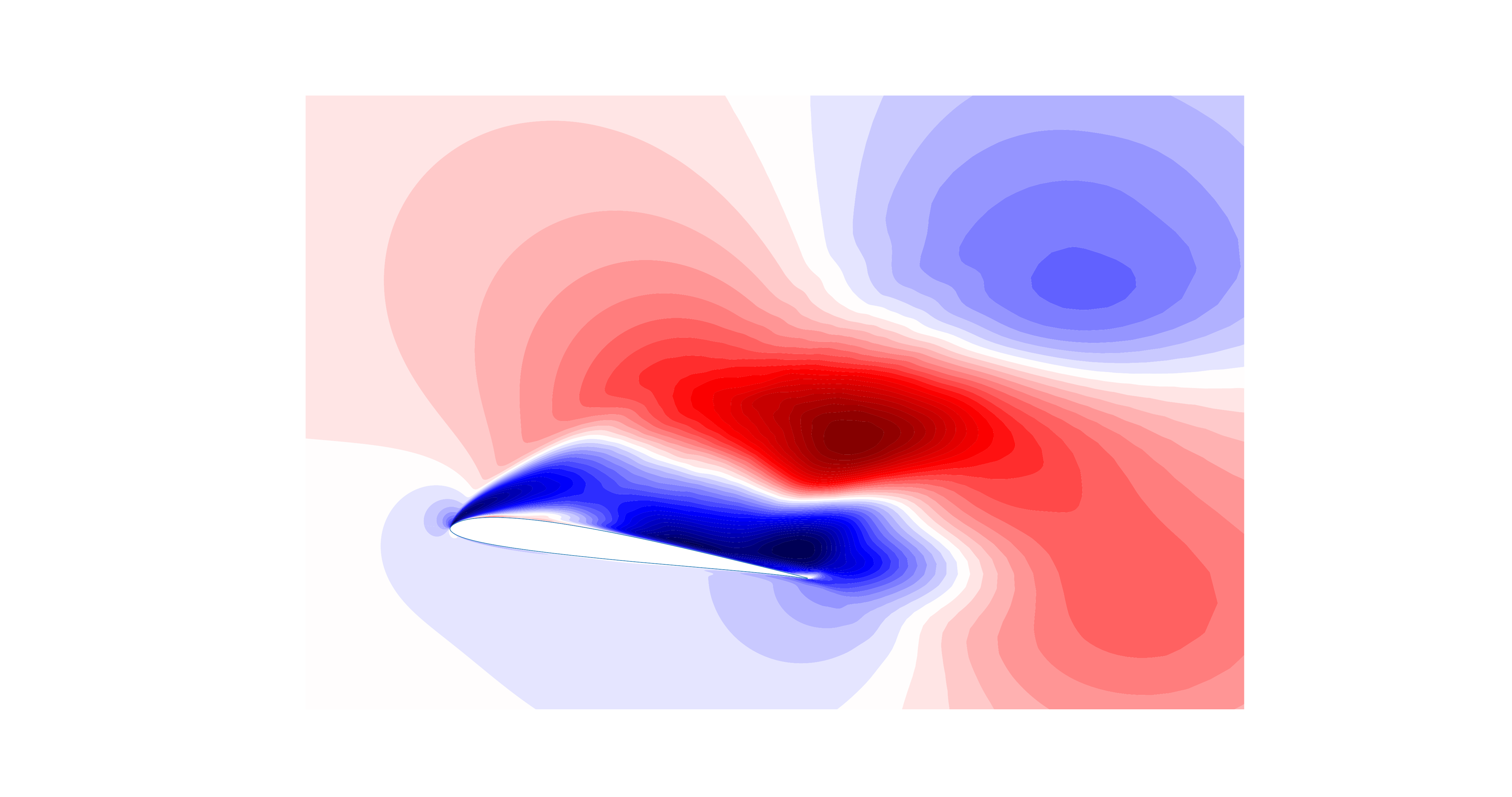}
    \end{subfigure}
    ~
    \begin{subfigure}[H]{.3\textwidth}
        \centering
        \includegraphics[width=1.\textwidth,trim={100mm 30mm 100mm 30mm},clip]{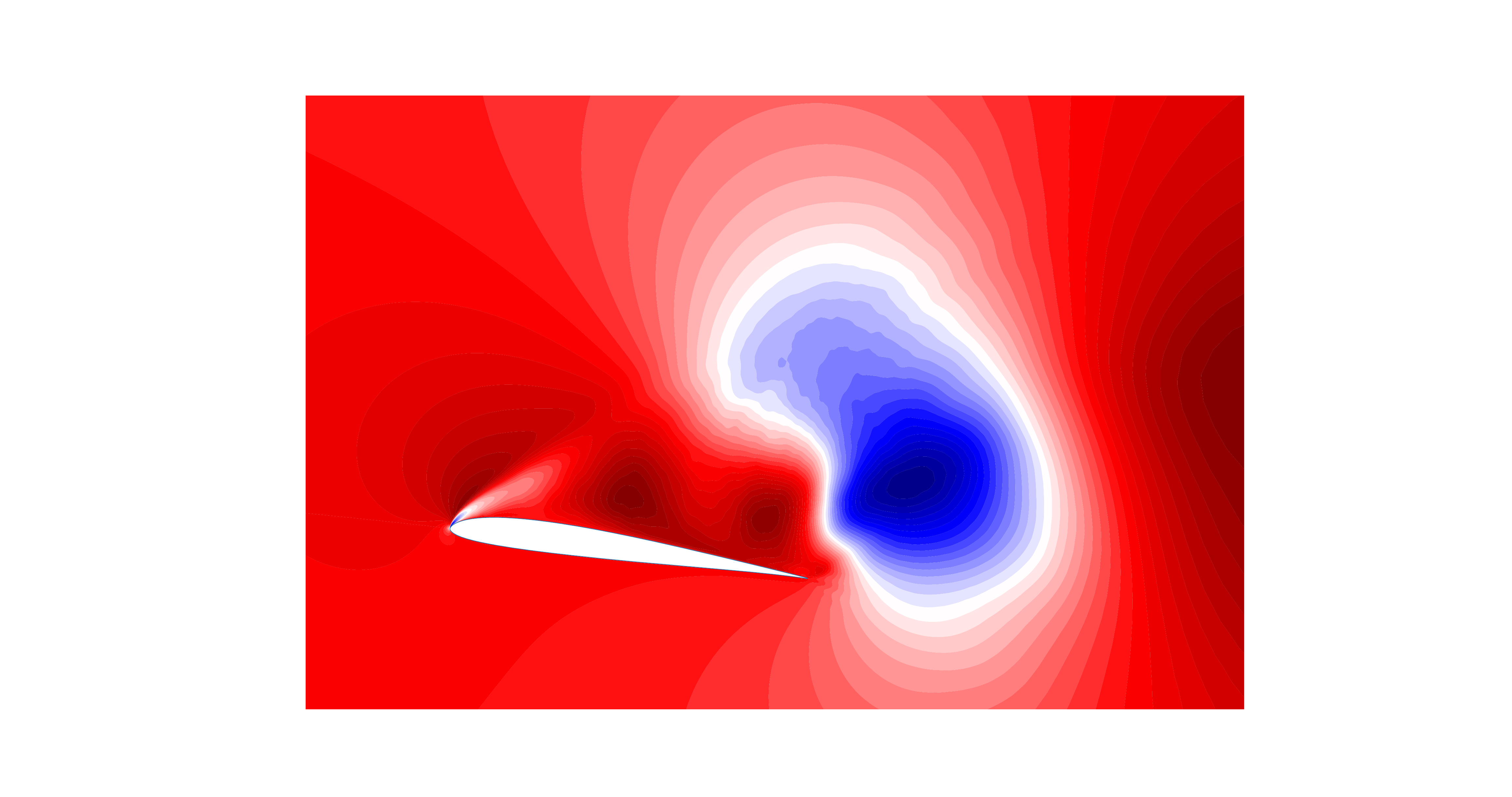}
    \end{subfigure}
    ~
    \begin{subfigure}[H]{.3\textwidth}
        \centering
        \includegraphics[width=1.\textwidth,trim={100mm 30mm 100mm 30mm},clip]{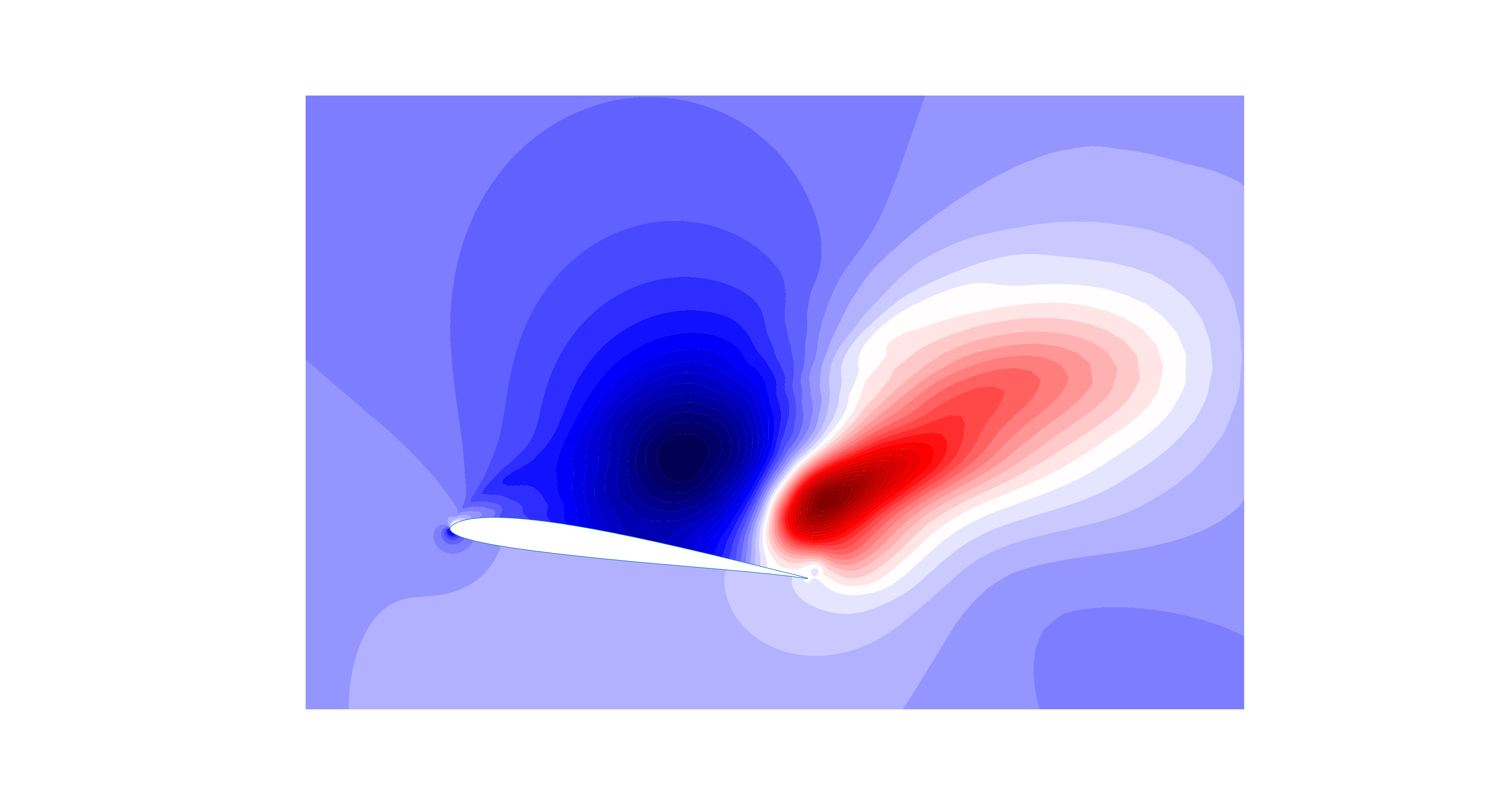}
    \end{subfigure}
    ~
    \begin{subfigure}[H]{.3\textwidth}
        \centering
        \includegraphics[width=1.\textwidth,trim={100mm 30mm 100mm 30mm},clip]{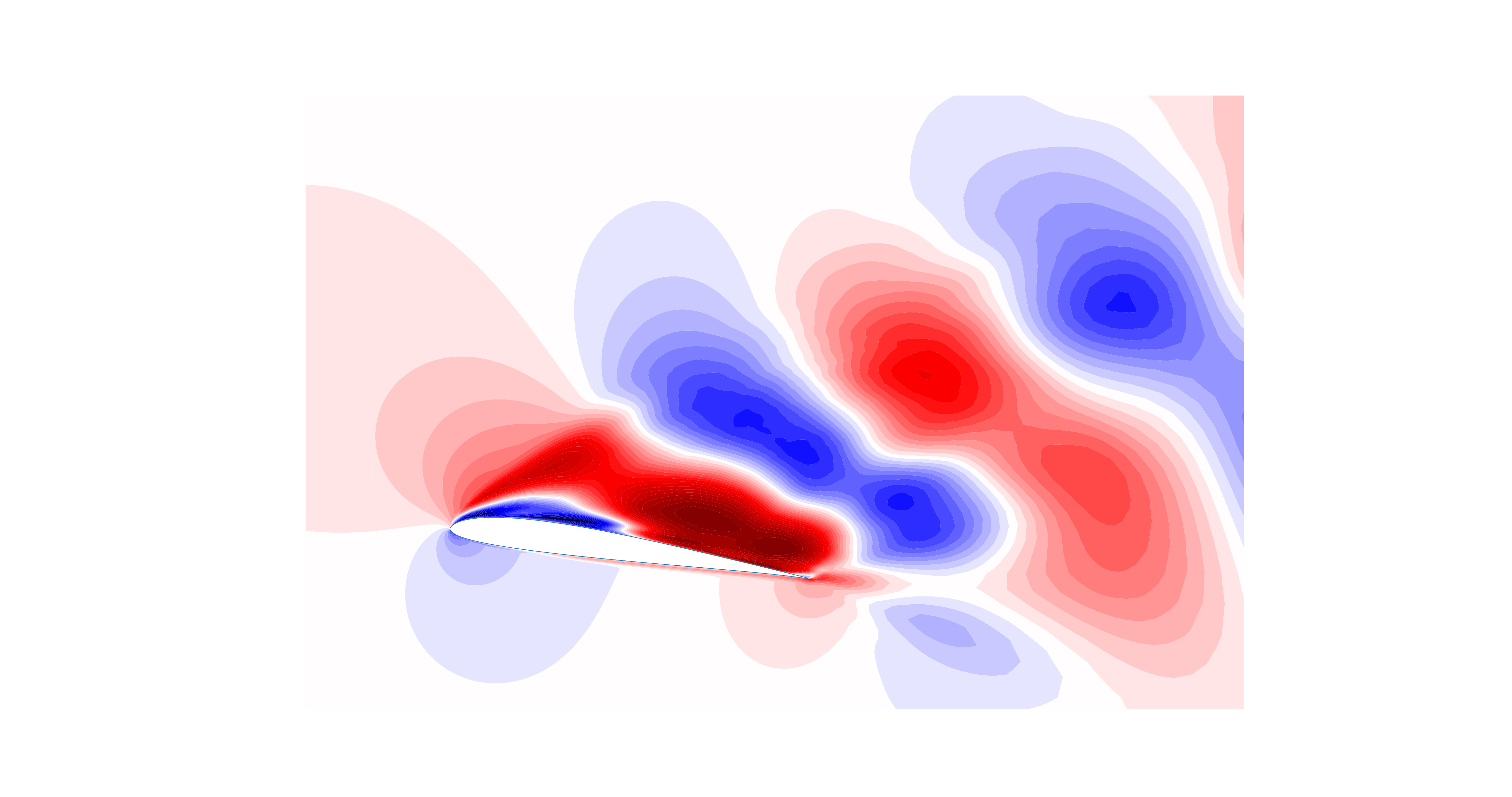}
    \end{subfigure}
    ~
    \begin{subfigure}[H]{.3\textwidth}
        \centering
        \includegraphics[width=1.\textwidth,trim={100mm 30mm 100mm 30mm},clip]{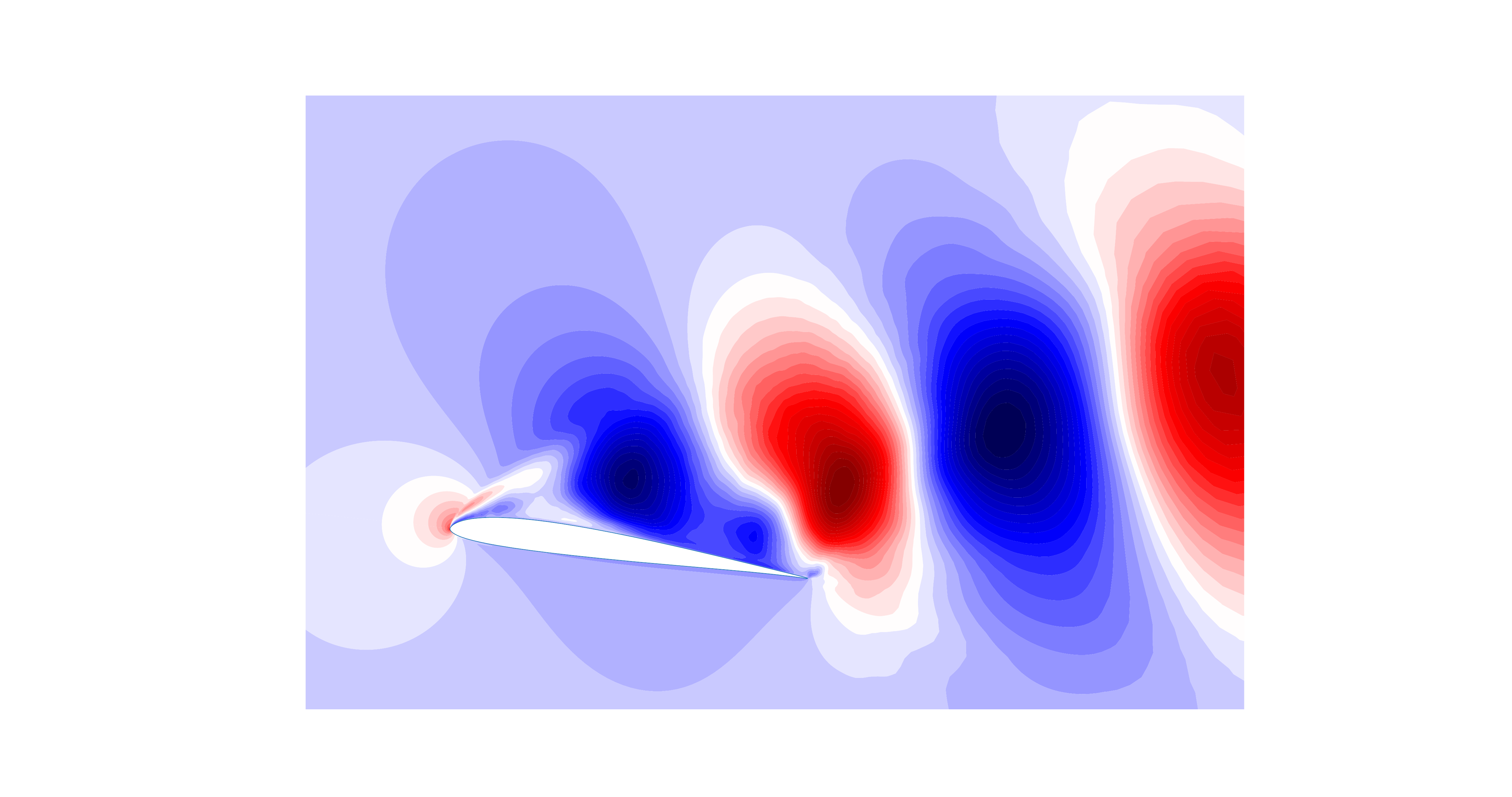}
    \end{subfigure}
    ~
    \begin{subfigure}[H]{.3\textwidth}
        \centering
        \includegraphics[width=1.\textwidth,trim={100mm 30mm 100mm 30mm},clip]{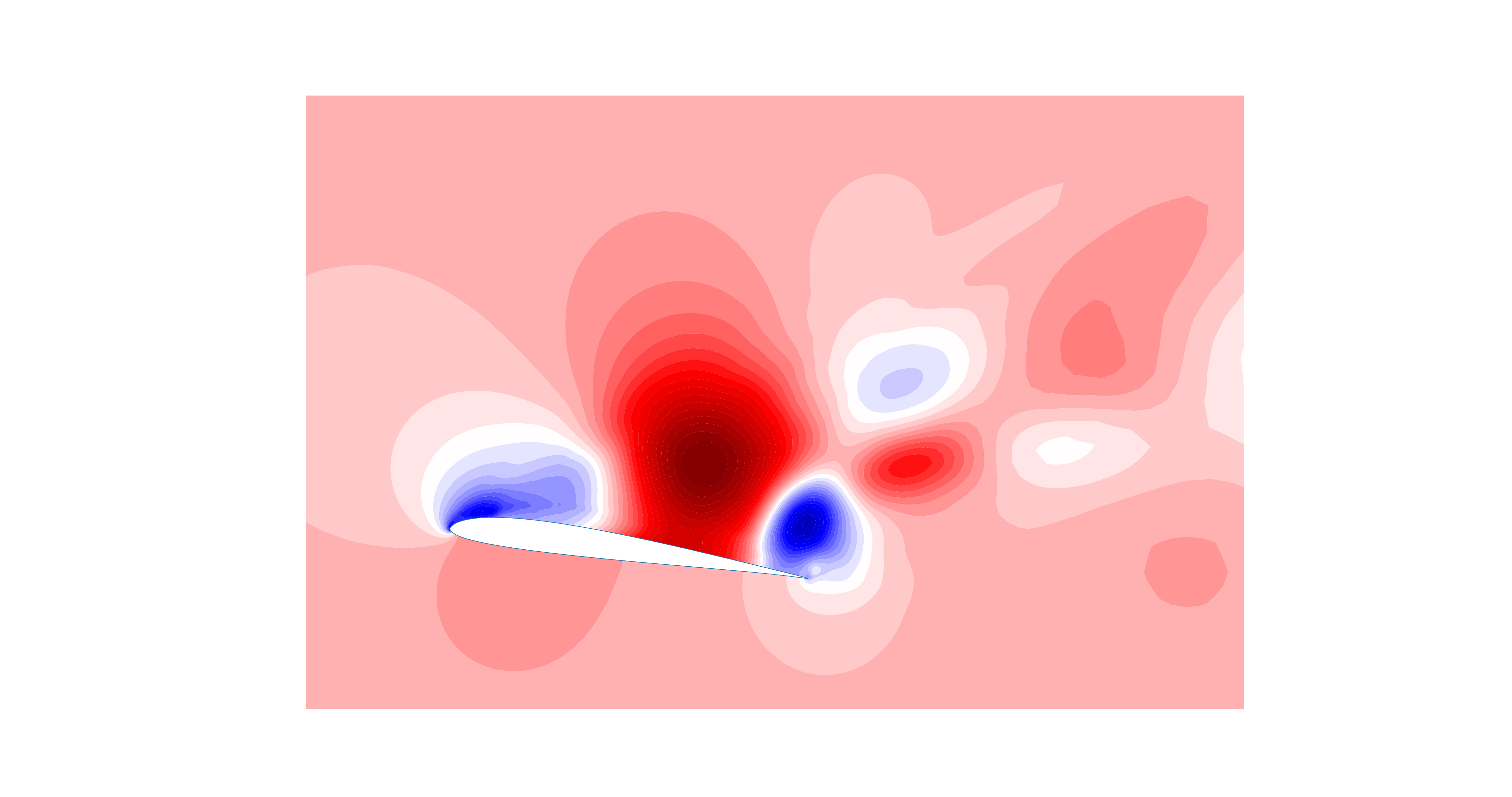}
    \end{subfigure}
    ~
    \begin{subfigure}[H]{.3\textwidth}
        \centering
        \includegraphics[width=1.\textwidth,trim={100mm 30mm 100mm 30mm},clip]{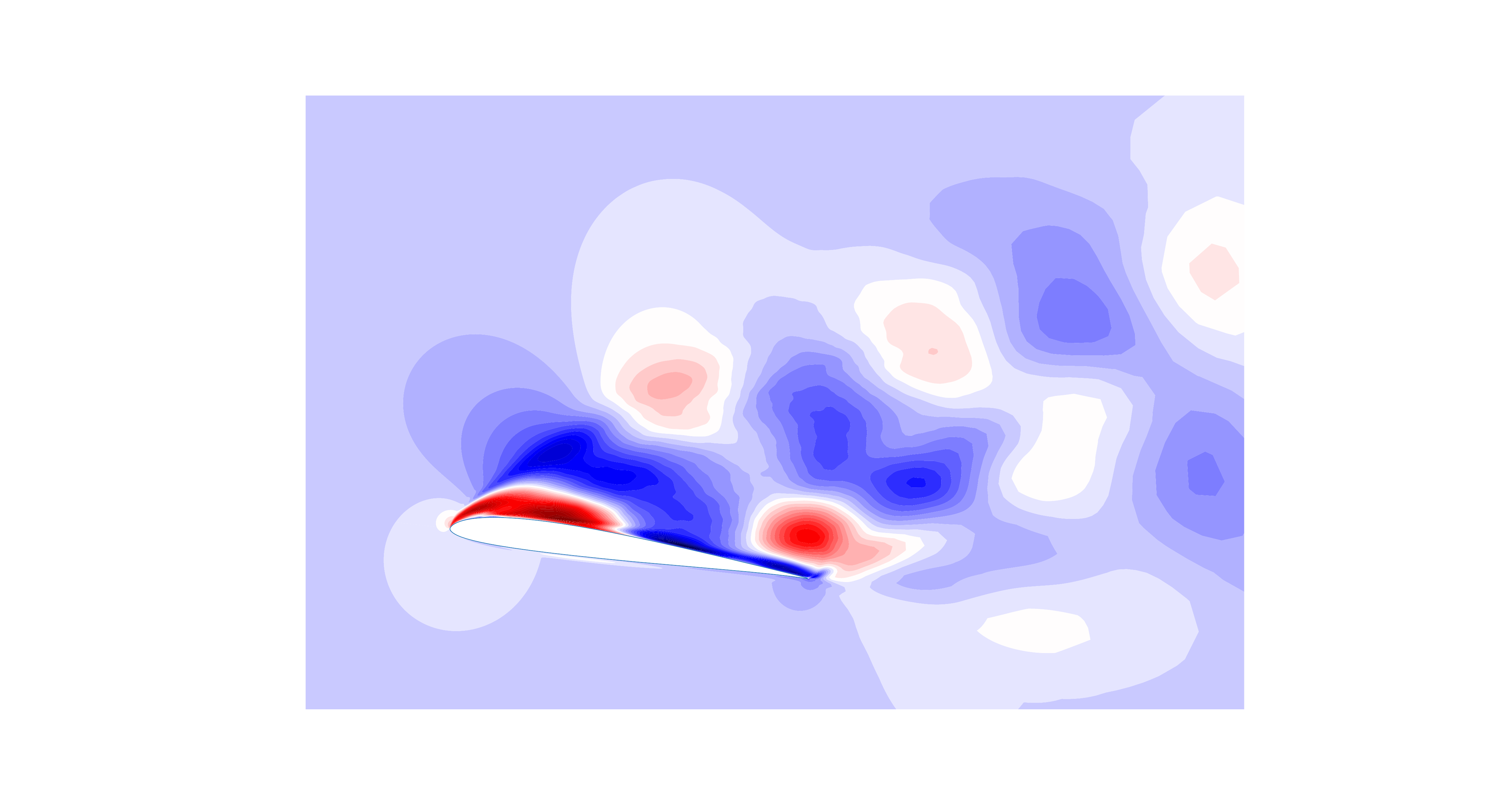}
    \end{subfigure}
    ~
    \begin{subfigure}[H]{.3\textwidth}
        \centering
        \includegraphics[width=1.\textwidth,trim={100mm 30mm 100mm 30mm},clip]{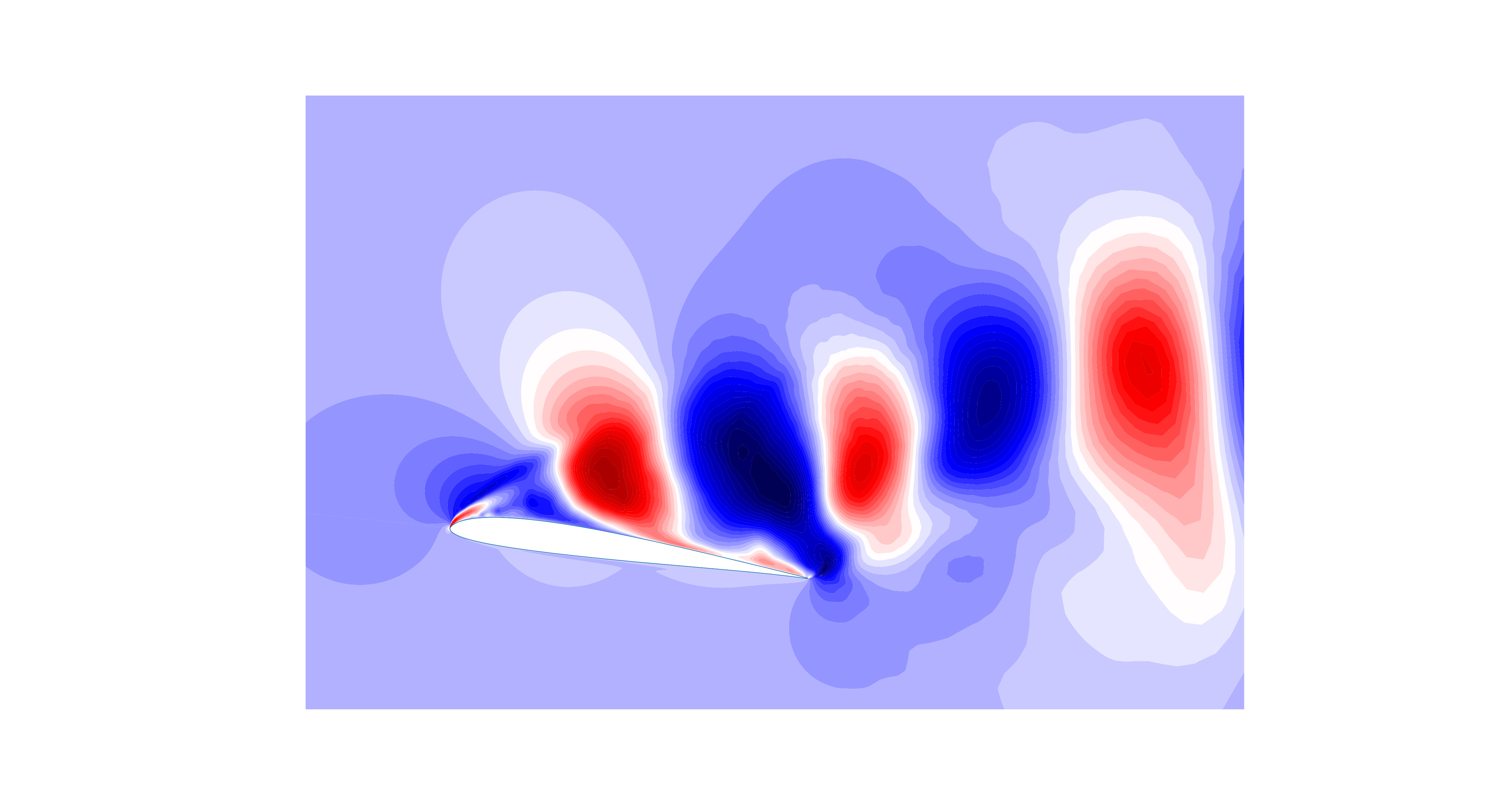}
    \end{subfigure}
    ~
    \begin{subfigure}[H]{.3\textwidth}
        \centering
        \includegraphics[width=1.\textwidth,trim={100mm 30mm 100mm 30mm},clip]{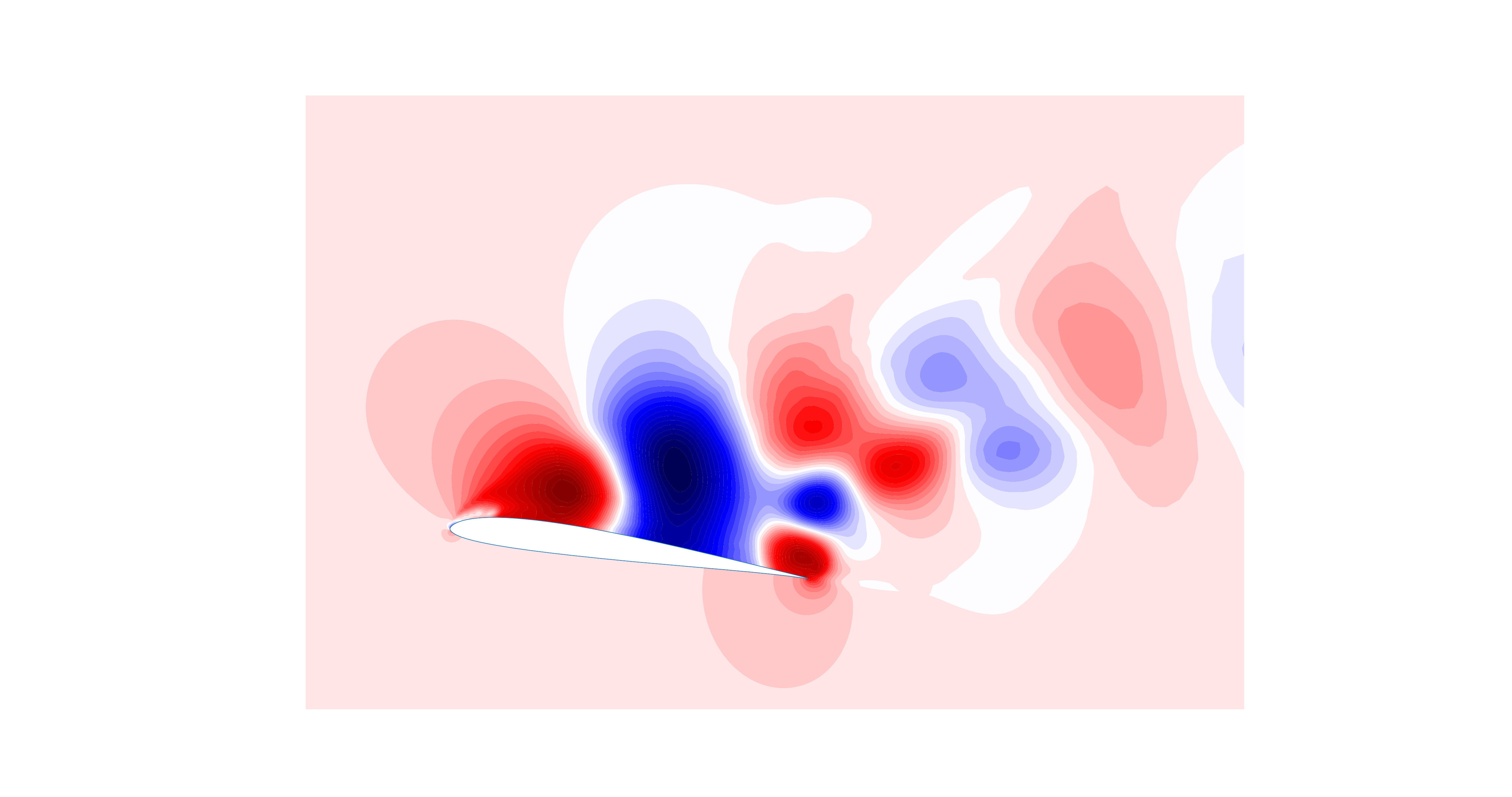}
    \end{subfigure}
    \caption{Contours of POD spatial modes for modes 1, 4, 8 and 12 (top to bottom) for u-velocity (left), v-velocity (center) and pressure fluctuations (right).}
    \label{fig:airfoil_modes_space}
\end{figure}

\begin{figure}[hbt!]
    \centering
    \begin{subfigure}[hbt!]{.90\textwidth}
        \centering
        \includegraphics[width=1.\textwidth,trim={10mm 5mm 20mm 6mm},clip]{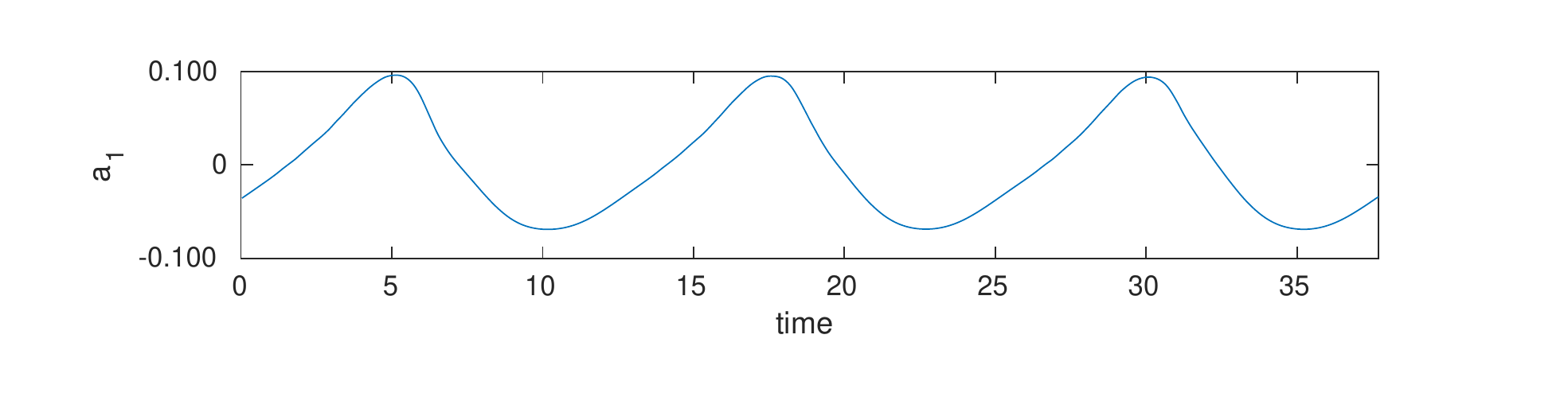}
    \end{subfigure}
    ~
    \begin{subfigure}[hbt!]{.9\textwidth}
        \centering
        \includegraphics[width=1.\textwidth,trim={10mm 5mm 20mm 6mm},clip]{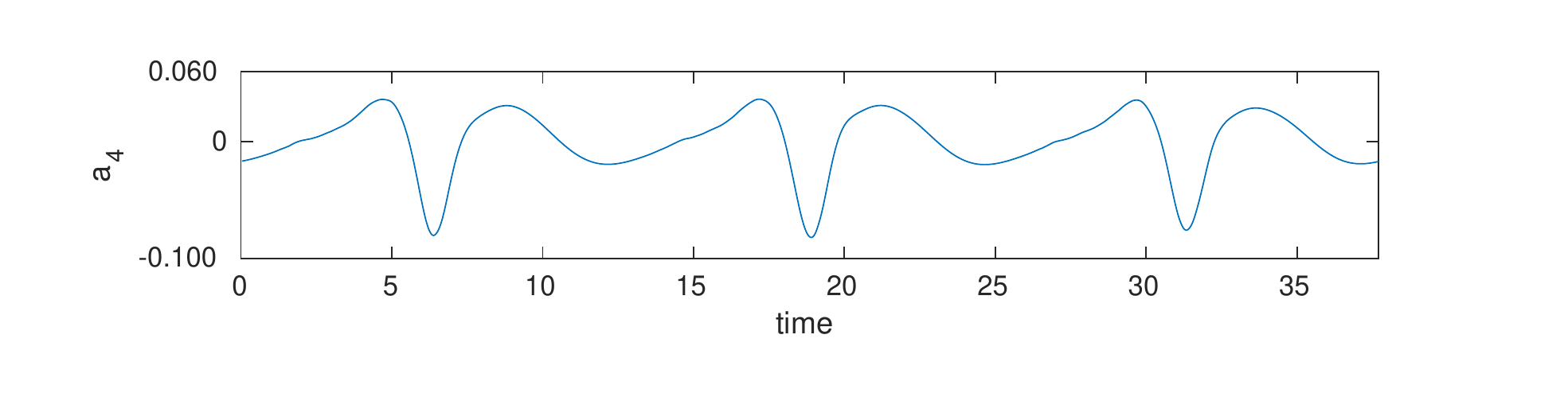}
    \end{subfigure}
    ~
    \begin{subfigure}[hbt!]{.9\textwidth}
        \centering
        \includegraphics[width=1.\textwidth,trim={10mm 5mm 20mm 6mm},clip]{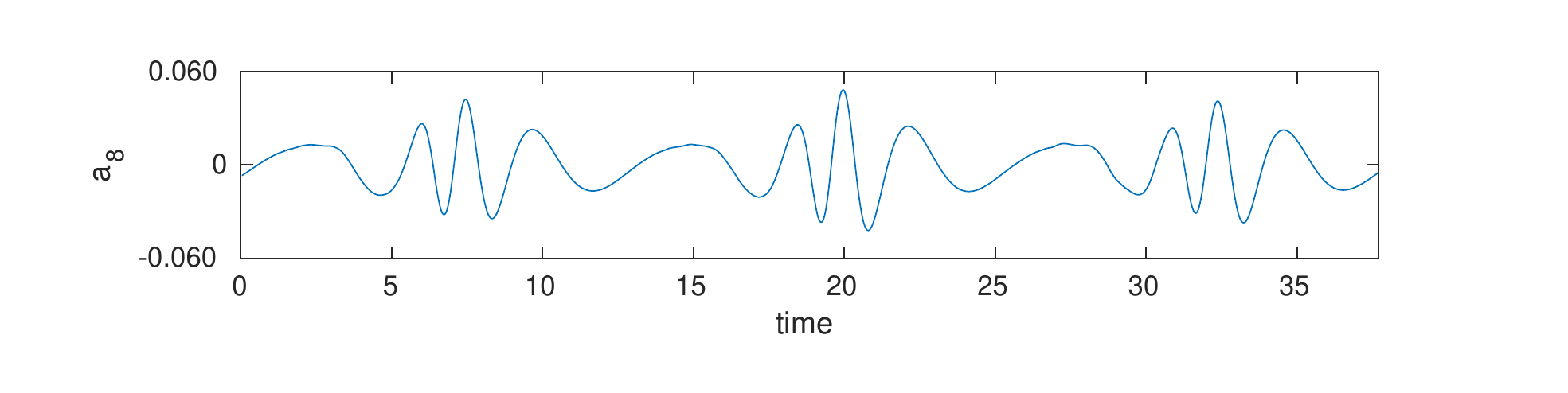}
    \end{subfigure}
    ~
    \begin{subfigure}[hbt!]{.9\textwidth}
        \centering
        \includegraphics[width=1.\textwidth,trim={10mm 5mm 20mm 6mm},clip]{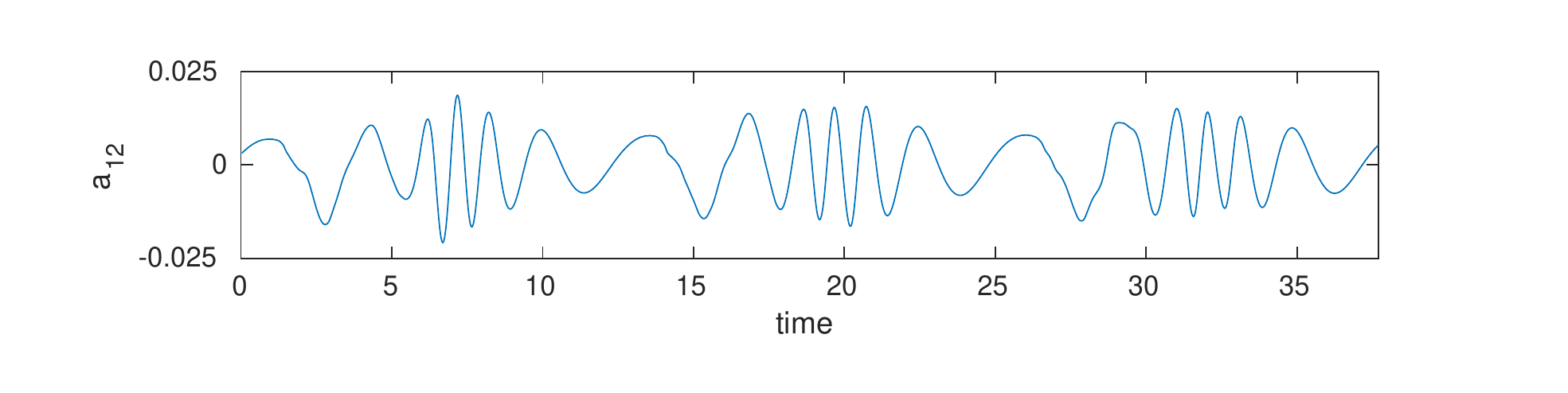}
    \end{subfigure}
    \caption{Temporal dynamics of POD modes 1, 4, 8 and 12.}
    \label{fig:airfoil_modes_time}
\end{figure}

Some POD spatial eigenfunctions are shown in Fig. \ref{fig:airfoil_modes_space}. These modes are presented for u and v velocity fluctuations, and pressure ($\zeta$ has similar spatial distribution as the latter variable). The first mode pair contains the large-scale structure related to the dynamic stall vortex. The other modes also contain information related to advection of the dynamic stall vortex combined with the trailing edge vortex and some higher frequency content from the turbulent flow. The higher POD modes are composed of high-frequency information and their convergence is not guaranteed due to limitations in the data acquisition frequency. The corresponding POD temporal modes are shown in Fig. \ref{fig:airfoil_modes_time}. The first mode has a quasi-sinusoidal behavior related to the plunging motion. Higher POD temporal modes display a more complex dynamics including multiple frequencies in a wavepacket-like content and some inherent noise from the turbulent simulation.

For this flow problem, the ROMs are constructed using 4, 8 and 12 POD modes. Similarly to the prevous test case, the spatial derivatives of the POD modes are computed using a $10th$-order central finite difference scheme and the equations are integrated in time with the first-order implicit Euler method for both Galerkin and LSPG techniques.

Coefficients arising from the projection are pre-computed for two conditions: without further approximation (gapless) and with application of hyper-reduction. In the previous case, modes are sampled by the accelerated greedy MPE for 1,000 points ($0.594\%$ of total points) to render pre-computation grid size independent and, hence, to evaluate the impact of hyper-reduction on the calibration techniques.
Furthermore, attempts to use additional modes led to ROMs being either unstable or inaccurate despite calibration. This can be attributed, at least in part, to the spatio-temporal resolution of the data. For this turbulent flow, the temporal and spatial resolutions are lower than those considered for the previous cylinder flow or in the test cases discussed in \cite{zucatti_01,zucatti_jcp}. In \cite{zucatti_02_scitech}, ROMs with 16 modes and similar linear calibration terms showed good results in both training and testing windows, but were unstable for long-term prediction.

\begin{figure}[hbt!]
    \centering
    \begin{subfigure}[hbt!]{.45\textwidth}
        \centering
        \includegraphics[width=1.\textwidth,trim={0mm 9mm 0mm 12mm},clip]{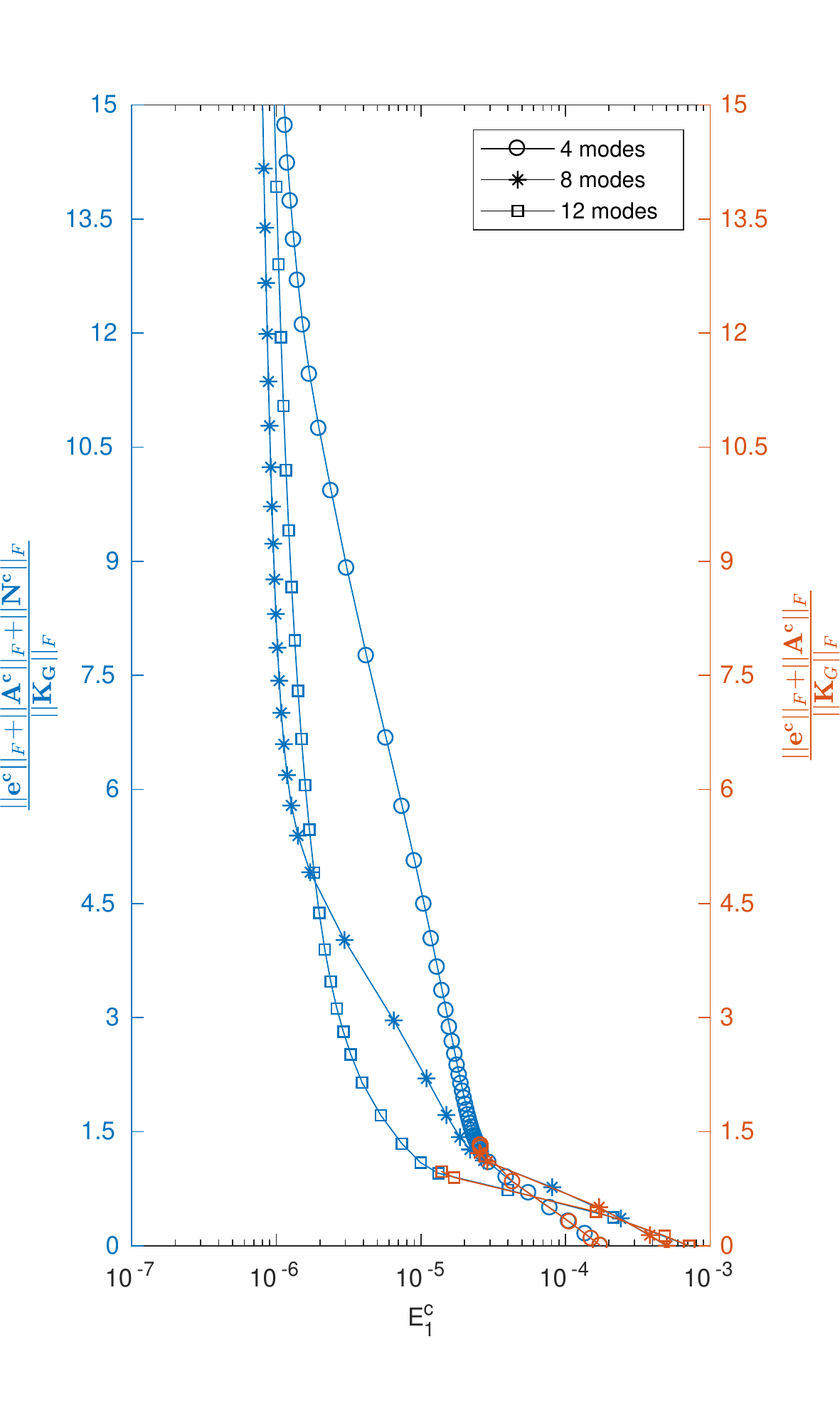}
        \caption{Galerkin curves.}
    \end{subfigure}
    ~
    \begin{subfigure}[hbt!]{.45\textwidth}
        \centering
        \includegraphics[width=1.\textwidth,trim={0mm 9mm 0mm 12mm},clip]{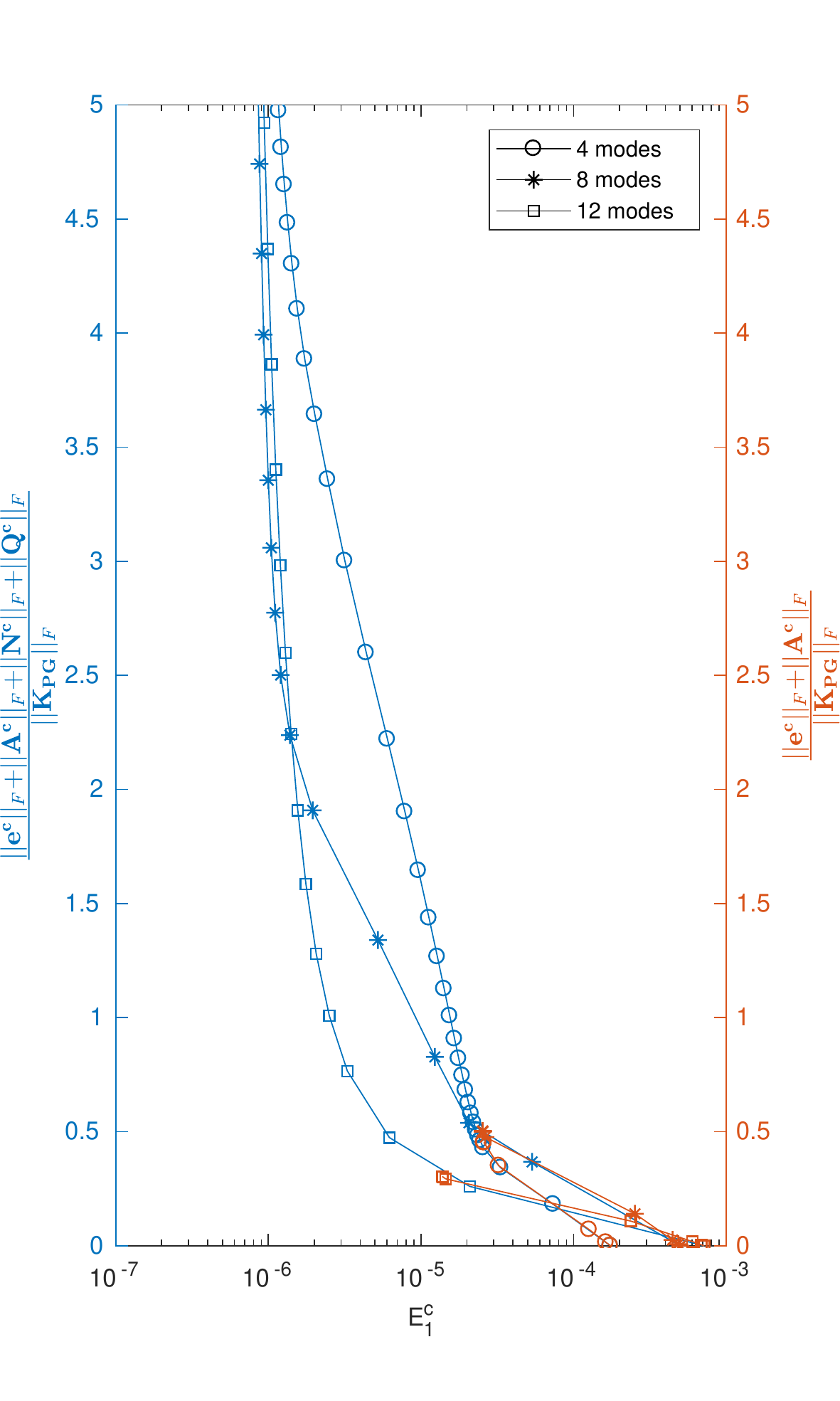}
        \caption{LSPG curves.}
    \end{subfigure}
    \caption{Ratio of Frobenius norms computed for calibrated and non-calibrated coefficients without hyper-reduction as a function of the approximation error using the iterative procedure described in section \ref{calibration}.}
    \label{fig:airfoil_Lcurves}
\end{figure}

\begin{figure}[hbt!]
    \centering
    \begin{subfigure}[hbt!]{.45\textwidth}
        \centering
        \includegraphics[width=1.\textwidth,trim={0mm 9mm 0mm 12mm},clip]{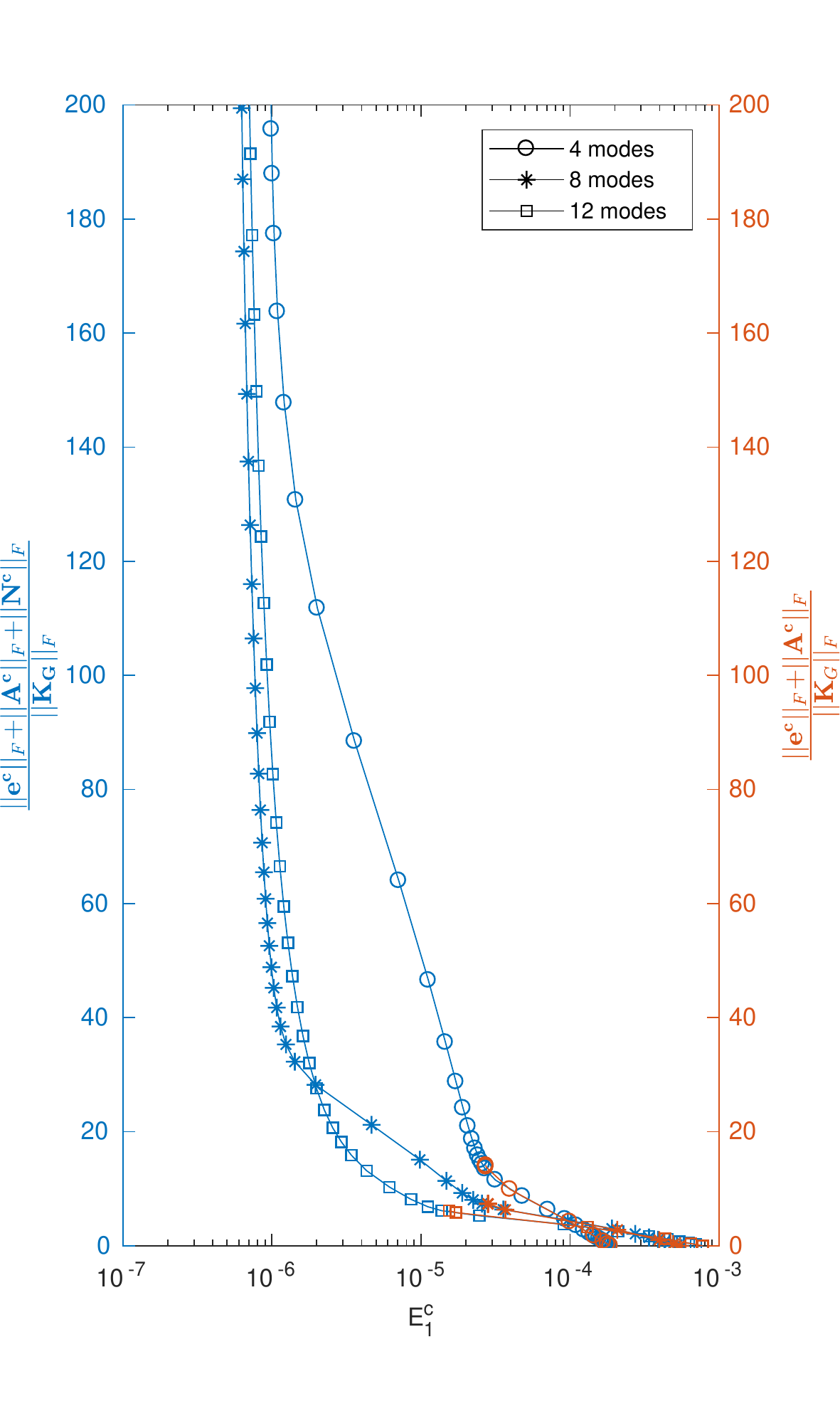}
        \caption{Galerkin curves.}
    \end{subfigure}
    ~
    \begin{subfigure}[hbt!]{.45\textwidth}
        \centering
        \includegraphics[width=1.\textwidth,trim={0mm 9mm 0mm 12mm},clip]{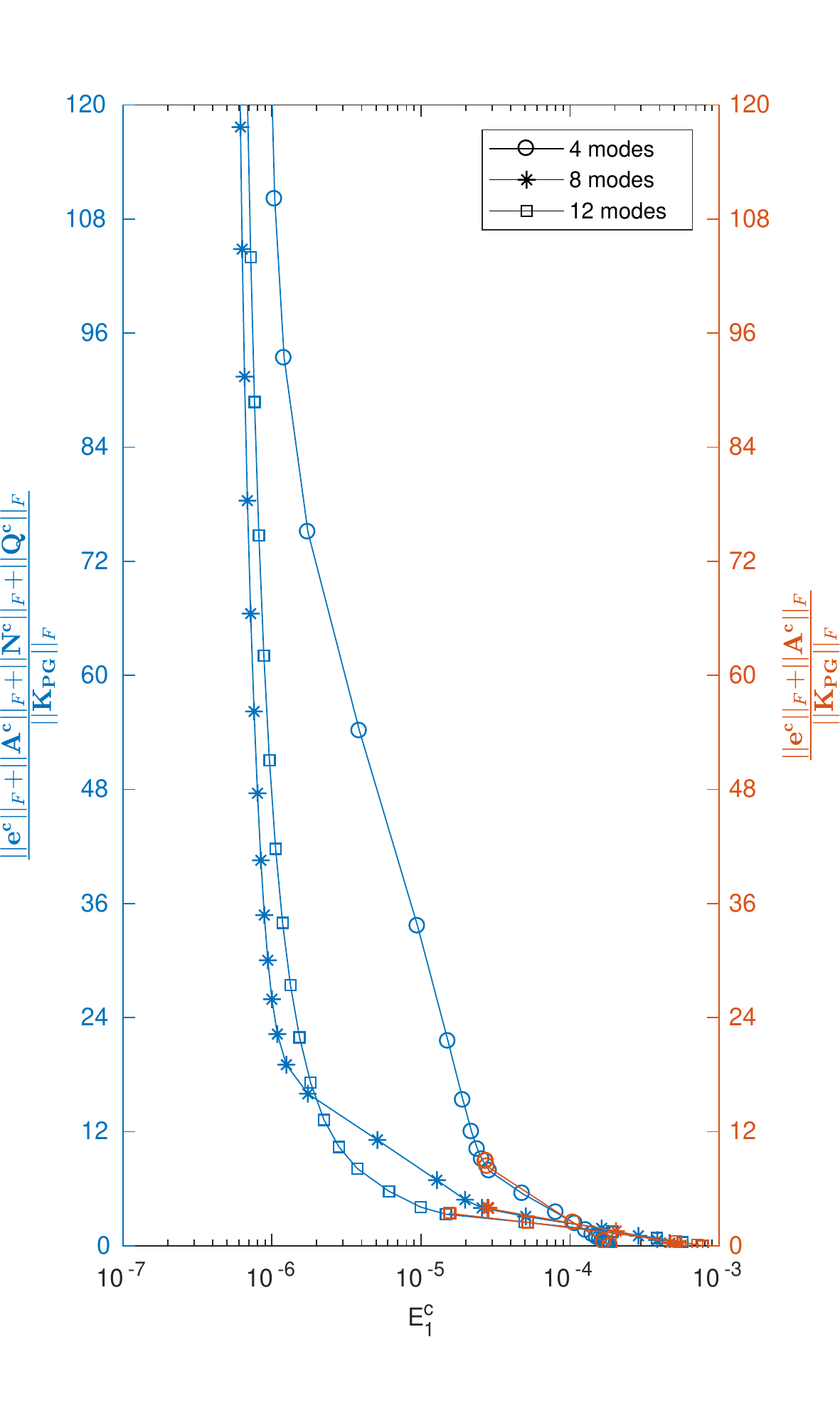}
        \caption{LSPG curves.}
    \end{subfigure}
    \caption{Ratio of Frobenius norms computed for calibrated and non-calibrated coefficients hyper-reduced to 1,000 points as a function of the approximation error using the iterative procedure described in section \ref{calibration}.}
    \label{fig:airfoil_Lcurves_hyper}
\end{figure}

\begin{figure}[hbt!]
    \centering
    \begin{subfigure}[hbt!]{.8\textwidth}
        \centering
        \includegraphics[width=1.\textwidth,trim={10mm 1mm 15mm 3mm},clip]{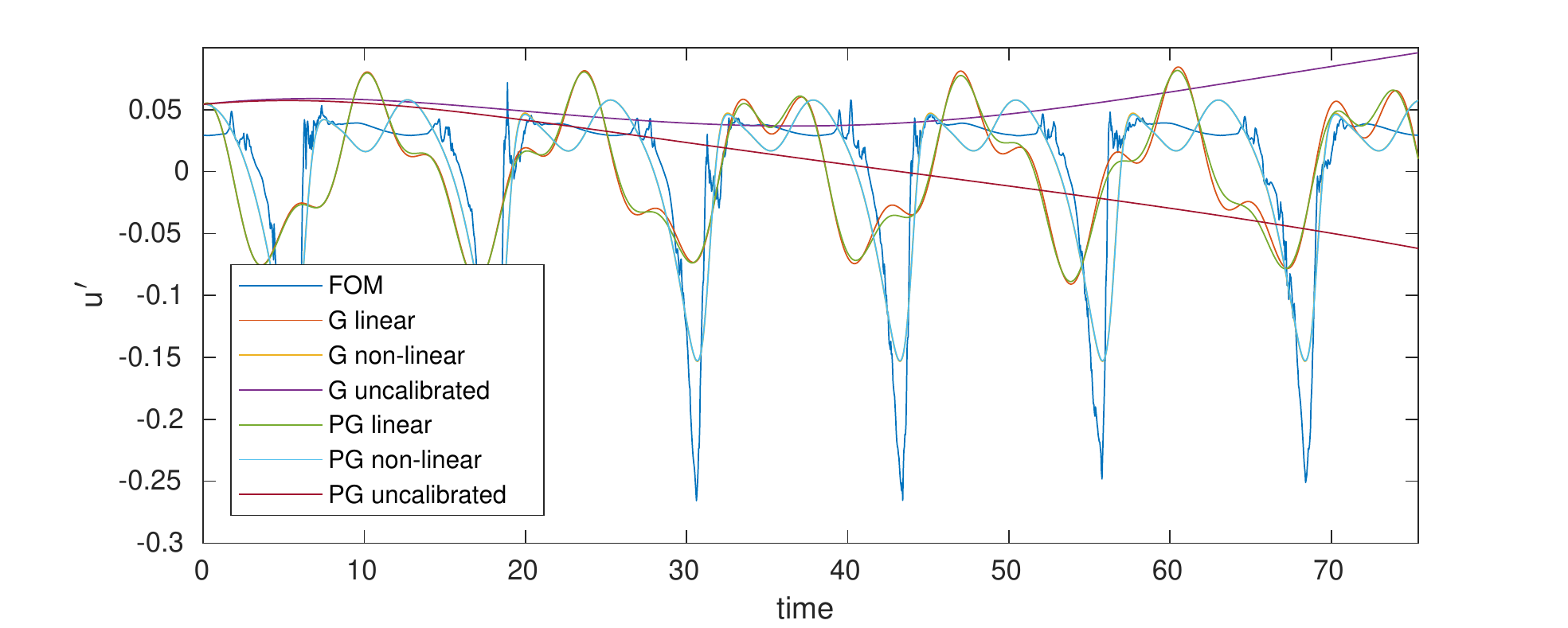}
        \caption{Probe in proximity of the leading edge.}
    \end{subfigure}
    ~
    \begin{subfigure}[hbt!]{.8\textwidth}
        \centering
        \includegraphics[width=1.\textwidth,trim={10mm 1mm 15mm 3mm},clip]{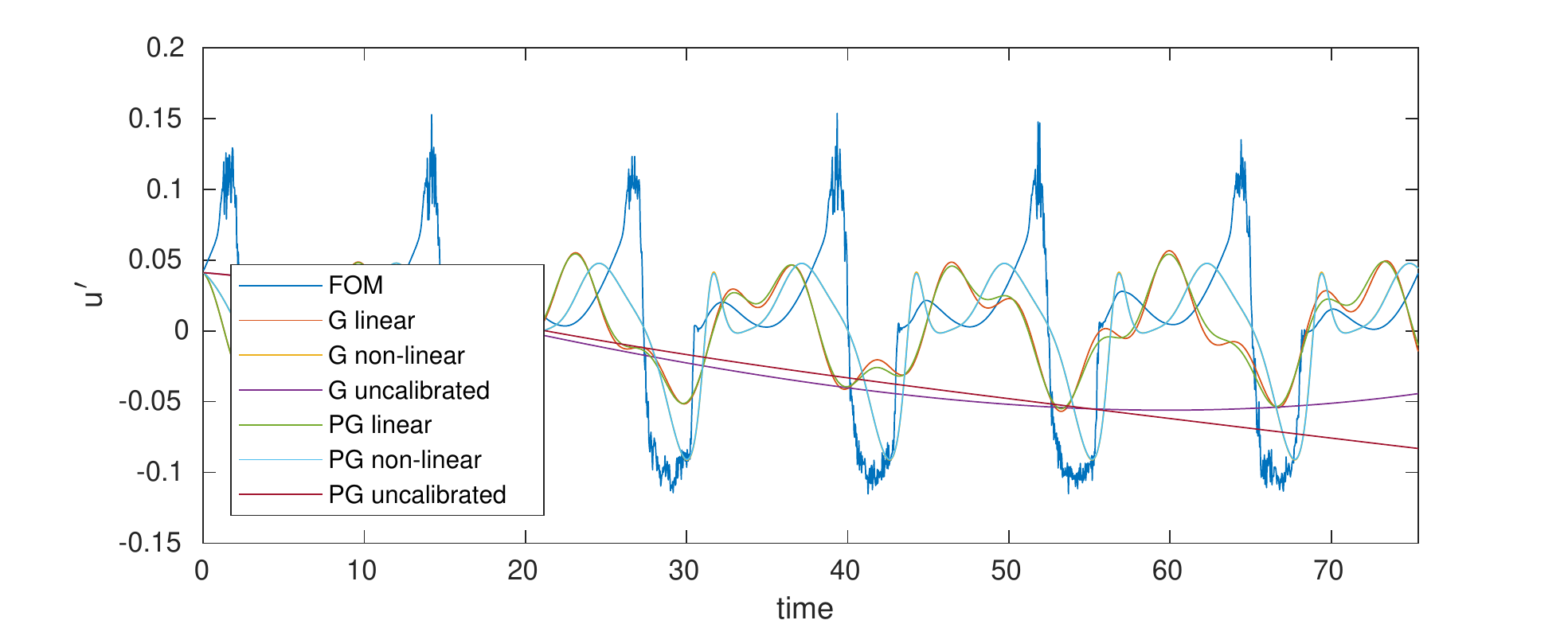}
        \caption{Probe in proximity of the trailing edge.}
    \end{subfigure}
    \caption{Fluctuation time histories computed by the FOM and gapless ROMs with 4 POD modes.}
    \label{fig:airfoil_probes_4modes}
\end{figure}

Figures \ref{fig:airfoil_Lcurves} and \ref{fig:airfoil_Lcurves_hyper} show the impact of the calibration coefficients relative to the ROMs' original coefficients without and with hyper-reduction, respectively. Similarly to the cylinder flow problem, the curves only exhibit a typical L-shape when non-linear calibration coefficients are considered. 
Differently from the cylinder flow, the present calibrated ROMs require significantly more invasive coefficients (at least the same order of magnitude) to provide substantial reduction of the error indicator $E_1^c$. This could be attributed to the lower quality of the original reduced order model coefficients, attested in Fig. \ref{fig:airfoil_probes_4modes}. This observation is even more evident when hyper-reduction is taken into consideration. In fact, hyper-reduced models produce very invasive calibration coefficients despite being similar in the shape of the L-curves compared to their gapless counterparts.
Another observation from Figs. \ref{fig:airfoil_Lcurves} and \ref{fig:airfoil_Lcurves_hyper} is that calibration of linear and non-linear terms present similar trajectories at first. Eventually, the linear coefficients converge in norm while the non-linear ones are capable of achieving further error reductions at the cost of more intrusive coefficients. As can be seen from the figures, without calibration, the models with 4 modes are intriguingly more precise compared to models with 8 and 12 modes. This shows that these additional modes are, at least initially, negatively impacting the models. On the other hand, the extra modes are capable of further reducing the error indicator with less intrusive calibration coefficients.

\begin{figure}[hbt!]
    \centering
    \begin{subfigure}[hbt!]{.8\textwidth}
        \centering
        \includegraphics[width=1.\textwidth,trim={10mm 1mm 15mm 3mm},clip]{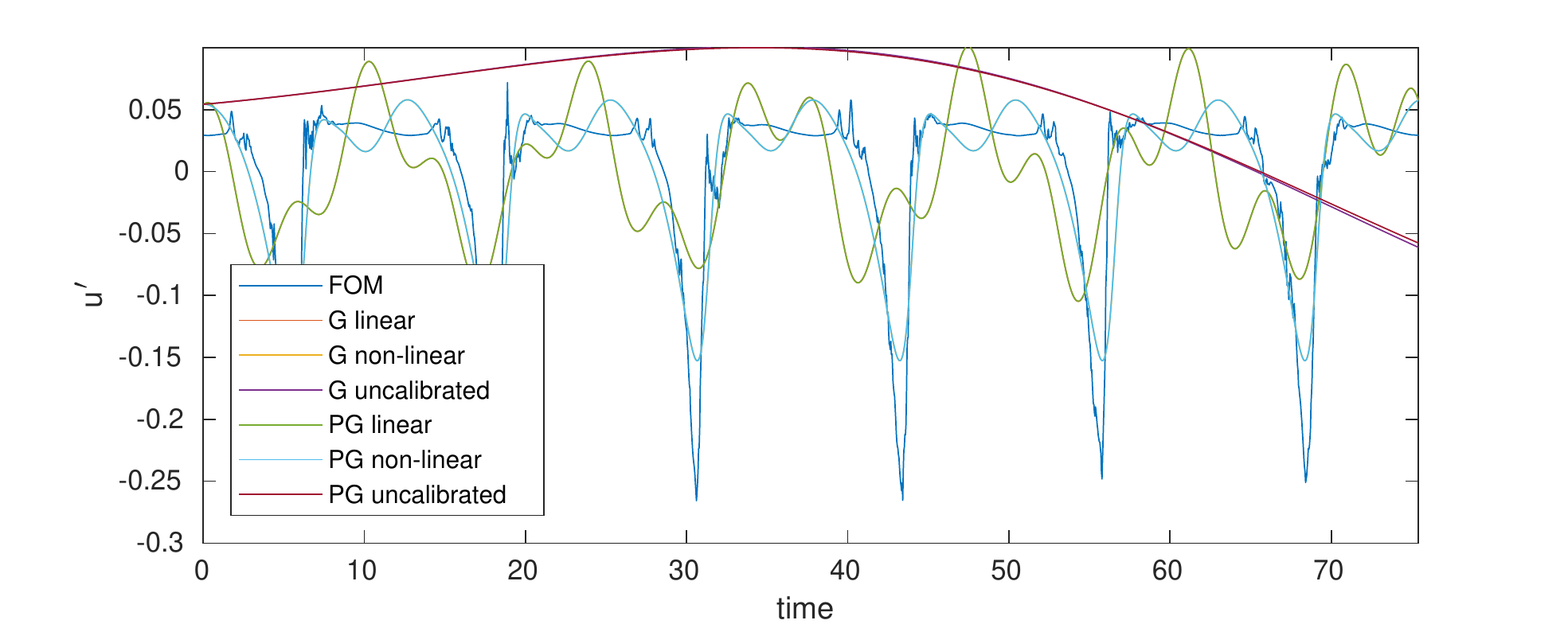}
        \caption{Probe in proximity of the leading edge.}
    \end{subfigure}
    ~
    \begin{subfigure}[hbt!]{.8\textwidth}
        \centering
        \includegraphics[width=1.\textwidth,trim={10mm 1mm 15mm 3mm},clip]{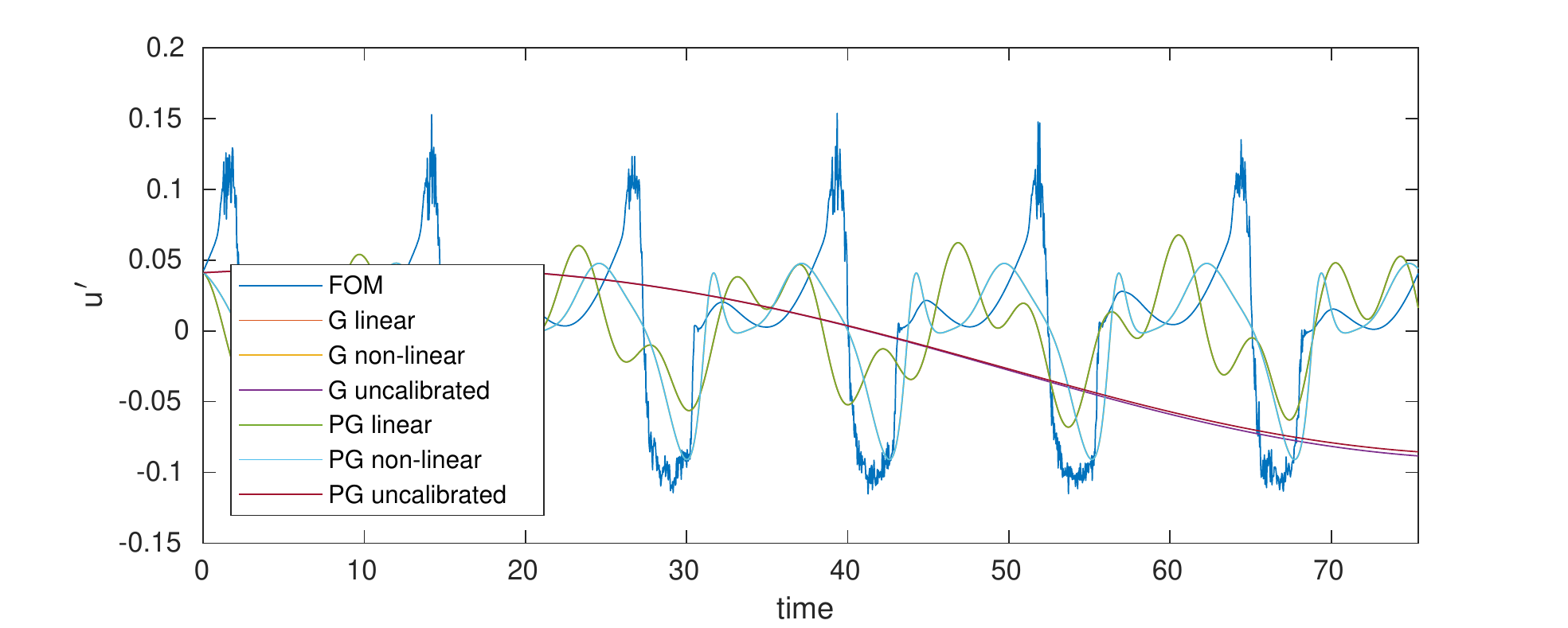}
        \caption{Probe in proximity of the trailing edge.}
    \end{subfigure}
    \caption{Fluctuation time histories computed by the FOM and hyper-reduced ROMs with 4 POD modes.}
    \label{fig:airfoil_probes_4modes_hyper}
\end{figure}

\begin{figure}[hbt!]
    \centering
    \begin{subfigure}[hbt!]{.8\textwidth}
        \centering
        \includegraphics[width=1.\textwidth,trim={10mm 1mm 15mm 3mm},clip]{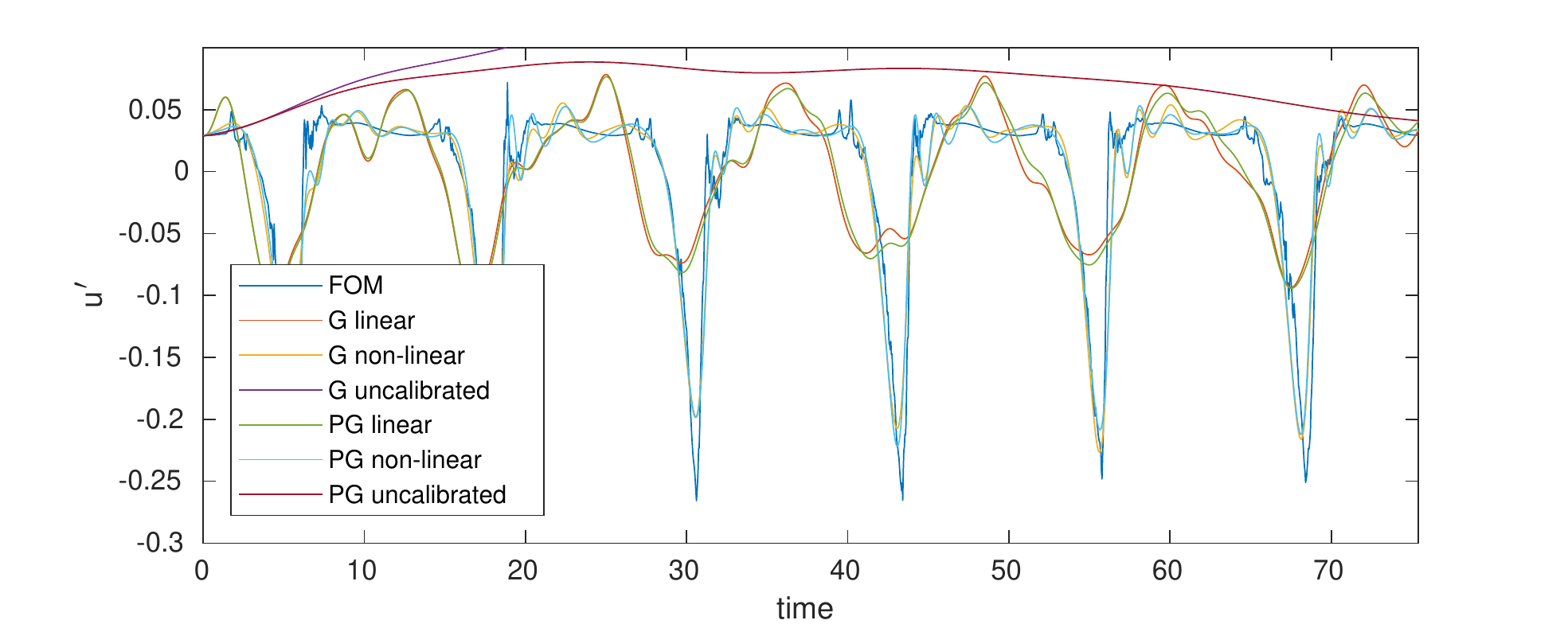}
        \caption{Probe in proximity of the leading edge.}
    \end{subfigure}
    ~
    \begin{subfigure}[hbt!]{.8\textwidth}
        \centering
        \includegraphics[width=1.\textwidth,trim={10mm 1mm 15mm 3mm},clip]{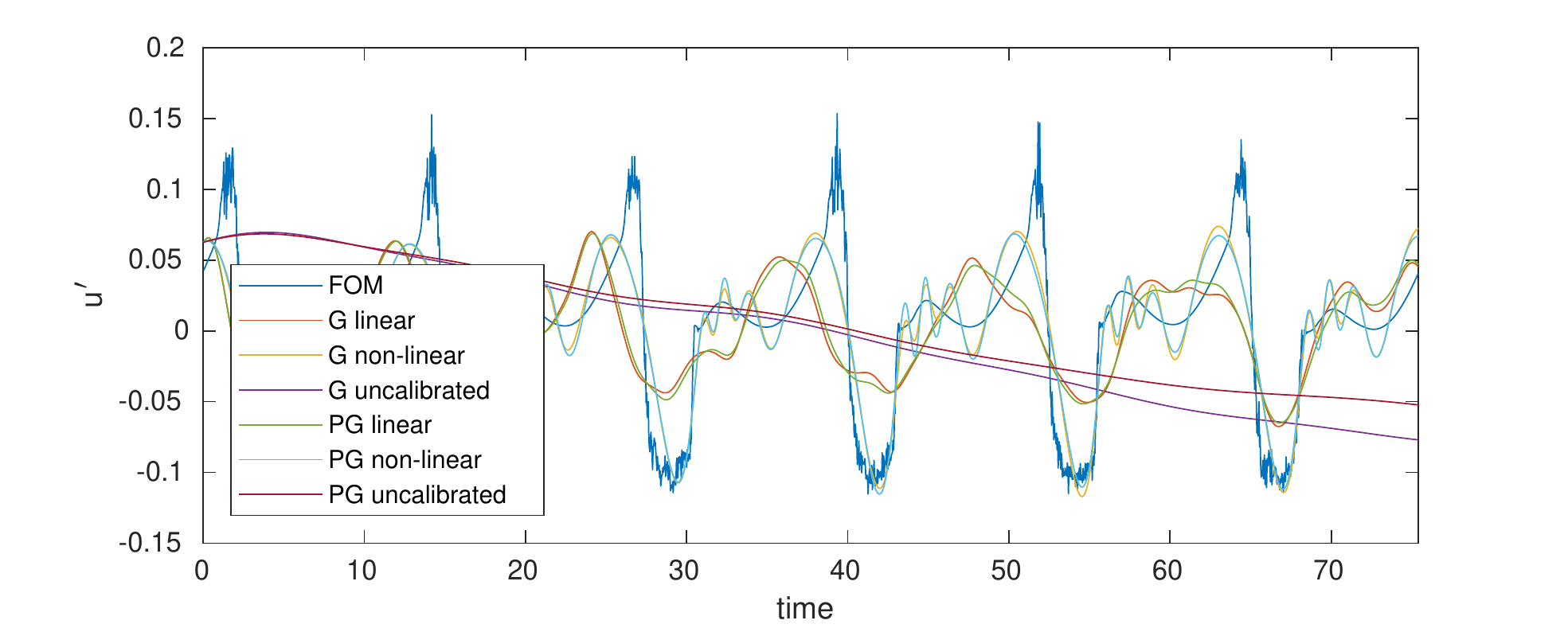}
        \caption{Probe in proximity of the trailing edge.}
    \end{subfigure}
    \caption{Fluctuation time histories computed by the FOM and gapless ROMs with 8 POD modes.}
    \label{fig:airfoil_probes_8modes}
\end{figure}

\begin{figure}[hbt!]
    \centering
    \begin{subfigure}[hbt!]{.8\textwidth}
        \centering
        \includegraphics[width=1.\textwidth,trim={10mm 1mm 15mm 3mm},clip]{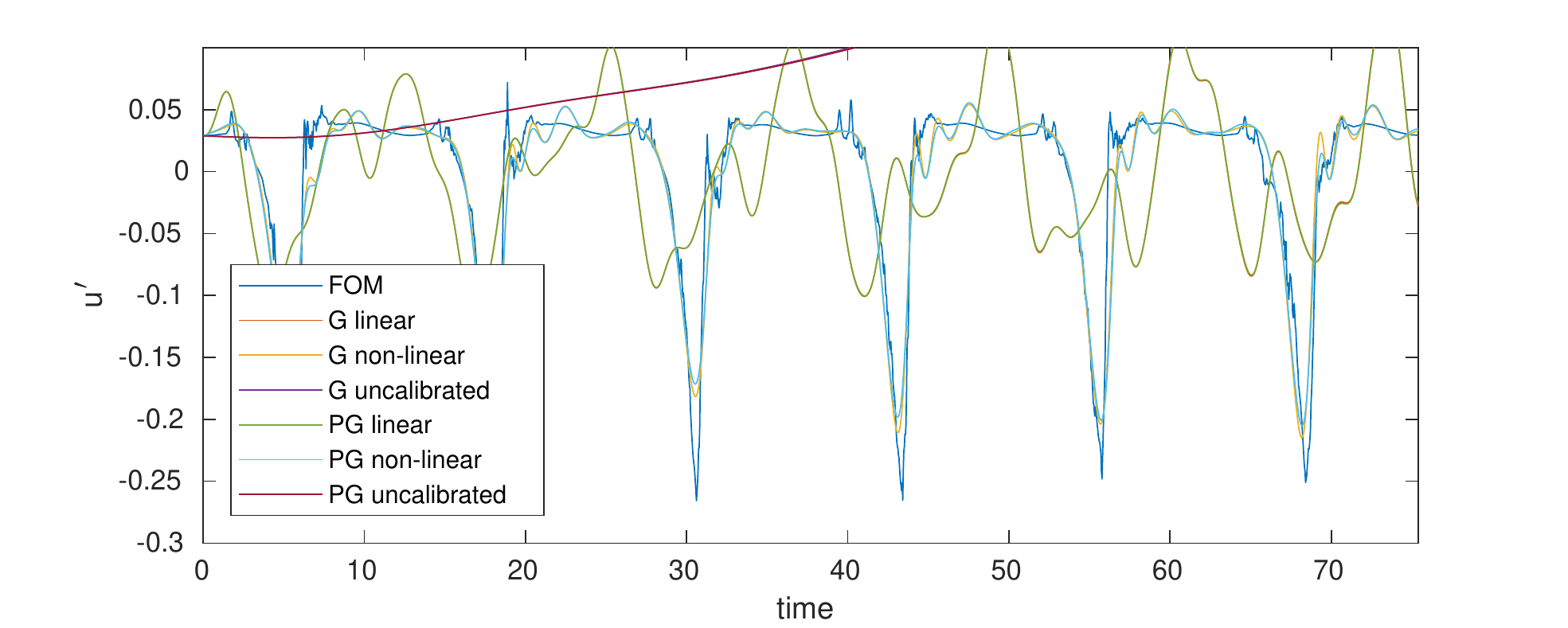}
        \caption{Probe in proximity of the leading edge.}
    \end{subfigure}
    ~
    \begin{subfigure}[hbt!]{.8\textwidth}
        \centering
        \includegraphics[width=1.\textwidth,trim={10mm 1mm 15mm 3mm},clip]{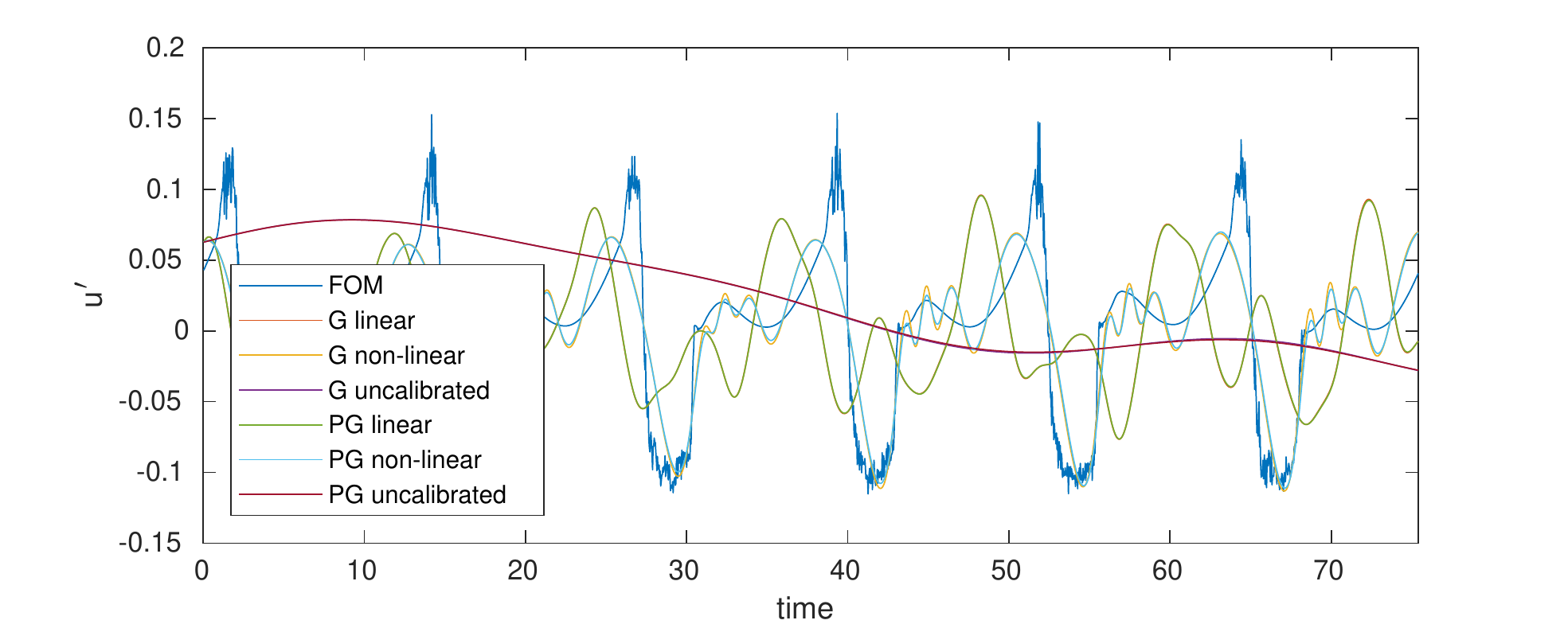}
        \caption{Probe in proximity of the trailing edge.}
    \end{subfigure}
    \caption{Fluctuation time histories computed by the FOM and hyper-reduced ROMs with 8 POD modes.}
    \label{fig:airfoil_probes_8modes_hyper}
\end{figure}

\begin{figure}[hbt!]
    \centering
    \begin{subfigure}[hbt!]{.8\textwidth}
        \centering
        \includegraphics[width=1.\textwidth,trim={10mm 1mm 15mm 3mm},clip]{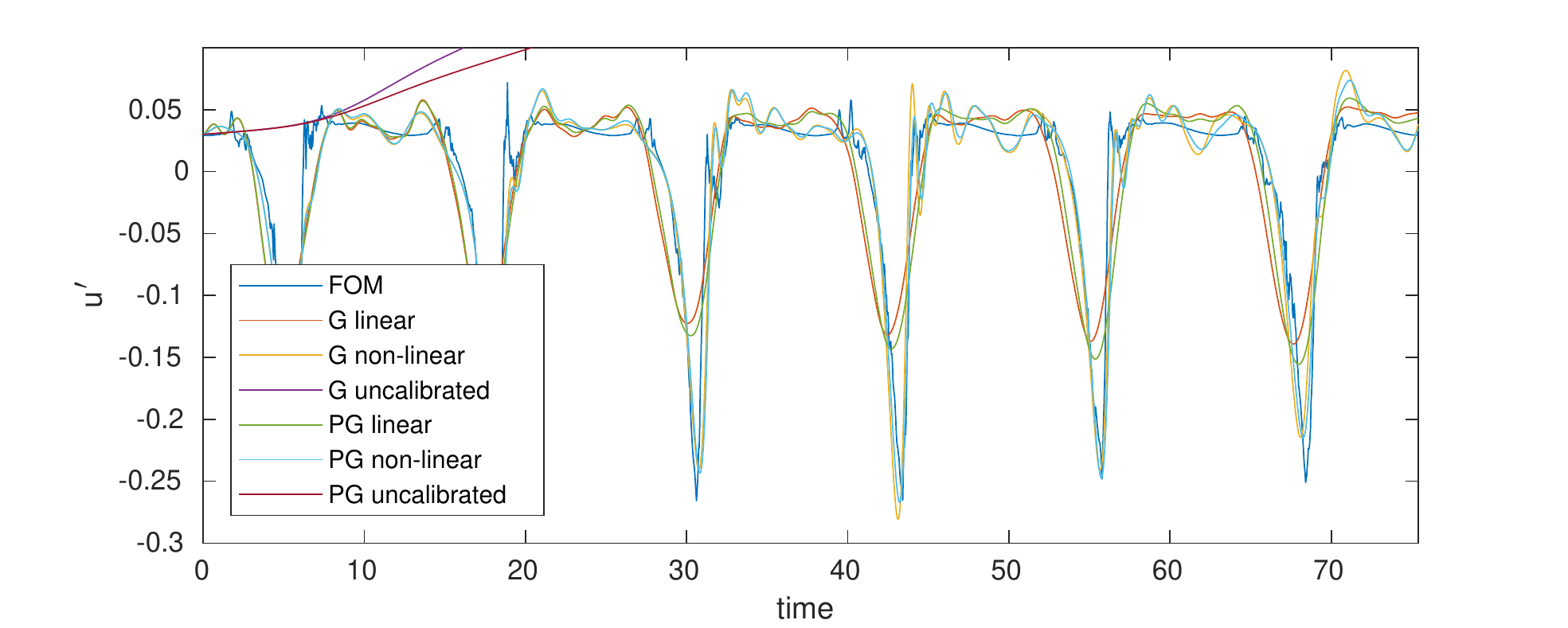}
        \caption{Probe in proximity of the leading edge.}
    \end{subfigure}
    ~
    \begin{subfigure}[hbt!]{.8\textwidth}
        \centering
        \includegraphics[width=1.\textwidth,trim={10mm 1mm 15mm 3mm},clip]{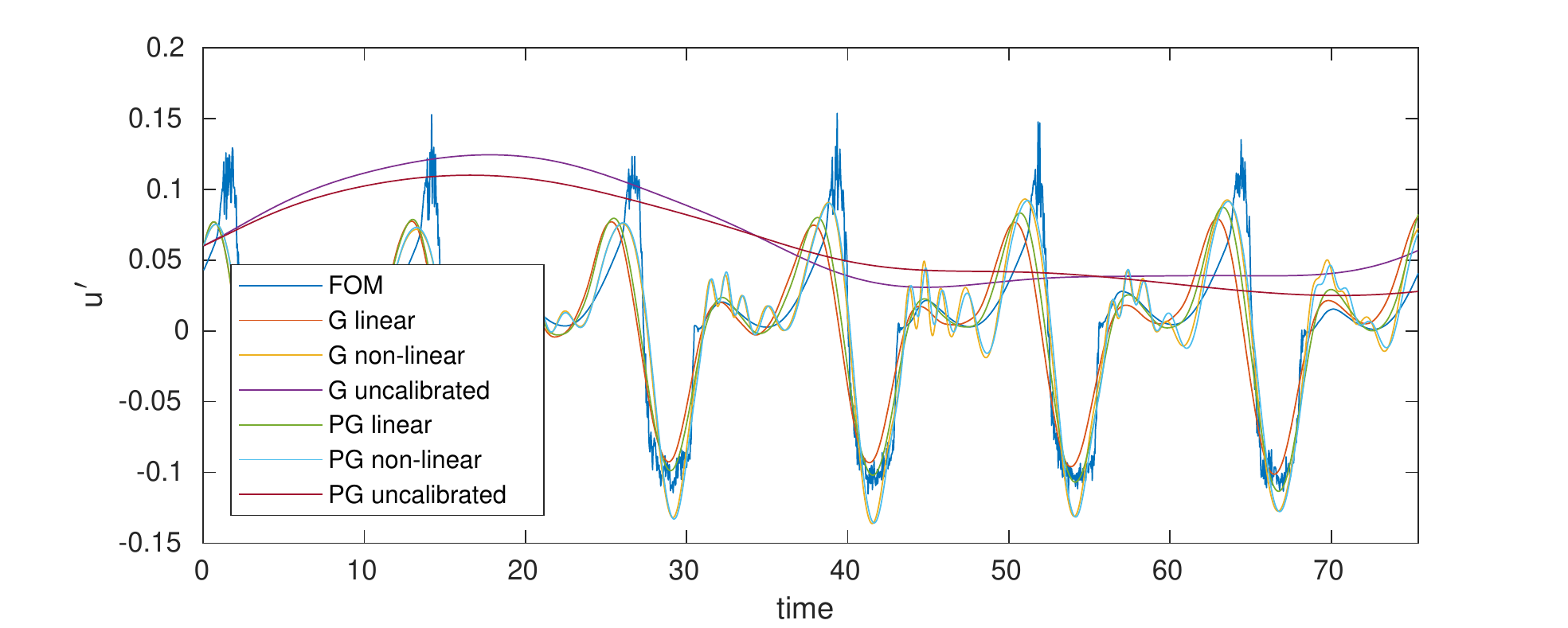}
        \caption{Probe in proximity of the trailing edge.}
    \end{subfigure}
    \caption{Fluctuation time histories computed by the FOM and gapless ROMs with 12 POD modes.}
    \label{fig:airfoil_probes_12modes}
\end{figure}

\begin{figure}[hbt!]
    \centering
    \begin{subfigure}[hbt!]{.8\textwidth}
        \centering
        \includegraphics[width=1.\textwidth,trim={10mm 1mm 15mm 3mm},clip]{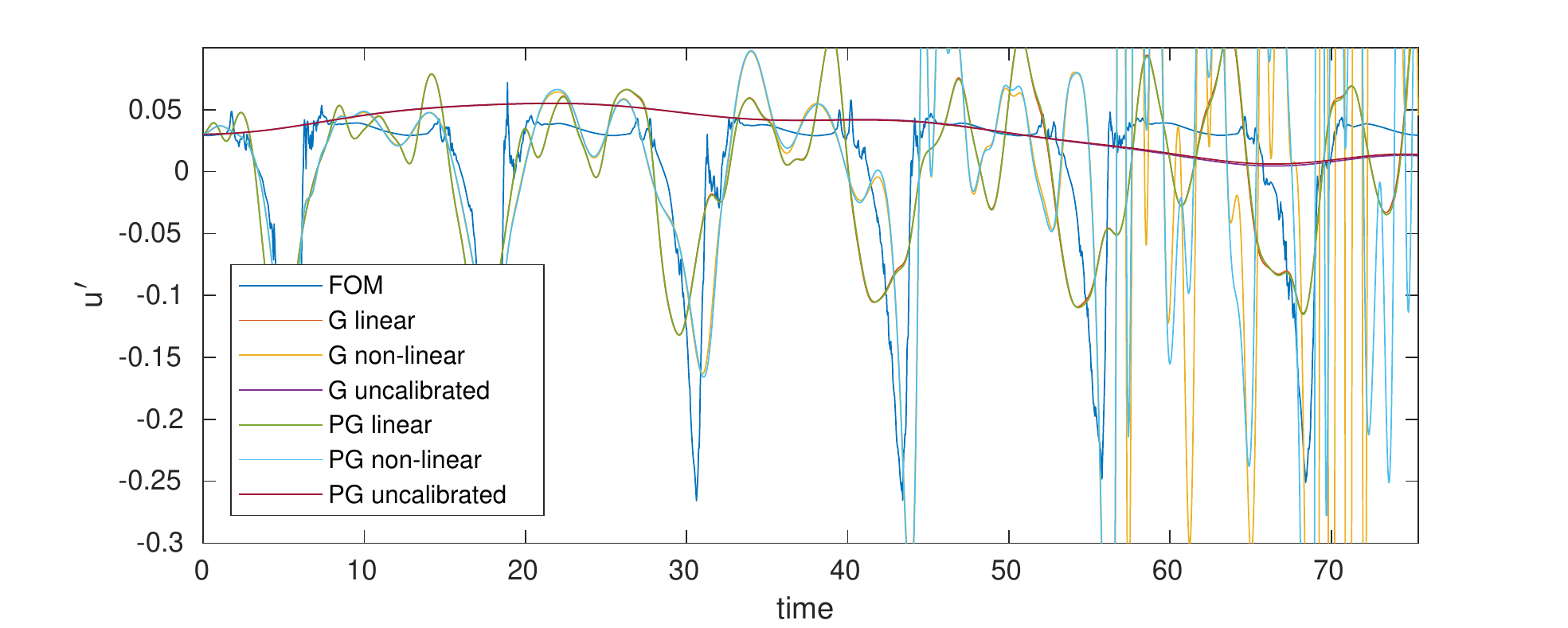}
        \caption{Probe in proximity of the leading edge.}
    \end{subfigure}
    ~
    \begin{subfigure}[hbt!]{.8\textwidth}
        \centering
        \includegraphics[width=1.\textwidth,trim={10mm 1mm 15mm 3mm},clip]{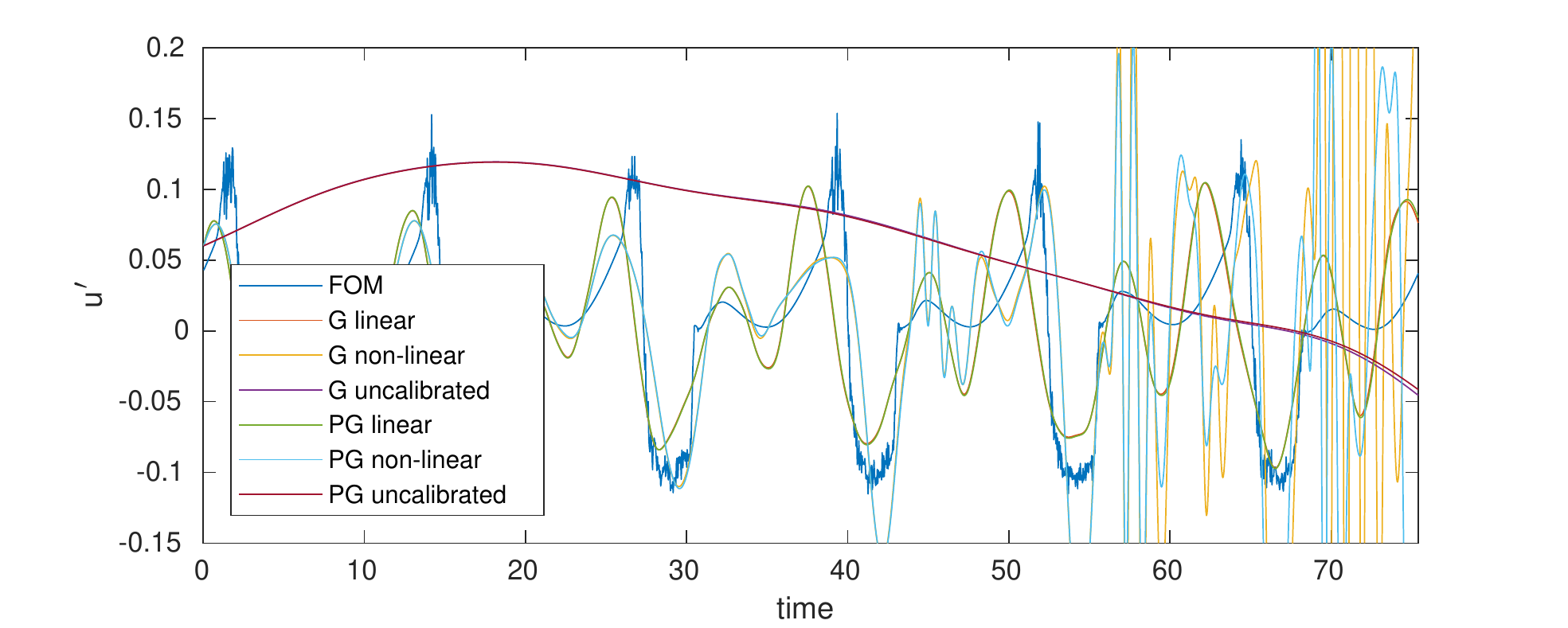}
        \caption{Probe in proximity of the trailing edge.}
    \end{subfigure}
    \caption{Fluctuation time histories computed by the FOM and hyper-reduced ROMs with 12 POD modes.}
    \label{fig:airfoil_probes_12modes_hyper}
\end{figure}

Time histories of u-velocity fluctuations computed by the FOM and ROMs are shown in Figs.  \ref{fig:airfoil_probes_4modes} -- \ref{fig:airfoil_probes_12modes_hyper} for probes placed near the airfoil leading and trailing edges. Additionally, Table \ref{tab:norm_error} contains the relative Frobenius norms $\rho$ and approximation errors $E_1^c$ for the ROMs. As can be see from the figures, uncalibrated ROMs are stable but completely inaccurate for both the Galerkin and LSPG methods, independently of the number of POD modes and hyper-reduction.
In Figs. \ref{fig:airfoil_probes_4modes} and \ref{fig:airfoil_probes_4modes_hyper} ROMs are constructed using 4 POD modes (RIC = $74.65 \%$).
For this number of modes, both linear and non-linear calibration models adopt fully modeled errors (i.e., no regularization required for stable and accurate ROMs). In particular, non-linear coefficients are capable of substantially reducing the error indicator but the resulting calibration coefficients are very invasive (specially when hyper-reduction is considered). This manifests itself visually by the gradual increase of the relative Frobenius norms shown in Table \ref{tab:norm_error}. For hyper-reduced models, purely linear coefficients have slightly higher errors when compared to gapless models.
As shown in the u-velocity fluctuations of Figs. \ref{fig:airfoil_probes_4modes} and \ref{fig:airfoil_probes_4modes_hyper}, solutions obtained with linear coefficients recover some of the lower frequency oscillations but suffer to learn finer flow features. On the other hand, the non-linear calibration solutions are capable of not only better recover the lower frequency fluctuations but also a significant part of the finer scales. For the gapless models, Galerkin and LSPG results are almost visually identical when non-linear calibration is employed. On the other hand, both linear and non-linear calibrated hyper-reduced models are visually identical.

Models constructed with 8 modes can be seen in Figs. \ref{fig:airfoil_probes_8modes} and \ref{fig:airfoil_probes_8modes_hyper} (RIC = $92.42 \%$). Results for linear calibration coefficients are shown for models with norm convergence and still suffer to reproduce the finer flow scales. Compared to the models with four POD modes, the gapless models show a small improvement of the error indicator after calibration while the error worsens when hyper-reduction is considered (see table \ref{tab:norm_error}). For the non-linear coefficients, the L-curves are steep and it is possible to obtain less intrusive coefficients with smaller errors. On the other hand, the additional modes used in the reconstruction makes regularization a requirement to obtain stable and accurate models. Stable models with good accuracy could be obtained using the guidance of the L-curve and running a few models to check for stability. Probes show results for stable models with less intrusive calibration coefficients but marginally larger error indicators (see Table \ref{tab:norm_error}).

The last set of ROMs analysed is obtained using a 12-mode basis (RIC = $97.34 \%$) and probed results can be seen in Figs. \ref{fig:airfoil_probes_12modes} and \ref{fig:airfoil_probes_12modes_hyper}. Again, linear calibration coefficients are fully modeled while non-linear calibration coefficients require regularization. Discrepancies between gapless and hyper-reduced models are considerably larger in this case.
Gapless models obtained with linear calibration coefficients benefit the most from having a more complete basis of POD modes. In fact, the error indicator $E_1^c$ is almost halved compared to the previous cases as shown in Table \ref{tab:norm_error}. Furthermore, regularization becomes an even bigger issue for models with non-linear calibration operators. Again, stable solutions are obtained after studying the L-curve and running a few models for a stability check. As can be observed in Table \ref{tab:norm_error}, the ratio of the Frobenius norms of these models is smaller compared to those obtained with fewer modes and this leads to an increase in $E_1^c$.
On the other hand, hyper-reduced models perform poorly for both linear and non-linear calibration coefficients. The performance of the hyper-reduced linear models is very different compared to the gapless linear models despite the small difference between error proxies. As shown by the figures, the performance of the former deteriorates with 12 modes. In a similar fashion, the hyper-reduced non-linear calibrations models present almost the same errors as their gapless counterparts. However, hyper-reduced solutions are unstable and  attempting to obtain more accurate ROMs using less intrusive non-linear coefficients was unsuccessful. In fact, this led to non-linear models performing similarly to the linear ones.
\begin{table}[hbt!]
    \centering
    \begin{adjustbox}{width=\columnwidth,center}
    \begin{tabular}{|c|c|c|c|c|c|c|c|c|c|}
        \hline
        Hyper-reduction & \multirow{2}{*}{Modes} & \multicolumn{4}{c|}{Galerkin} & \multicolumn{4}{c|}{Petrov-Galerkin}  \\ \cline{3-10}
        (Grid points) & & $\rho_{G}^{L}$ & $E_1^c \cdot 10^{7}$ & $\rho_{G}^{NL}$ & $E_1^c \cdot 10^{7}$ & $\rho_{PG}^{L}$ & $E_1^c \cdot 10^{7}$ & $\rho_{PG}^{NL}$ & $E_1^c \cdot 10^{7}$ \\ \hline
        \multirow{3}{*}{\adjustbox{stack=cc}{Yes\\(1,000)}} & 4 & 14.058 & 273.29 & 228.18 & 9.7357 & 8.9025 & 273.35 & 144.46 & 9.7198 \\ \cline{2-10}
        & 8 & 7.3549 & 281.78 & 29.586 & 17.354 & 3.9919 & 281.89 & 15.255 & 19.811 \\ \cline{2-10}
        & 12 & 6.1504 & 154.67 & 16.431 & 32.661 & 3.4044 & 154.91 & 9.2299 & 32.329 \\ \hline
        \multirow{3}{*}{\adjustbox{stack=cc}{No\\(168,350)}} & 4 & 1.3182 & 260.35 & 21.7529 & 9.7341 & 0.4519 & 259.34 & 7.1836 & 9.7600 \\ \cline{2-10}
        & 8 & 1.2191 & 254.04 & 4.9123 & 17.143 & 0.5018 & 253.86 & 2.2382 & 13.860 \\ \cline{2-10}
        & 12 & 0.9751 & 139.13 & 2.5158 & 32.734 & 0.3025 & 138.67 & 0.7660 & 32.966 \\ \hline
    \end{tabular}
    \end{adjustbox}
    \caption{Ratio of the Frobenius norms $\rho$ and approximation error $E_1^c$ of the ROMs. Here, the superscript specifies if calibration coefficients are linear (L) or non-linear (NL) and the subscript specifies if the projection method is Galerkin (G) or Petrov-Galerkin (PG).}
    \label{tab:norm_error}
\end{table}

\begin{figure}[hbt!]
    \centering
    \begin{subfigure}[hbt!]{.3\textwidth}
        \centering
        \includegraphics[width=.9\textwidth,trim={100mm 20mm 80mm 30mm},clip]{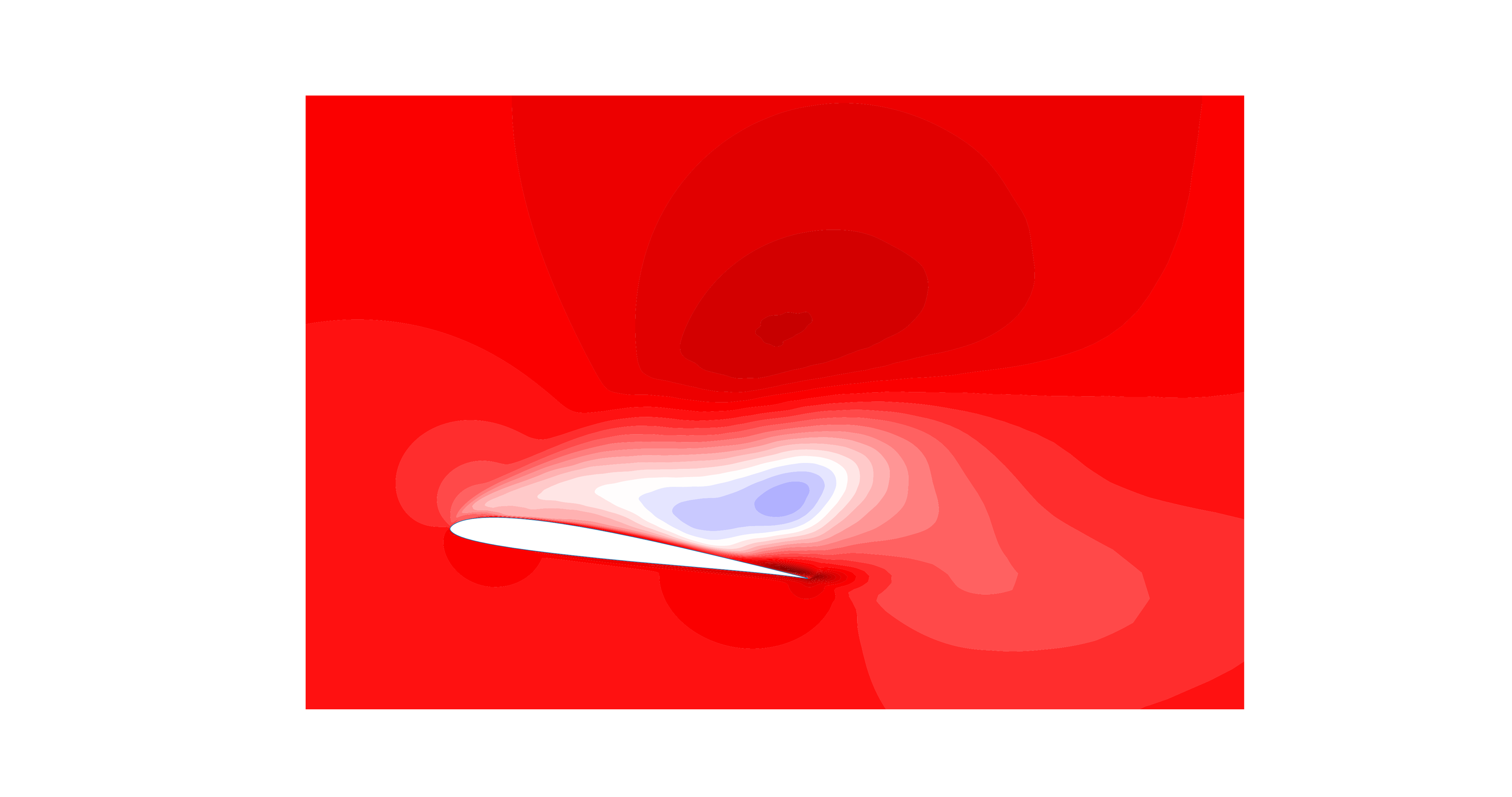}
        \caption{Linear 4-mode G-ROM.}
    \end{subfigure}
    ~
    \begin{subfigure}[hbt!]{.3\textwidth}
        \centering
        \includegraphics[width=.9\textwidth,trim={100mm 20mm 80mm 30mm},clip]{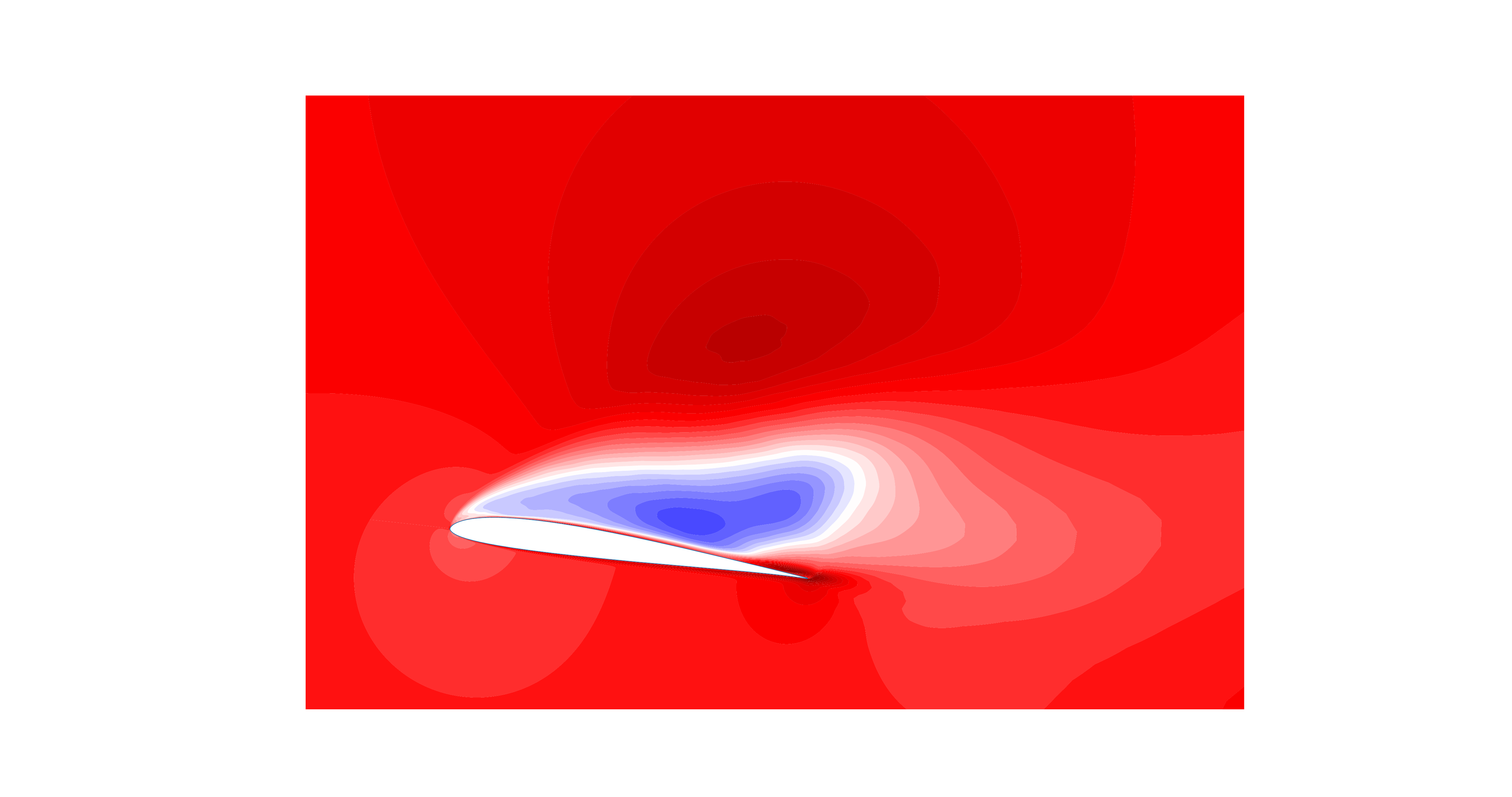}
        \caption{Linear 8-mode G-ROM.}
    \end{subfigure}
    ~
    \begin{subfigure}[hbt!]{.3\textwidth}
        \centering
        \includegraphics[width=.9\textwidth,trim={100mm 20mm 80mm 30mm},clip]{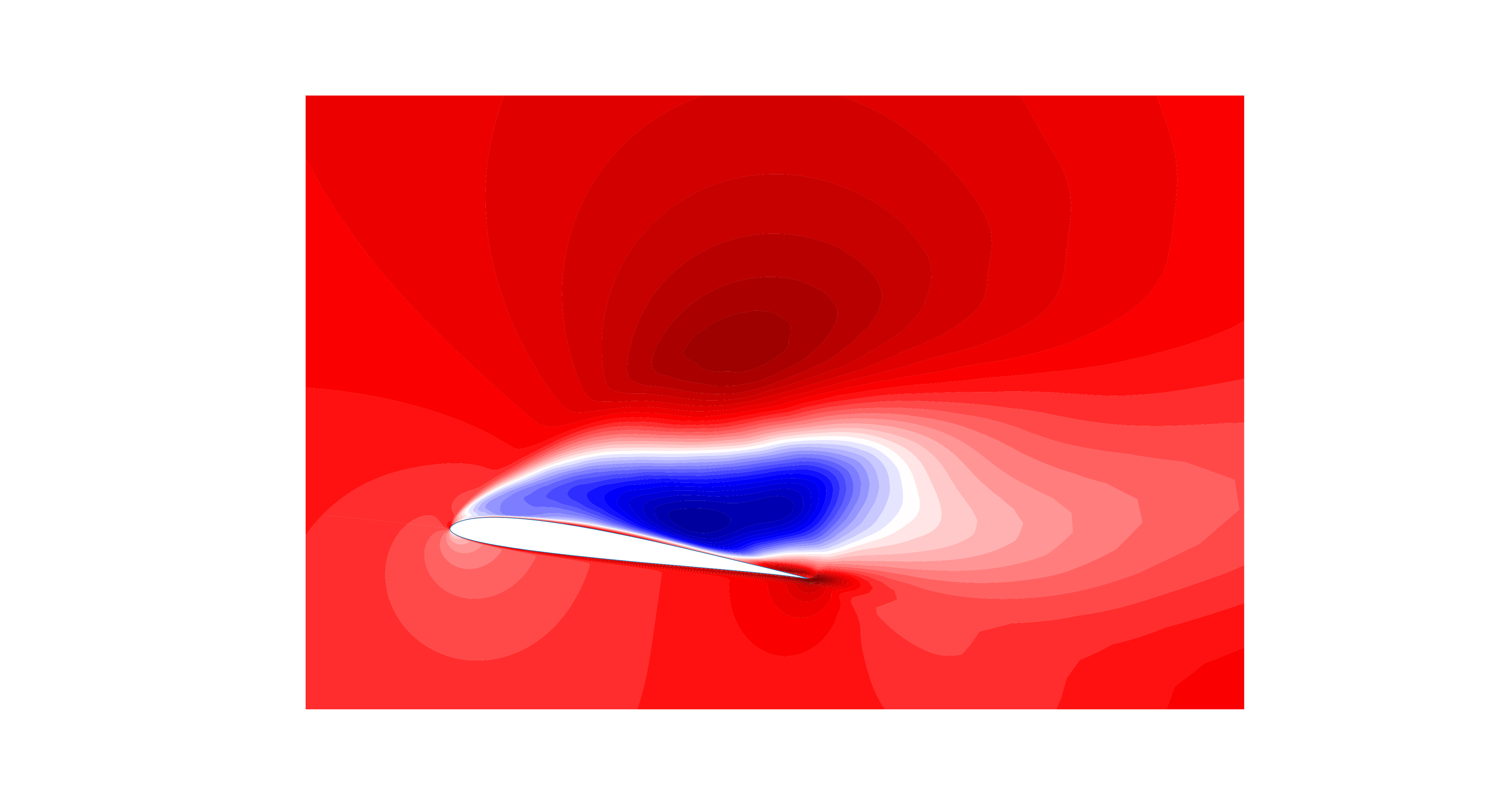}
        \caption{Linear 12-mode G-ROM.}
    \end{subfigure}
    \begin{subfigure}[hbt!]{.30\textwidth}
        \centering
        \includegraphics[width=.9\textwidth,trim={100mm 20mm 80mm 30mm},clip]{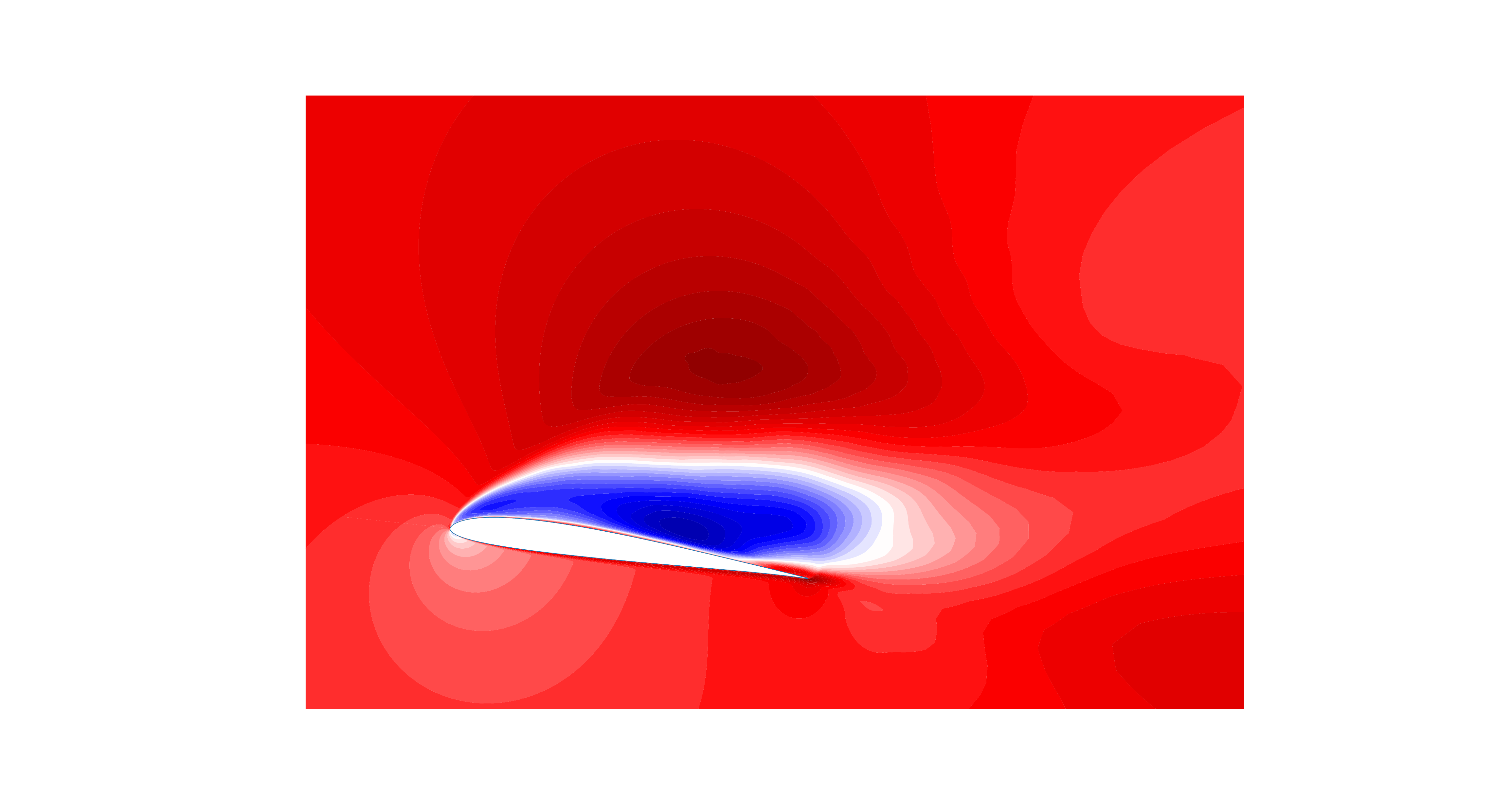}
        \caption{Non-linear 4-mode G-ROM.}
    \end{subfigure}
    ~
    \begin{subfigure}[hbt!]{.30\textwidth}
        \centering
        \includegraphics[width=.9\textwidth,trim={100mm 20mm 80mm 30mm},clip]{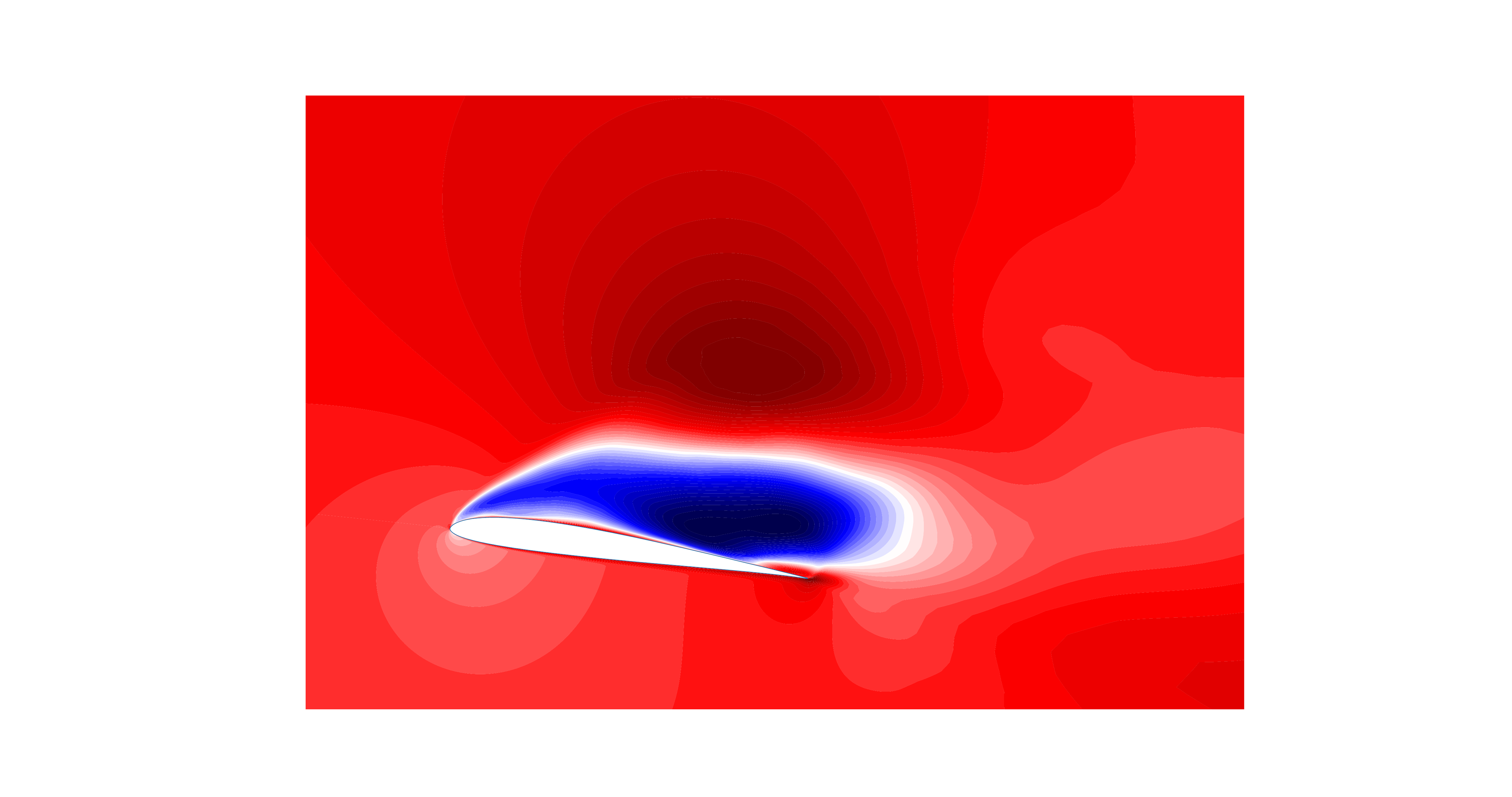}
        \caption{Non-linear 8-mode G-ROM.}
    \end{subfigure}
    ~
    \begin{subfigure}[hbt!]{.30\textwidth}
        \centering
        \includegraphics[width=.9\textwidth,trim={100mm 20mm 80mm 30mm},clip]{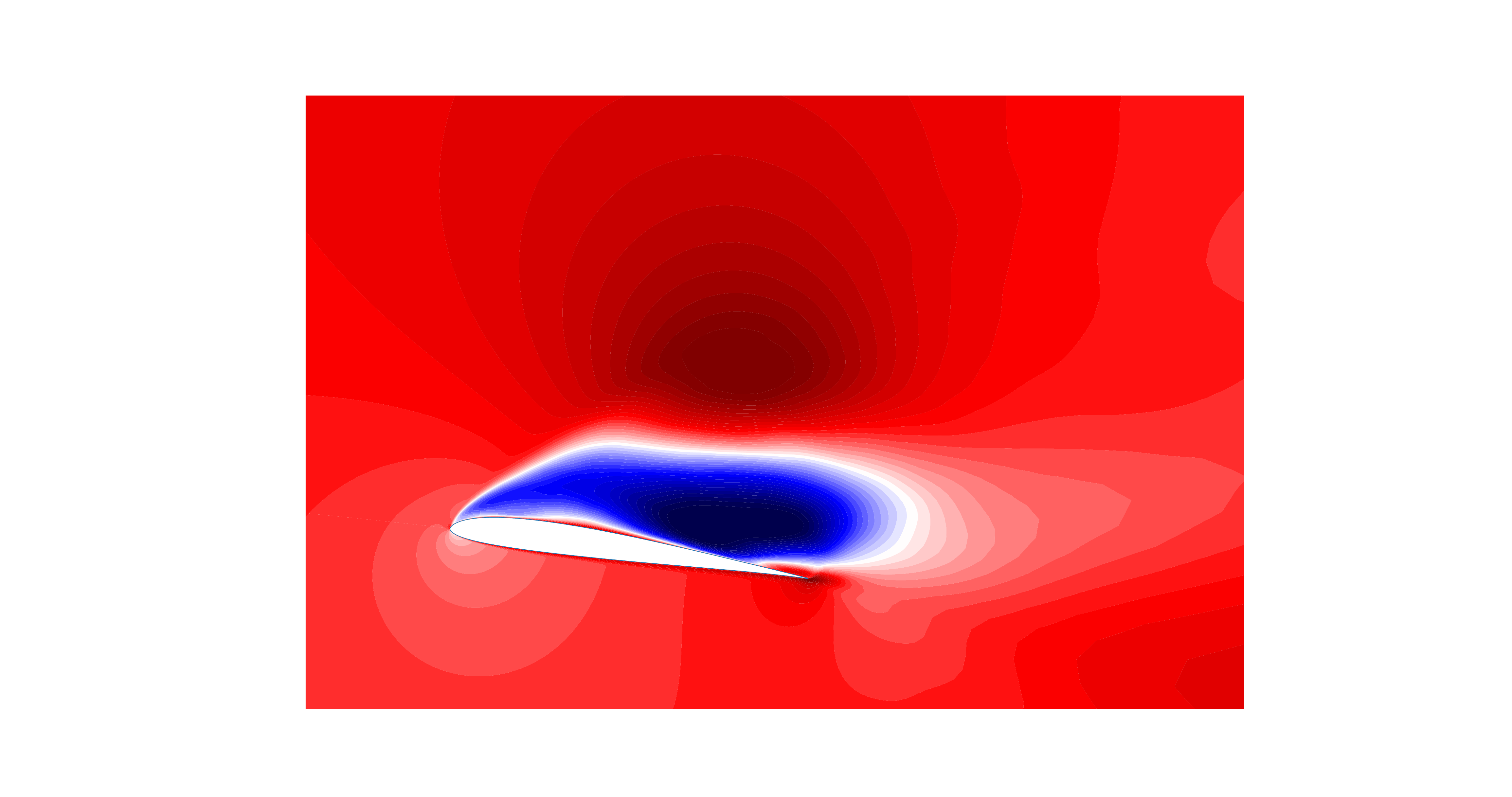}
        \caption{Non-linear 12-mode G-ROM.}
    \end{subfigure}
    \begin{subfigure}[hbt!]{.30\textwidth}
        \centering
        \includegraphics[width=.9\textwidth,trim={100mm 20mm 80mm 30mm},clip]{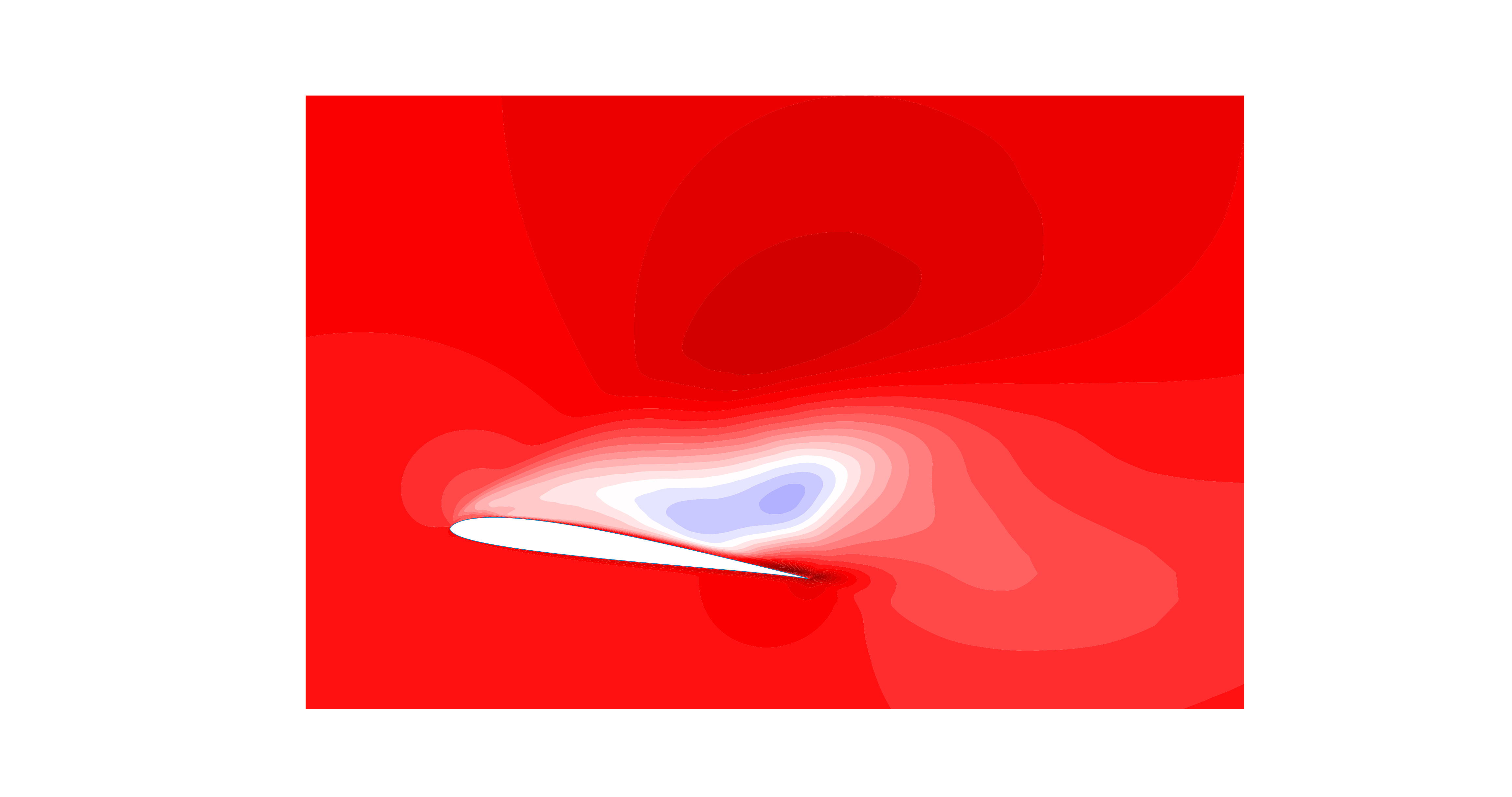}
        \caption{Linear 4-modes PG-ROM.}
    \end{subfigure}
    ~
    \begin{subfigure}[hbt!]{.30\textwidth}
        \centering
        \includegraphics[width=.9\textwidth,trim={100mm 20mm 80mm 30mm},clip]{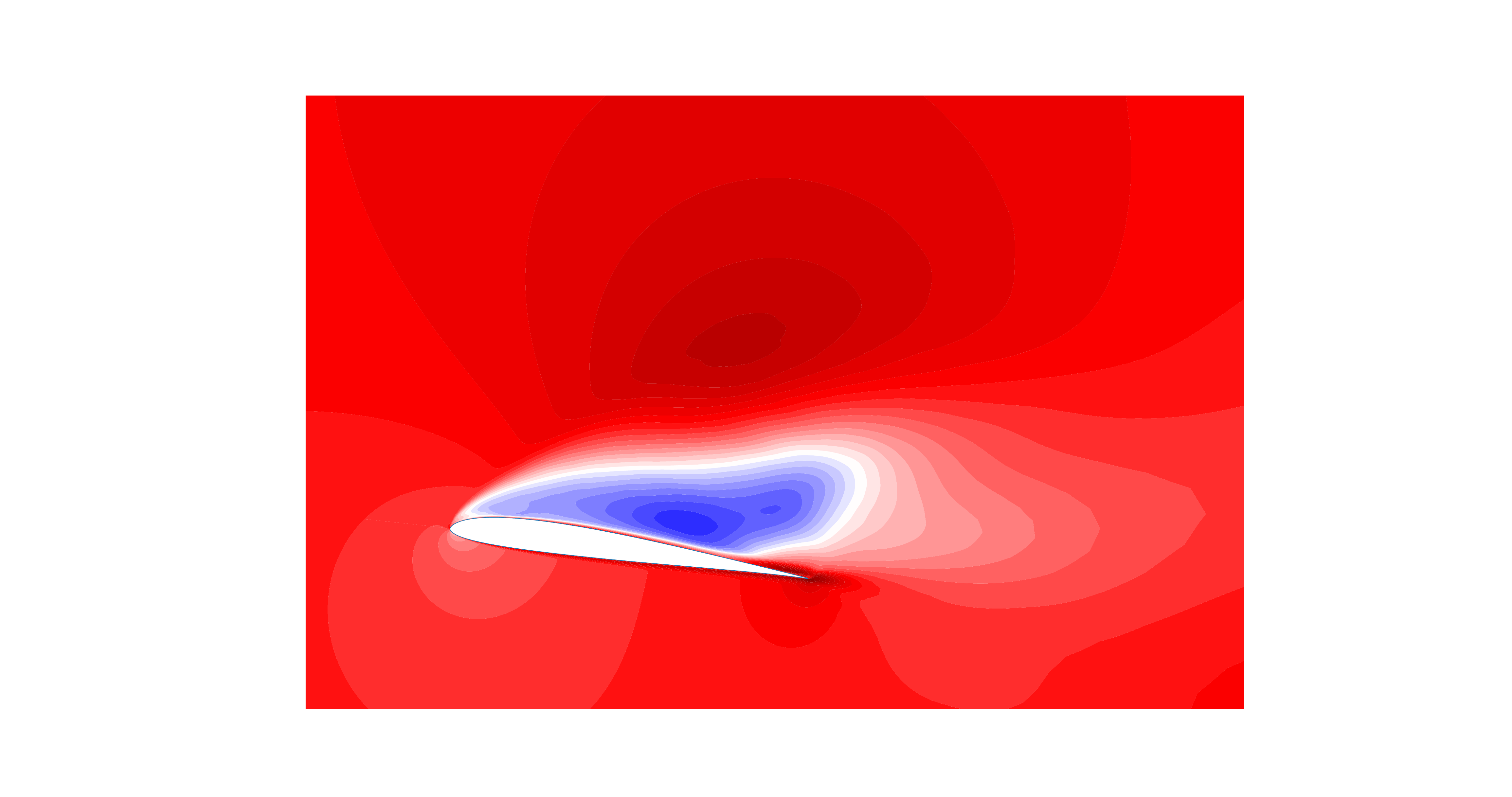}
        \caption{Linear 8-modes PG-ROM.}
    \end{subfigure}
    ~
    \begin{subfigure}[hbt!]{.30\textwidth}
        \centering
        \includegraphics[width=.9\textwidth,trim={100mm 20mm 80mm 30mm},clip]{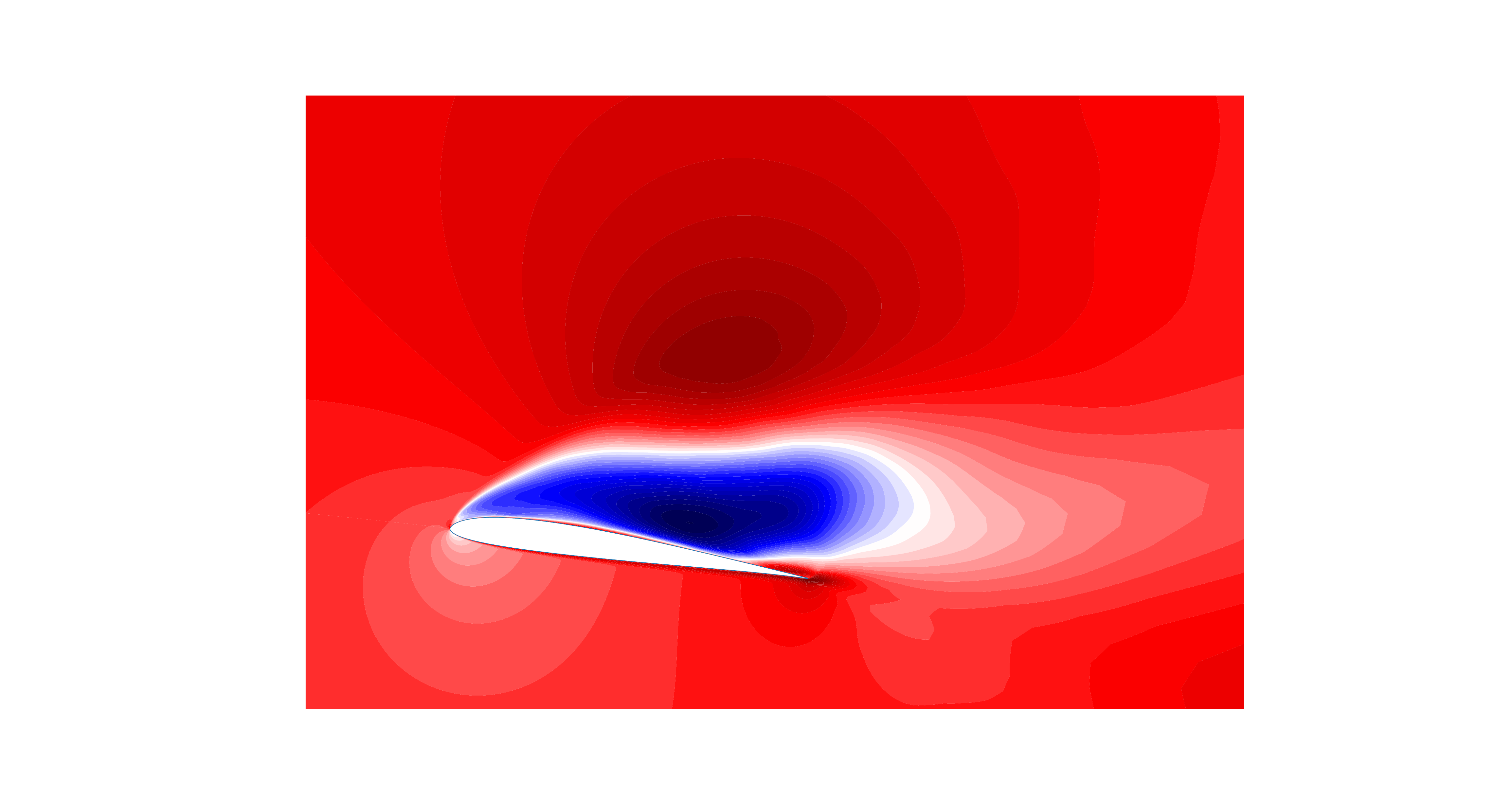}
        \caption{Linear 12-modes PG-ROM.}
    \end{subfigure}
    \begin{subfigure}[hbt!]{.30\textwidth}
        \centering
        \includegraphics[width=.9\textwidth,trim={100mm 20mm 80mm 30mm},clip]{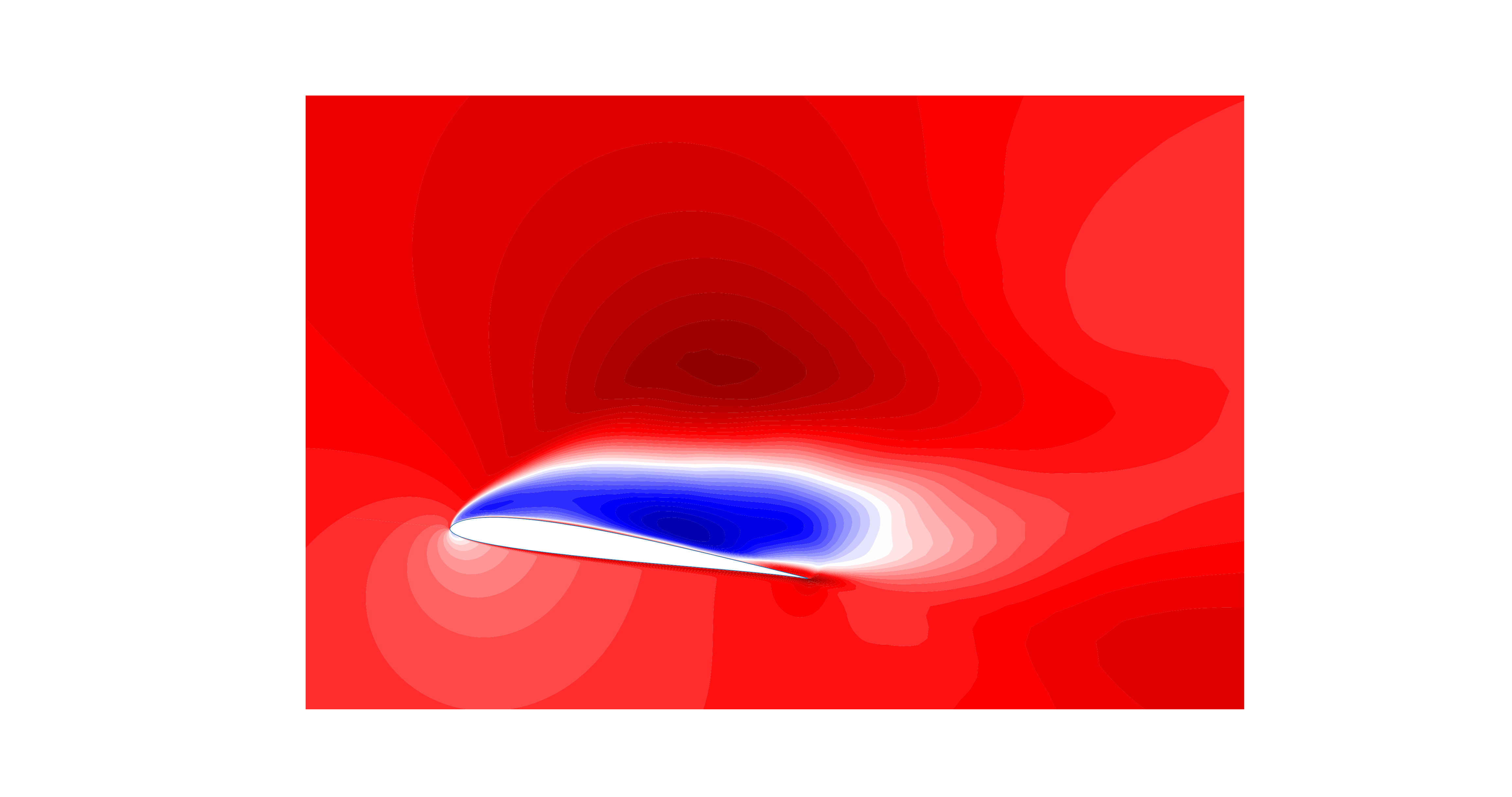}
        \caption{Non-linear 4-modes PG-ROM.}
    \end{subfigure}
    ~
    \begin{subfigure}[hbt!]{.30\textwidth}
        \centering
        \includegraphics[width=.9\textwidth,trim={100mm 20mm 80mm 30mm},clip]{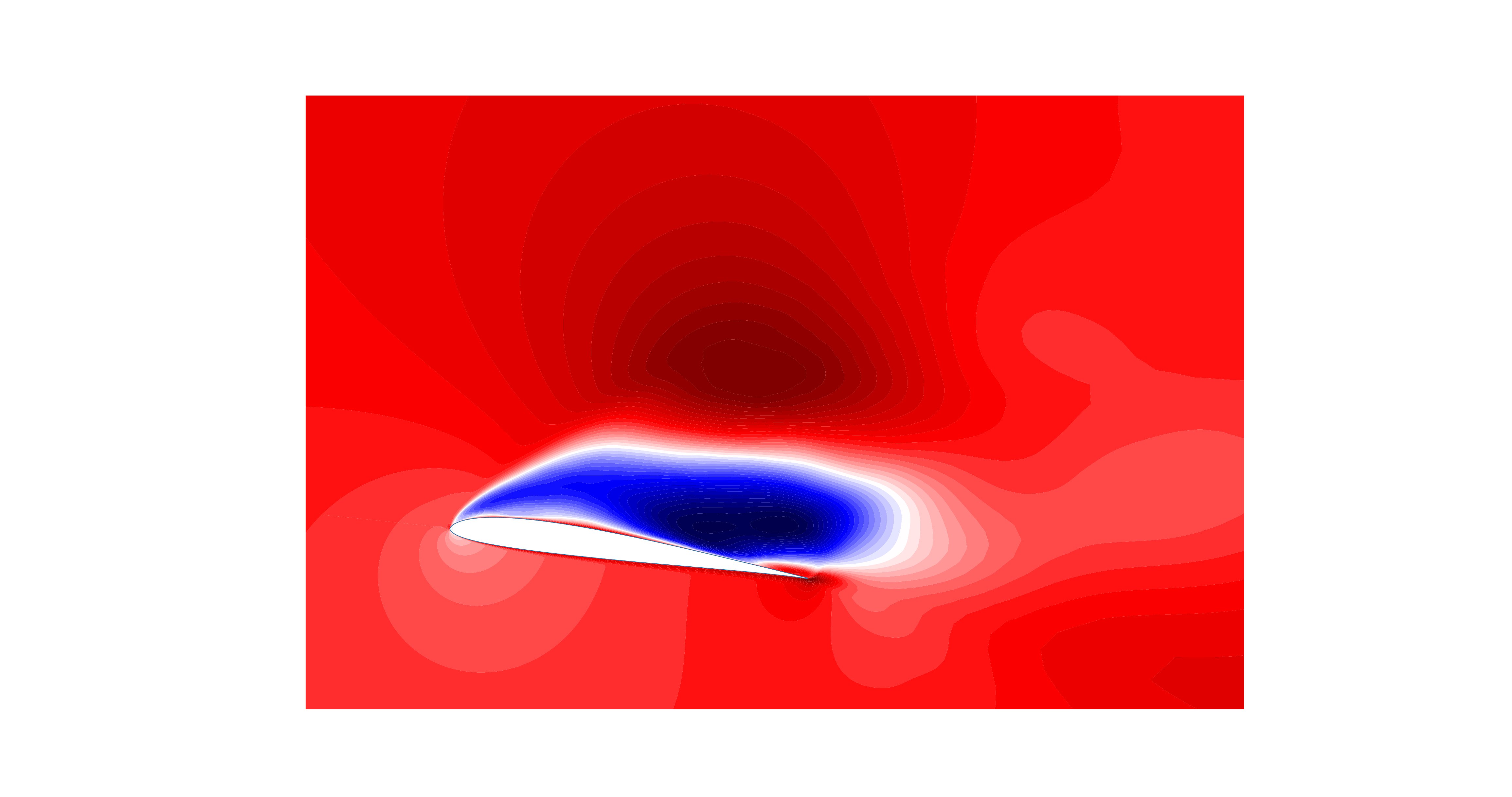}
        \caption{Non-linear 8-modes PG-ROM.}
    \end{subfigure}
    ~
    \begin{subfigure}[hbt!]{.30\textwidth}
        \centering
        \includegraphics[width=.9\textwidth,trim={100mm 20mm 80mm 30mm},clip]{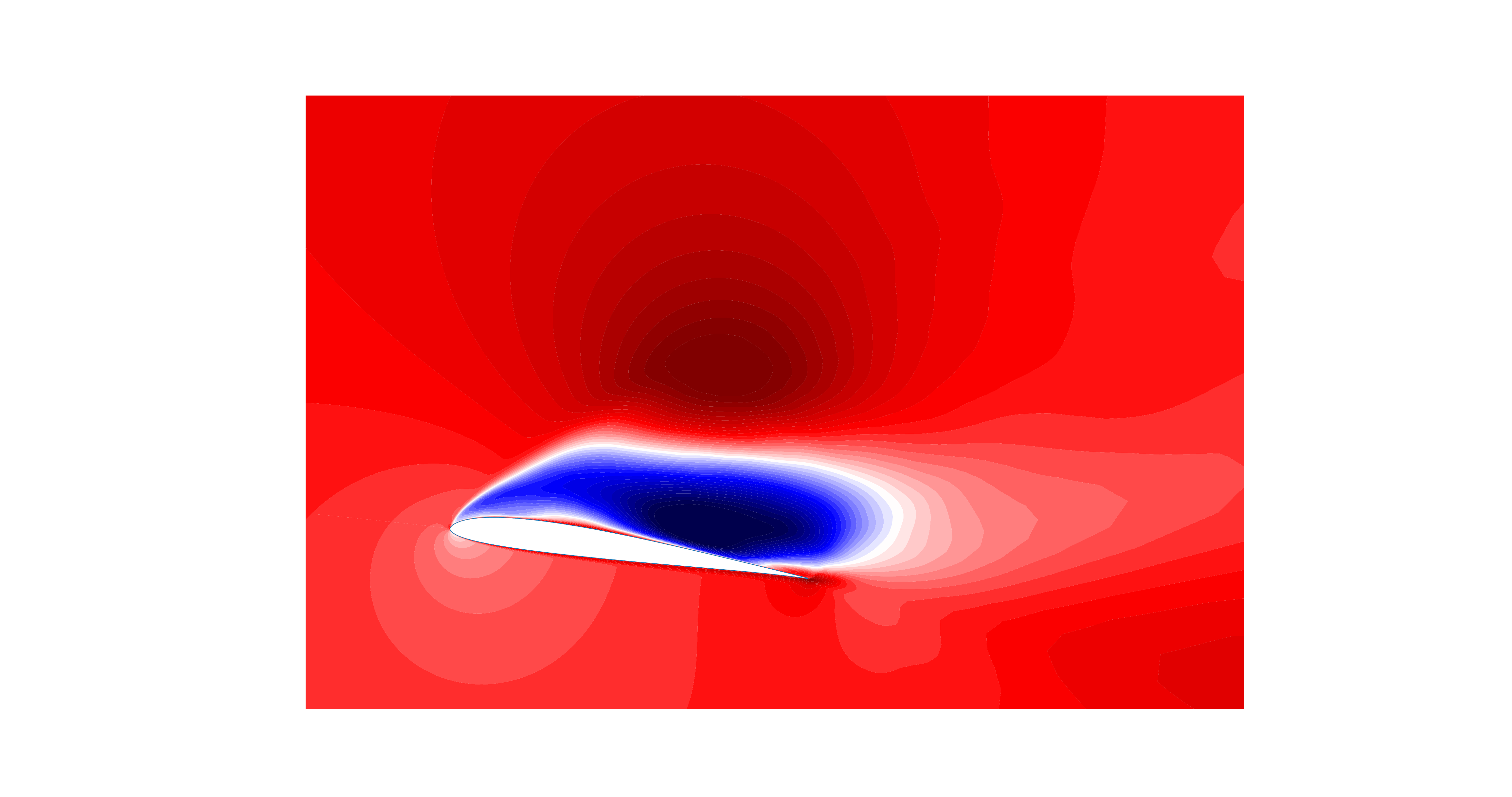}
        \caption{Non-linear 12-modes PG-ROM.}
    \end{subfigure}
    \begin{subfigure}[hbt!]{.30\textwidth}
        \centering
        \includegraphics[width=.9\textwidth,trim={100mm 20mm 80mm 30mm},clip]{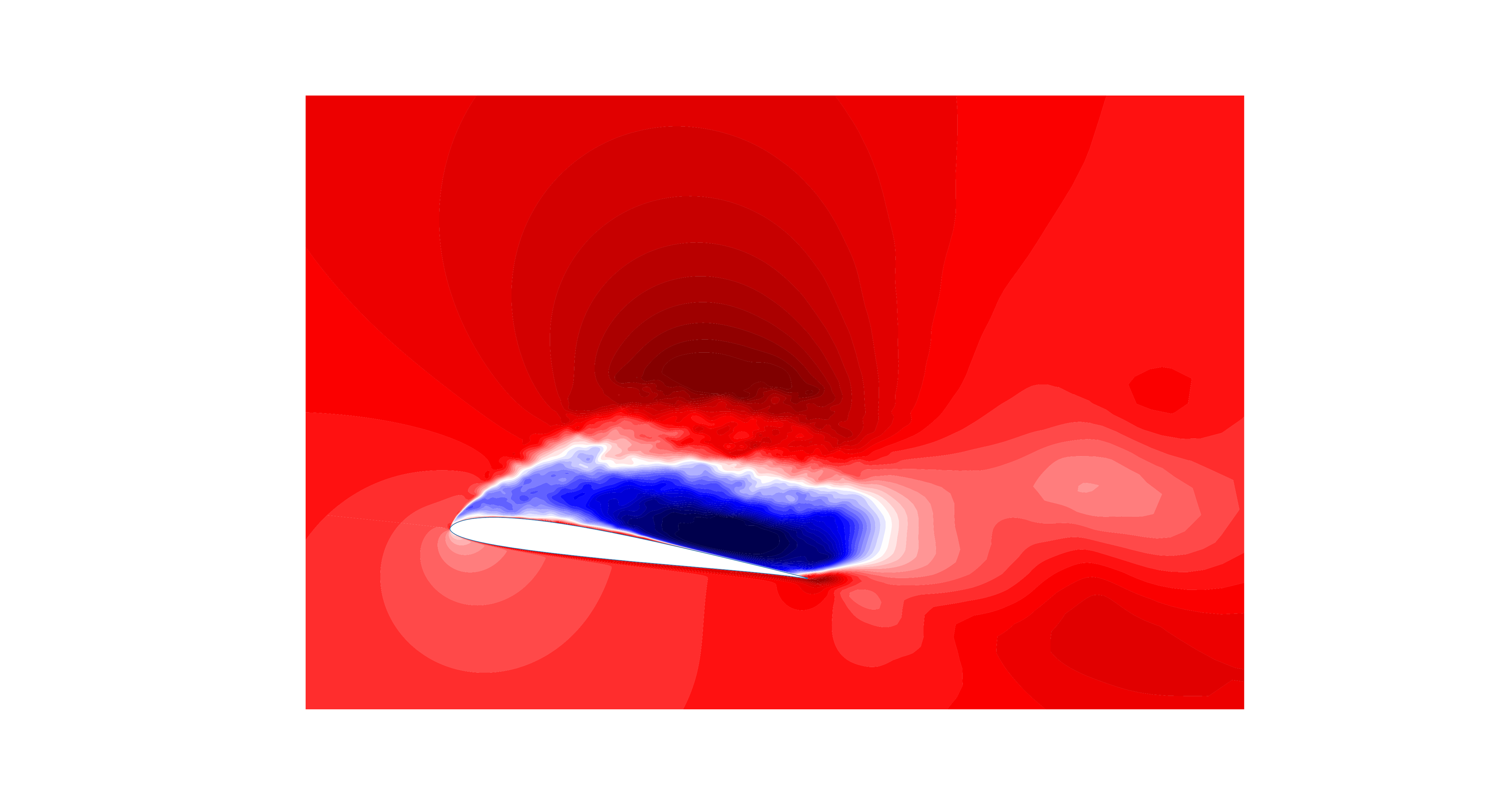}
        \caption{Full order model.}
    \end{subfigure}
    \caption{Contours of u-velocity fluctuations computed at $t = 67.5$ using calibrated gapless models with implicit Euler integration. The terms ``G'' and ``PG'' stand for Galerkin and LSPG models, respectively.}
    \label{fig:airfoil_snapshots}
\end{figure}

\begin{figure}[hbt!]
    \centering
    \begin{subfigure}[hbt!]{.30\textwidth}
        \centering
        \includegraphics[width=.9\textwidth,trim={100mm 20mm 80mm 30mm},clip]{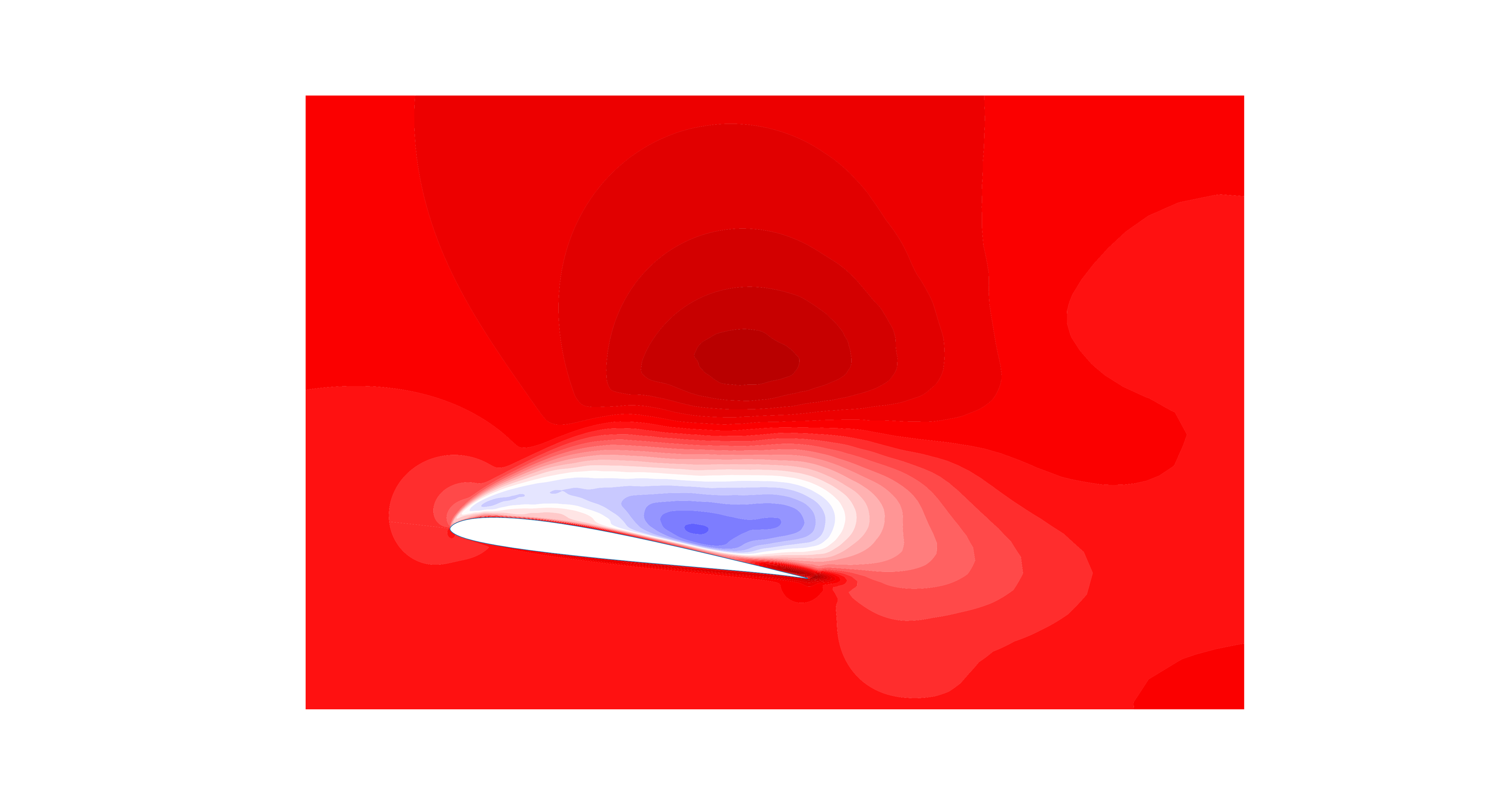}
        \caption{Linear 4-mode G-ROM.}
    \end{subfigure}
    ~
    \begin{subfigure}[hbt!]{.30\textwidth}
        \centering
        \includegraphics[width=.9\textwidth,trim={100mm 20mm 80mm 30mm},clip]{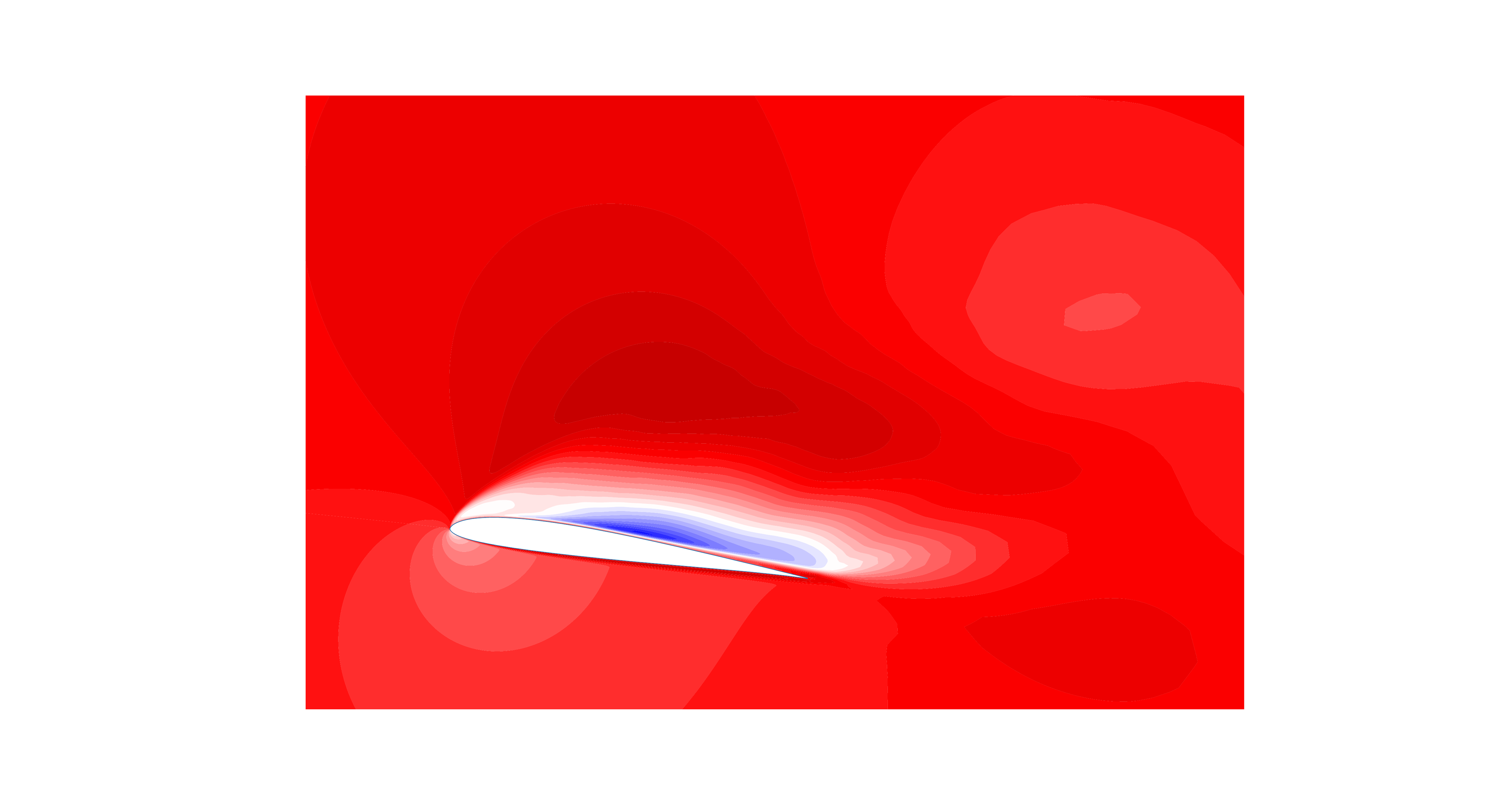}
        \caption{Linear 8-mode G-ROM.}
    \end{subfigure}
    ~
    \begin{subfigure}[hbt!]{.30\textwidth}
        \centering
        \includegraphics[width=.9\textwidth,trim={100mm 20mm 80mm 30mm},clip]{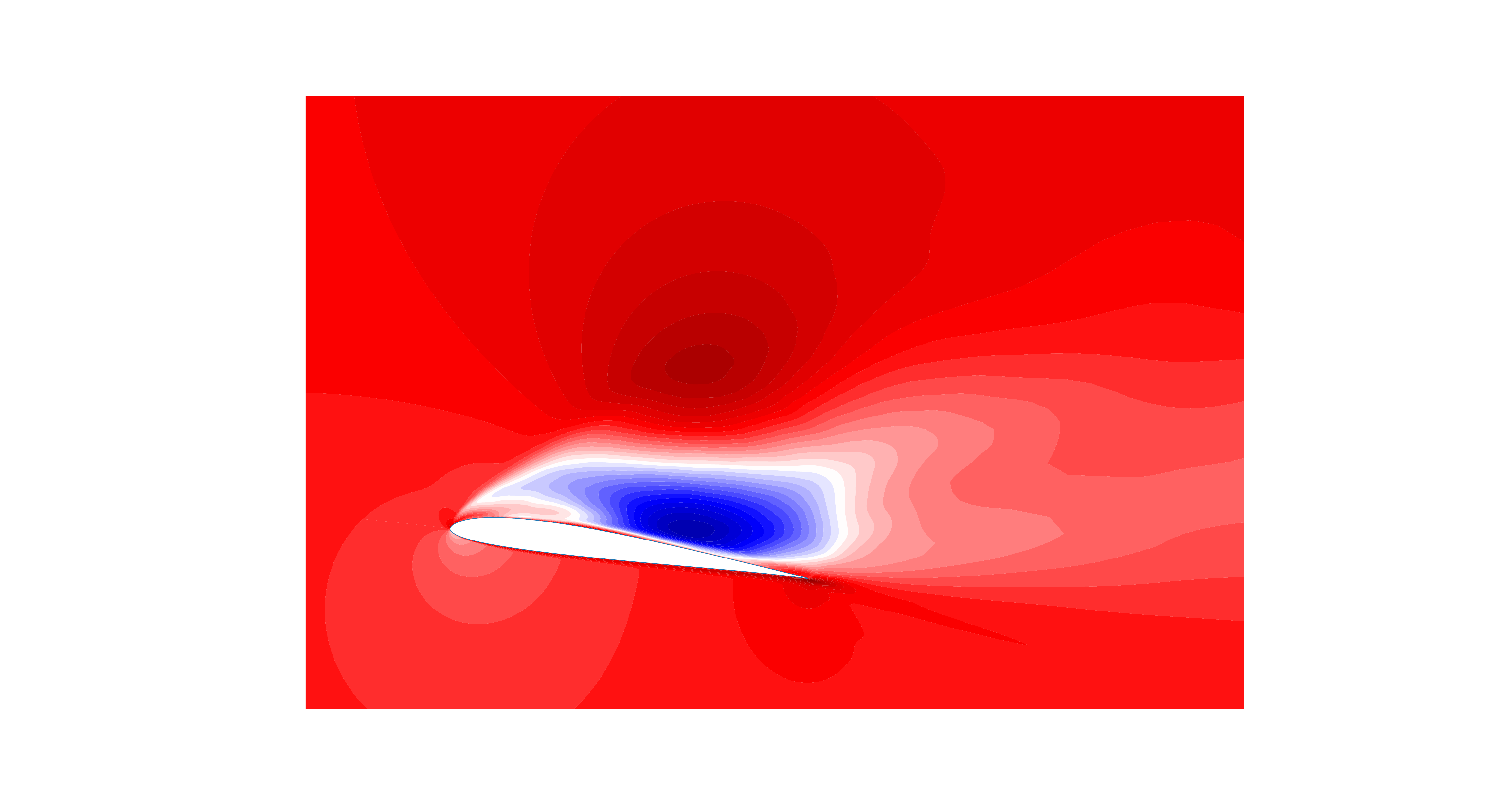}
        \caption{Linear 12-mode G-ROM.}
    \end{subfigure}
    \begin{subfigure}[hbt!]{.30\textwidth}
        \centering
        \includegraphics[width=.9\textwidth,trim={100mm 20mm 80mm 30mm},clip]{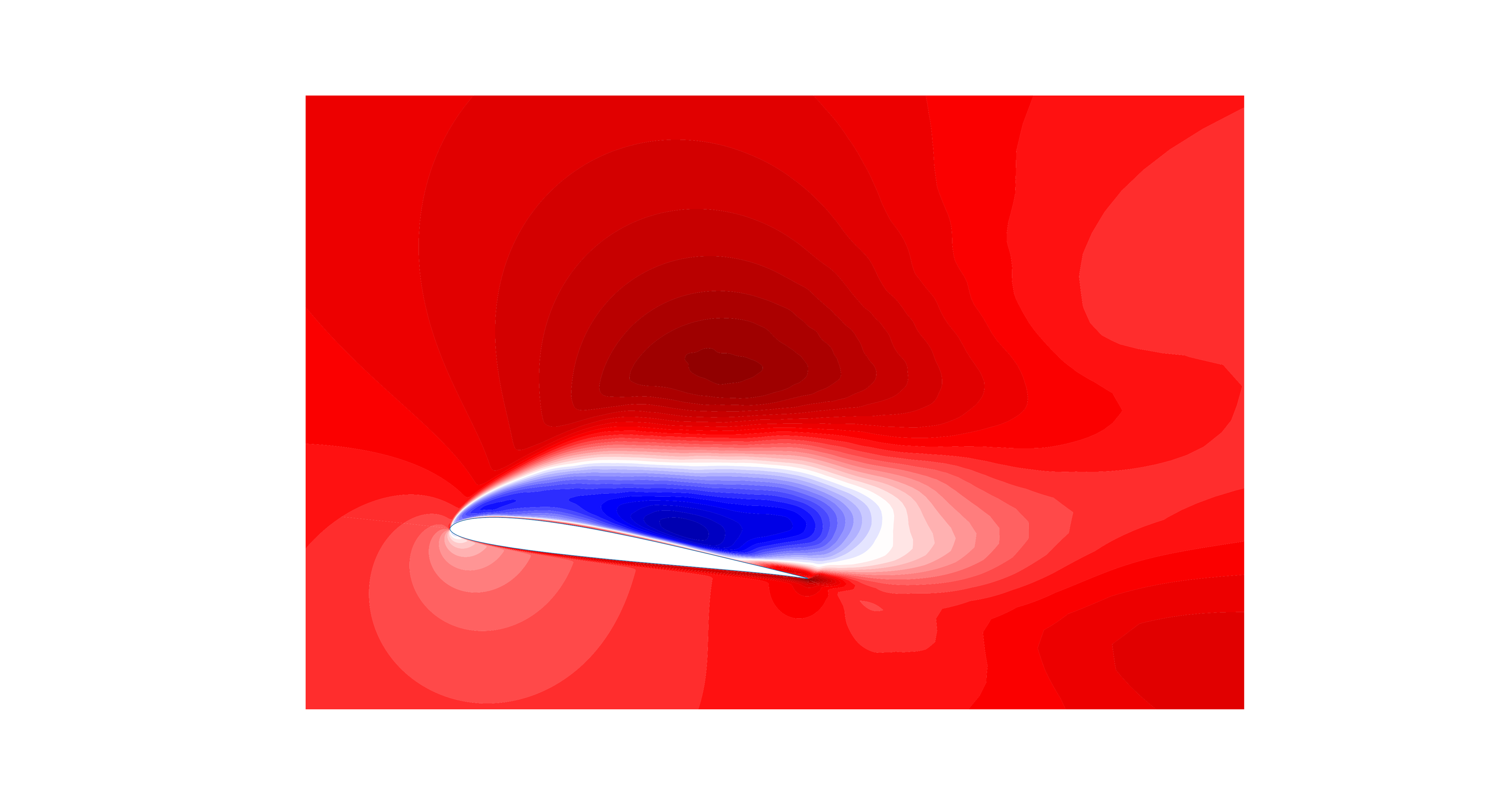}
        \caption{Non-linear 4-mode G-ROM.}
    \end{subfigure}
    ~
    \begin{subfigure}[hbt!]{.30\textwidth}
        \centering
        \includegraphics[width=.9\textwidth,trim={100mm 20mm 80mm 30mm},clip]{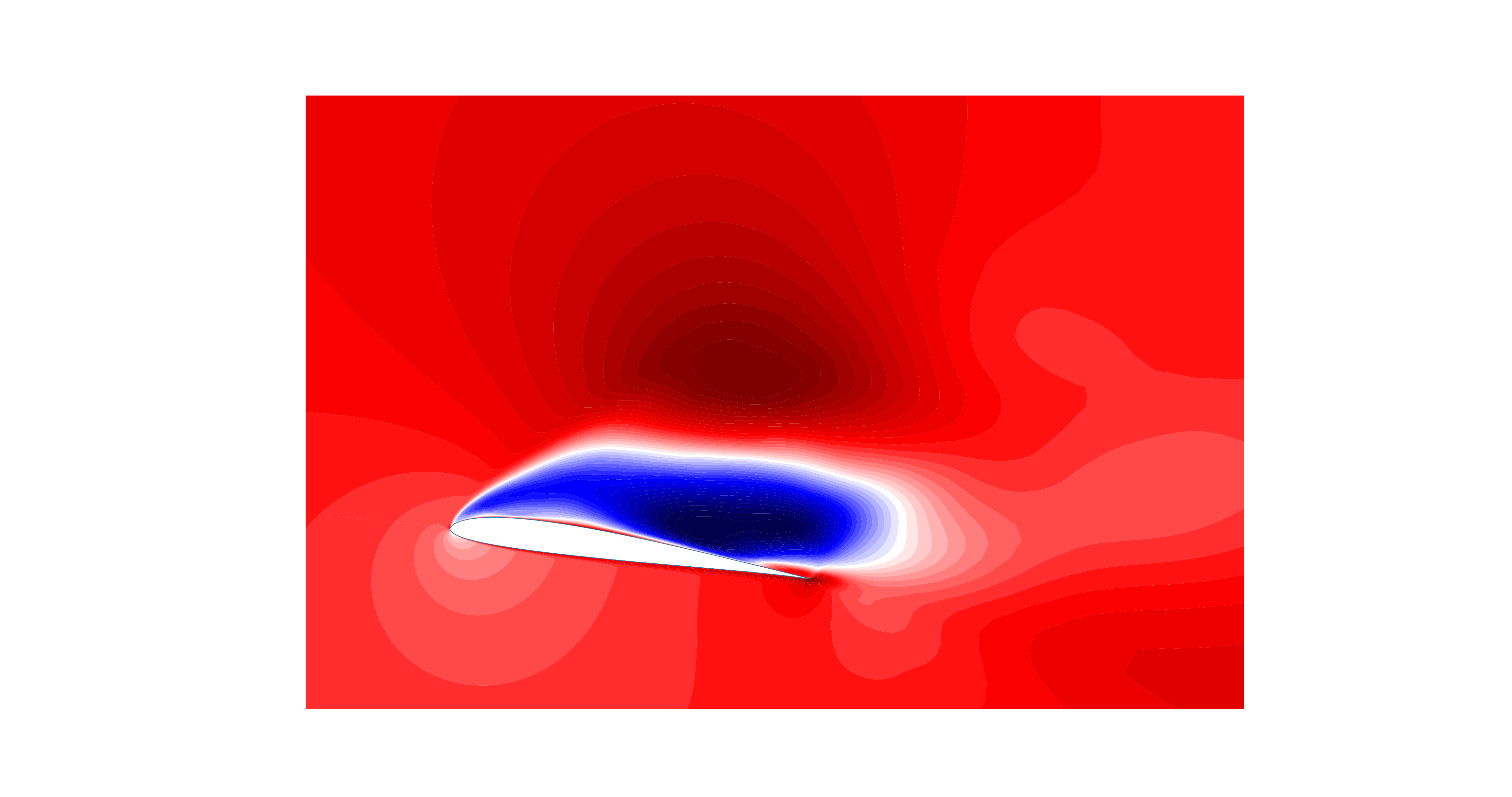}
        \caption{Non-linear 8-mode G-ROM.}
    \end{subfigure}
    ~
    \begin{subfigure}[hbt!]{.30\textwidth}
        \centering
        \includegraphics[width=.9\textwidth,trim={100mm 20mm 80mm 30mm},clip]{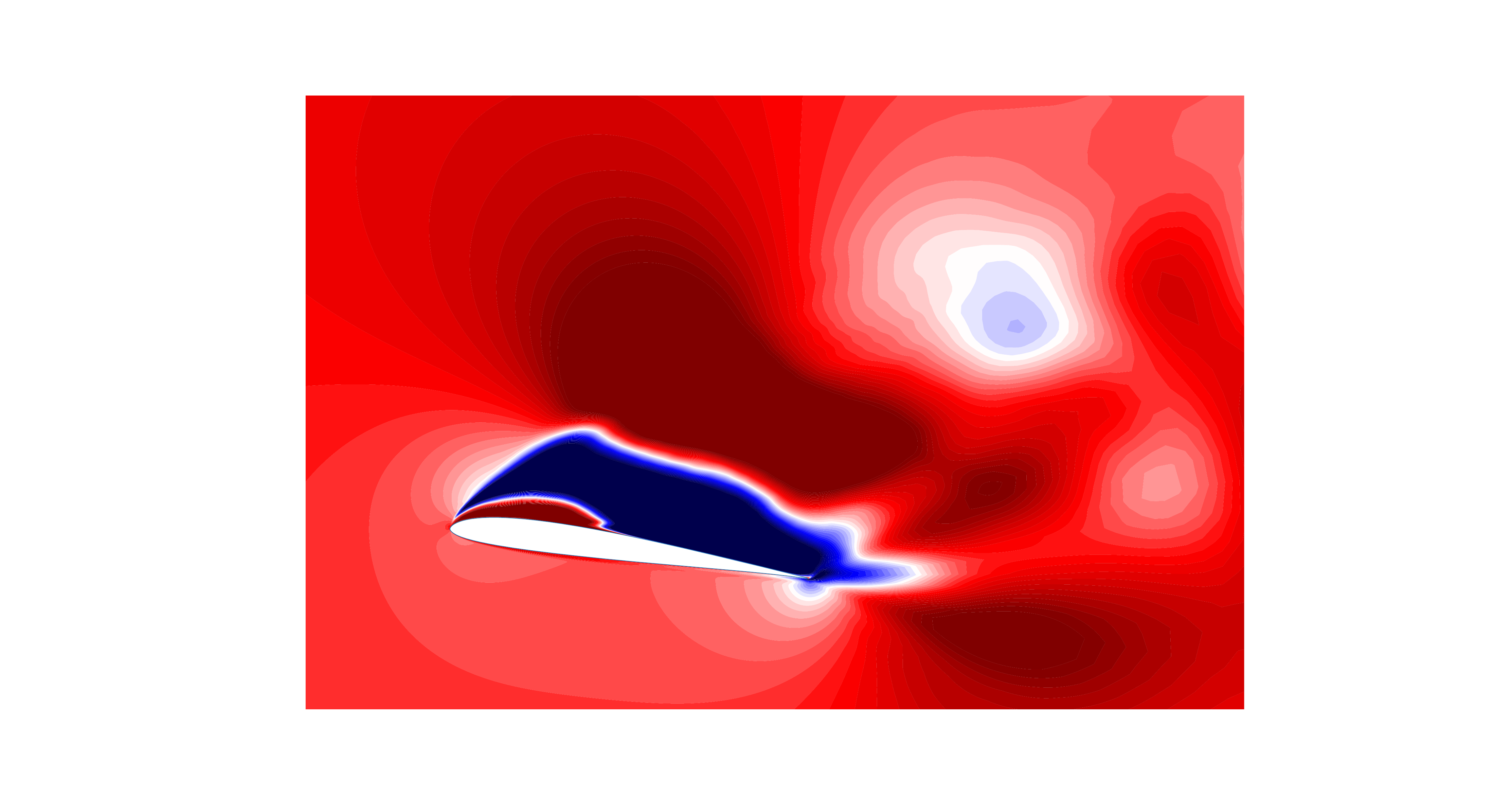}
        \caption{Non-linear 12-mode G-ROM.}
    \end{subfigure}
    \begin{subfigure}[hbt!]{.30\textwidth}
        \centering
        \includegraphics[width=.9\textwidth,trim={100mm 20mm 80mm 30mm},clip]{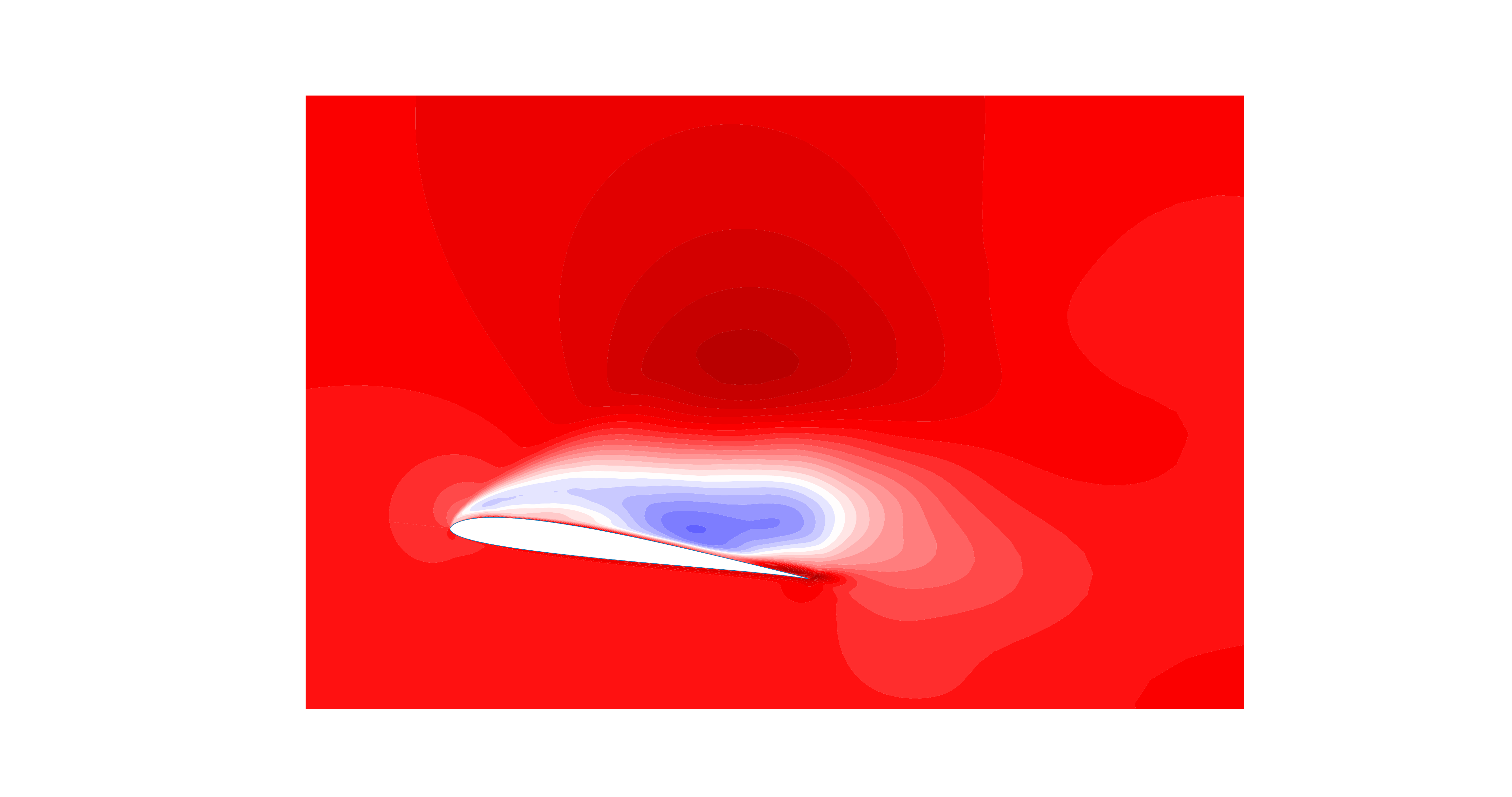}
        \caption{Linear 4-modes PG-ROM.}
    \end{subfigure}
    ~
    \begin{subfigure}[hbt!]{.30\textwidth}
        \centering
        \includegraphics[width=.9\textwidth,trim={100mm 20mm 80mm 30mm},clip]{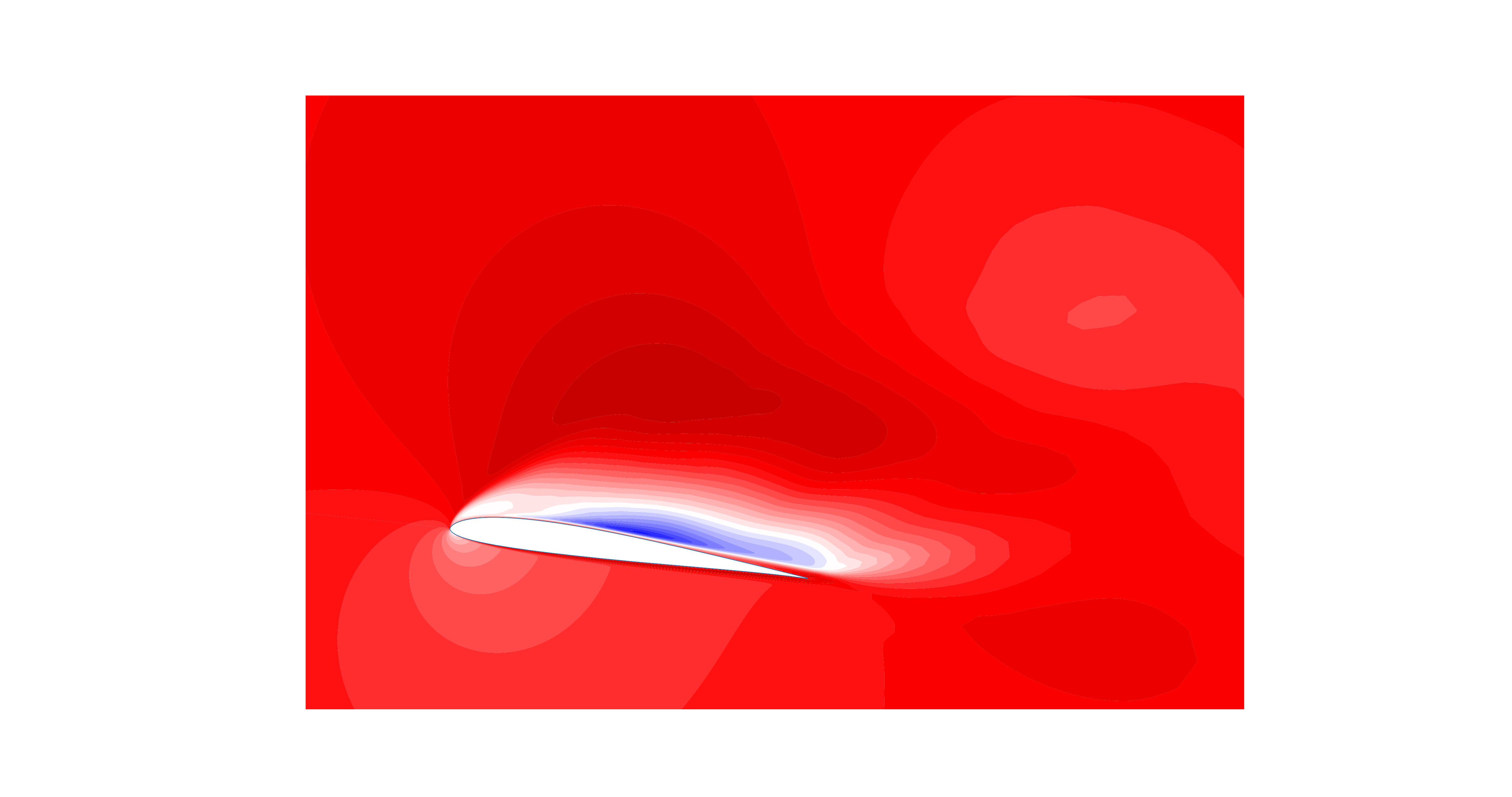}
        \caption{Linear 8-modes PG-ROM.}
    \end{subfigure}
    ~
    \begin{subfigure}[hbt!]{.30\textwidth}
        \centering
        \includegraphics[width=.9\textwidth,trim={100mm 20mm 80mm 30mm},clip]{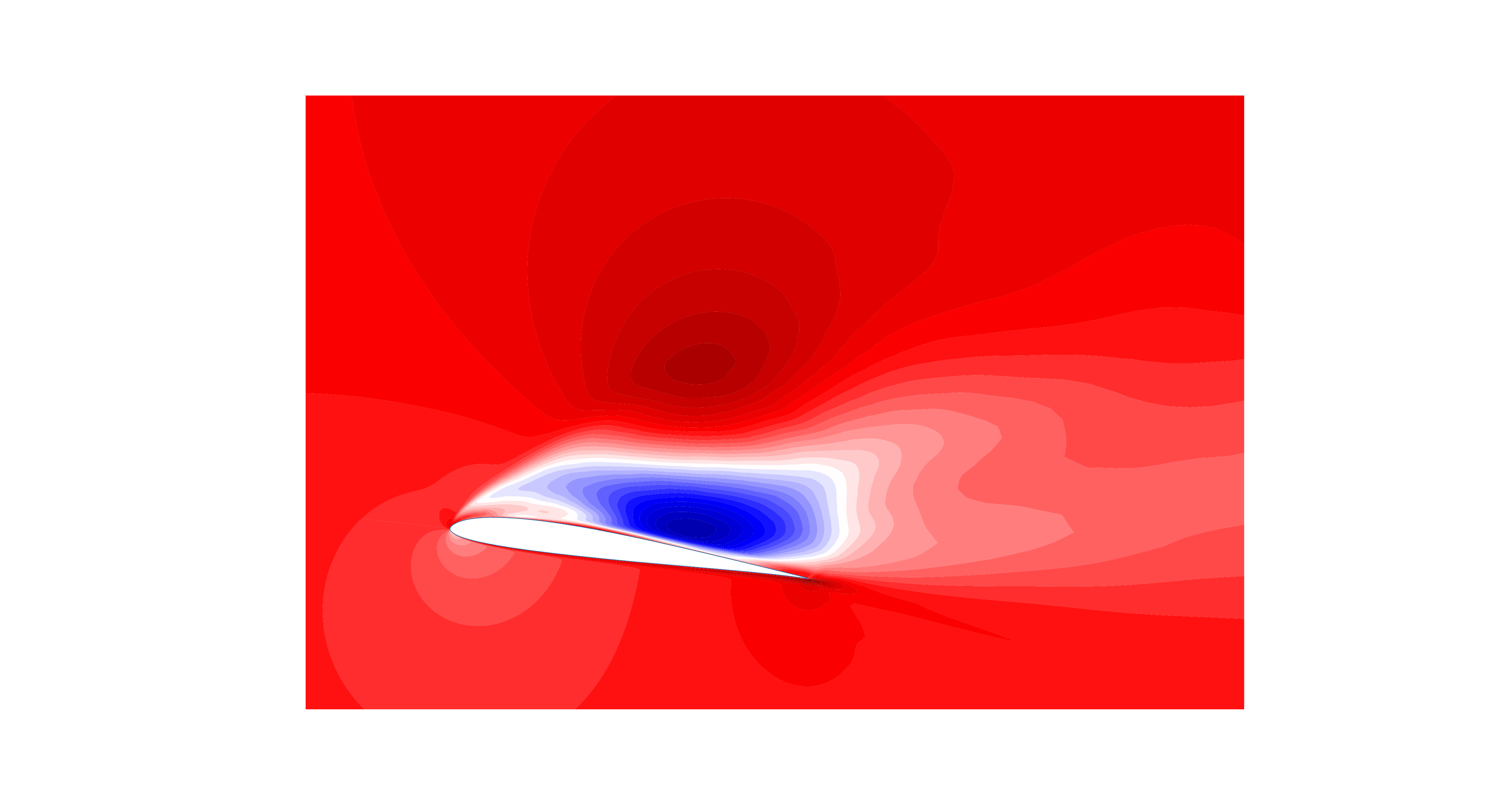}
        \caption{Linear 12-modes PG-ROM.}
    \end{subfigure}
    \begin{subfigure}[hbt!]{.30\textwidth}
        \centering
        \includegraphics[width=.9\textwidth,trim={100mm 20mm 80mm 30mm},clip]{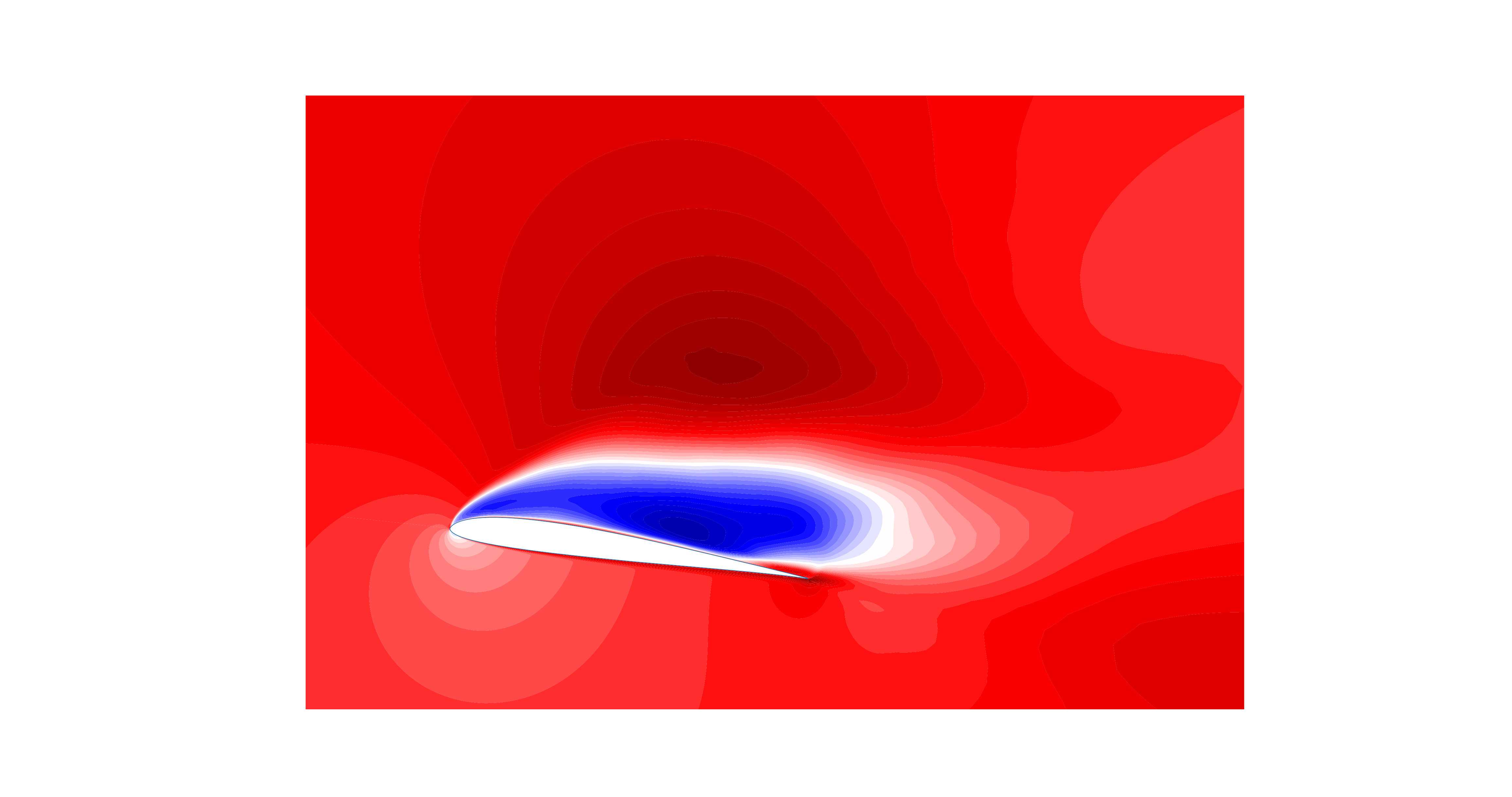}
        \caption{Non-linear 4-modes PG-ROM.}
    \end{subfigure}
    ~
    \begin{subfigure}[hbt!]{.30\textwidth}
        \centering
        \includegraphics[width=.9\textwidth,trim={100mm 20mm 80mm 30mm},clip]{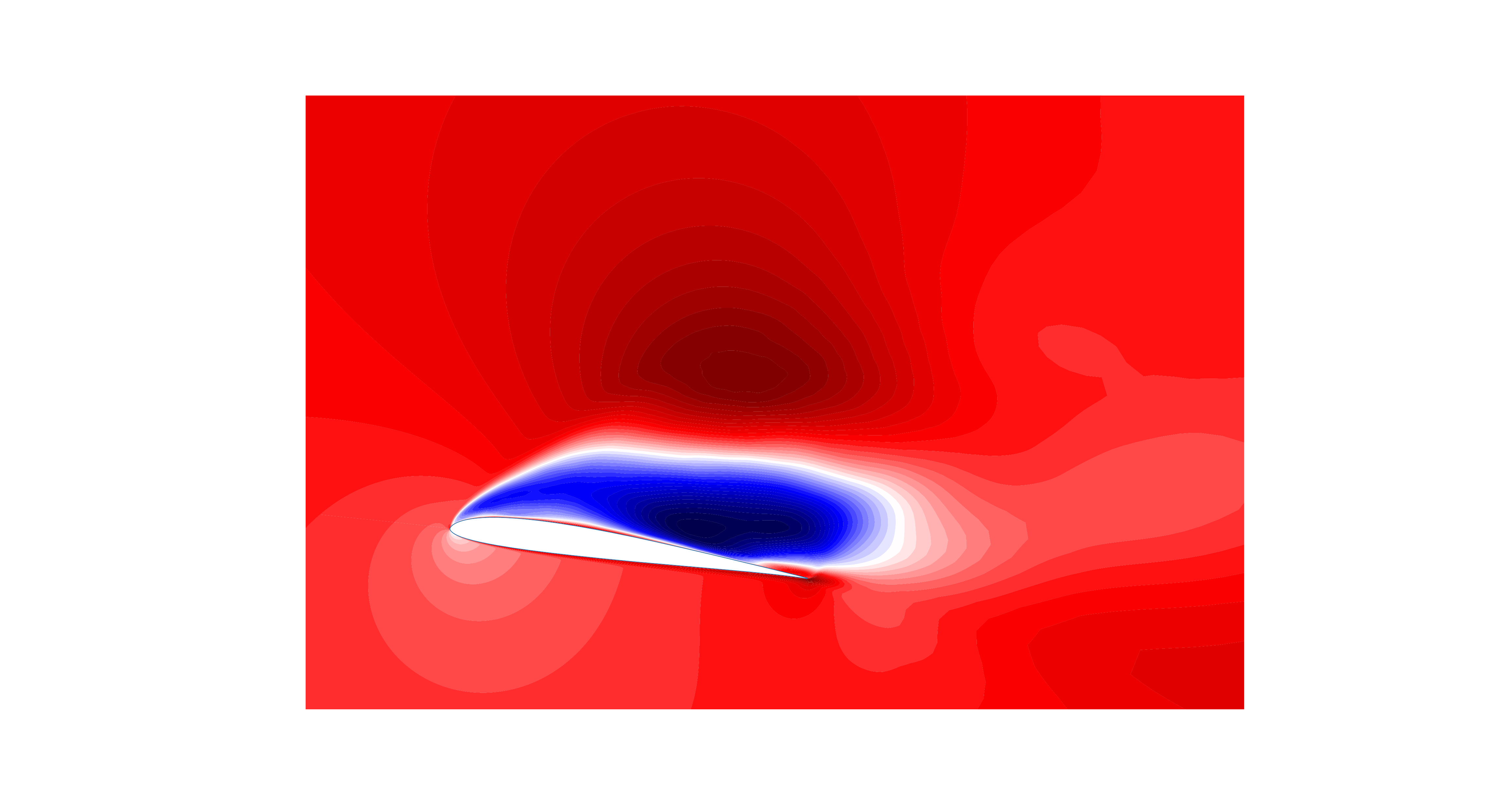}
        \caption{Non-linear 8-modes PG-ROM.}
    \end{subfigure}
    ~
    \begin{subfigure}[hbt!]{.30\textwidth}
        \centering
        \includegraphics[width=.9\textwidth,trim={100mm 20mm 80mm 30mm},clip]{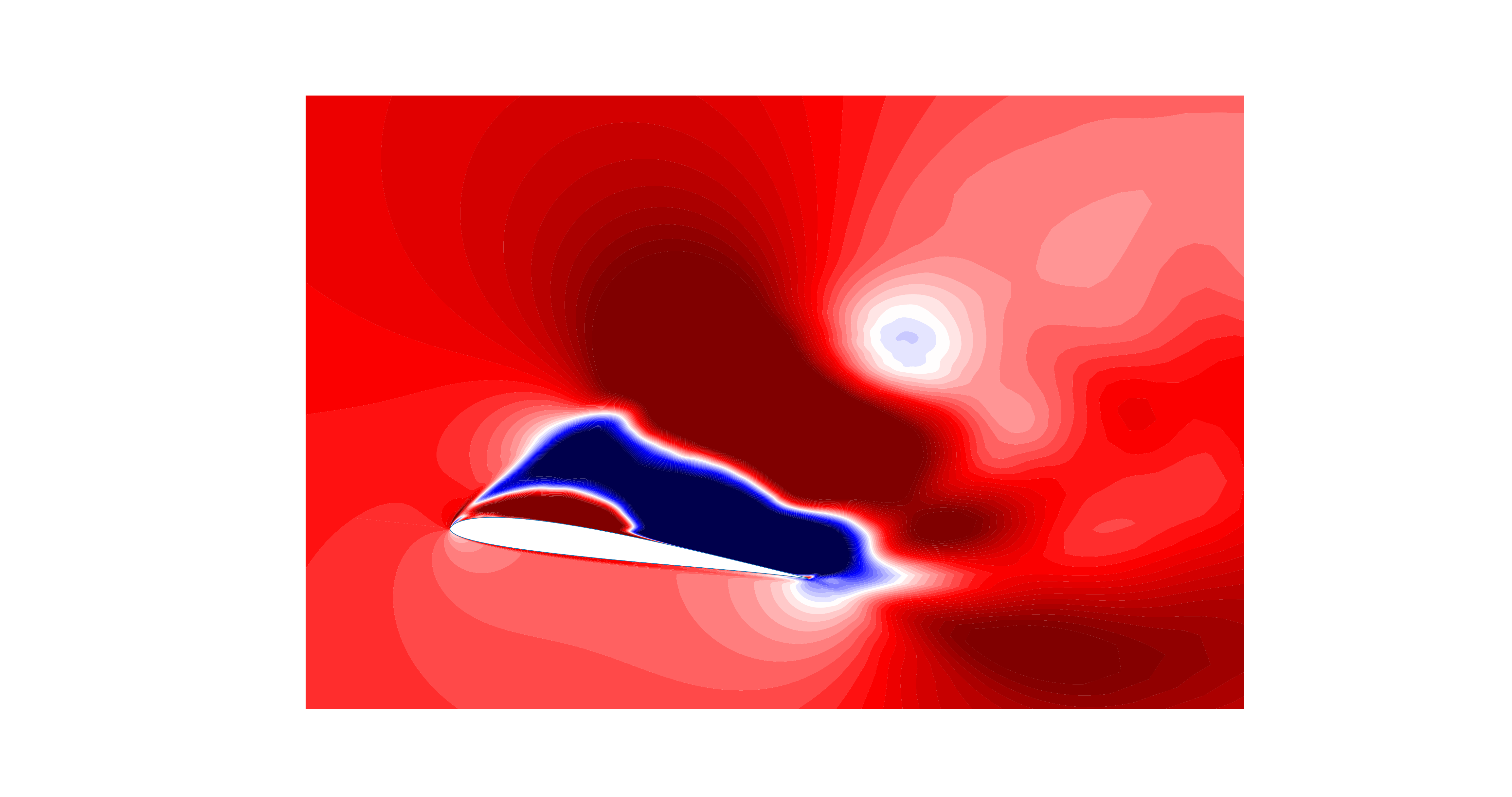}
        \caption{Non-linear 12-modes PG-ROM.}
    \end{subfigure}
    \begin{subfigure}[hbt!]{.30\textwidth}
        \centering
        \includegraphics[width=.9\textwidth,trim={100mm 20mm 80mm 30mm},clip]{u_snap_fom.pdf}
        \caption{Full order model.}
    \end{subfigure}
    \caption{Contours of u-velocity fluctuations computed at $t = 67.5$ using calibrated hyper-reduced models with implicit Euler integration. The terms ``G'' and ``PG'' stand for Galerkin and LSPG models, respectively.}
    \label{fig:airfoil_snapshots_hyper}
\end{figure}

Snapshots of u-velocity fluctuations at $t = 67.5$ are presented in Figs. \ref{fig:airfoil_snapshots} and \ref{fig:airfoil_snapshots_hyper}. The first figure allows a comparison of results between the FOM and the different gapless calibrated ROMs with an increasing number of POD modes used in the reconstruction. Similarly, the second figure can be used to observe some of the effects of hyper-reduction on the solutions.
For this time instant and, without employing hyper-reduction, the main features of the flow are recovered when using non-linear calibration coefficients for all cases considered. The further addition of POD modes to these ROMs improves accuracy to some degree but not significantly. On the other hand, solutions obtained with linear calibration coefficients display a poorer performance when only 4 modes are used in the reconstruction. However, results show that they benefit from a more complete basis of POD modes. In particular, they are capable of delivering results of similar accuracy compared to those of the non-linear method. For example, linear and non-linear calibration solutions are similar when 12 modes are used in the reconstruction with the advantage of linear models not requiring regularization. 
On the other hand, performance deteriorates for most cases when hyper-reduction is applied. Fortunately, cheap hyper-reduced models with non-linear calibration coefficients show good results for smaller POD bases (4 and 8 modes). However, adding further modes to these models is inefficient since they become inaccurate and unstable. Furthermore, linear methods perform poorly for all hyper-reduced cases considered.

\section{Conclusions}

In this work, a data-driven closure methodology is applied to projection-based ROMs. The closure approach is based on calibration of model solutions using the POD temporal modes. The performance of  calibration is assessed for both the Galerkin and least-squares Petrov-Galerkin techniques for two unsteady compressible flows. The first is the canonical compressible flow past a cylinder and it serves for testing and verification purposes. The second test case involves the turbulent flow over a plunging airfoil undergoing deep dynamic stall. This problem is particularly challenging because of the wide range of spatial and temporal flow scales.

The ROMs are obtained by projection of the non-conservative form of the Navier-Stokes equations in order to reduce the polynomial complexity of the non-linear ODEs arising in the models. This form of the Navier-Stokes equations facilitates an offline/online procedure where all grid-dependent calculations are performed only once during the offline step. In order to achieve further cost reductions in the ROMs' construction, hyper-reduction is also applied. The consequences of an additional layer of approximation by an accelerated greedy MPE method is implemented and evaluated for the deep dynamic stall problem. Results show that cheap and accurate ROMs can be obtained when hyper-reduction is combined with non-linear calibration coefficients and a smaller POD basis. However, for these cases, the calibration operators are an order of magnitude more intrusive than those obtained for gapless ROMs.

In the absence of closure, the cylinder flow ROMs display a high amplitude error. The model solutions are very dissipative, with the LSPG method being more accurate. This can be explained by the choice of a first-order implicit Euler time-marching scheme and can be diminished by time step refinement. Linear and non-linear calibration coefficients are both capable of correcting the amplitude error without changing the time step, and the resulting calibrated model solutions are visually indistinguishable from the full order model. In all calibration cases, resulting operators have relatively low intrusiveness. Also, non-linear calibration leads to L-curves with clear corners, resulting in an easy choice of adequate closure operators.

For the dynamic stall problem, non-calibrated ROMs fail completely in recovering the flow dynamics despite the stable solutions achieved by the implicit Euler time-marching scheme. Differently from the first problem, non-linear calibration is far superior to that performed exclusively with linear coefficients. This is specially true when a smaller POD basis is considered. The differences between linear and non-linear calibration solutions are reduced as more POD modes are added to the model bases. This is particularly observed for gapless ROMs since hyper-reduced calibrated models show stability issues when higher POD modes are used in their reconstruction. Eventually, further adding modes  makes all models unstable as this type of closure does not guarantee stability or accuracy for long-term temporal integration. One should also remind that the present dynamic stall problem is obtained by a large eddy simulation of a turbulent flow and, in this case, the limited sampling frequency resolution of snapshots directly impacts the convergence of higher POD modes, impairing the calibration procedure. For this case, determining an adequate level of regularization is also more complicated given that the L-curves do not display clear corners. Despite these issues, present results show that non-linear calibration with few POD modes can still deliver stable and accurate solutions for this complex unsteady flow problem.

\section*{Acknowledgments}

The authors of this work would like to acknowledge Fun\-da\-\c{c}\~{a}o de Amparo \`{a} Pesquisa do Estado de S\~{a}o Paulo, FAPESP, for supporting the present work under research grants No.\ 2013/08293-7, 2018/11410-9 and 2019/18809-7 and Conselho Nacional de Desenvolvimento Científico e Tecnológico, CNPq, for supporting this research under grants No.\ 407842/2018-7 and 304335/2018-5.
We also thank SDUMONT-LNCC (Project SimTurb) and CENAPAD-SP (Project 551) for providing the computational resources used in this work.

 \bibliographystyle{elsarticle-num} 
 \bibliography{mybibfile}





\end{document}